\documentclass[journal]{IEEEtran}
\usepackage[T1]{fontenc}
\usepackage[latin9]{inputenc}
\usepackage{amsthm}
\usepackage{amsmath} 
\usepackage{graphicx} 
\usepackage{color} 
\usepackage[usenames,dvipsnames]{xcolor}
\usepackage{cite}
\usepackage{algpseudocode}
\usepackage{algorithm}
\usepackage{amssymb}
\usepackage{subfigure}
\usepackage{esint}
\usepackage{algpseudocode}
\usepackage{epstopdf}
\usepackage{multirow}
\usepackage{makecell}
\usepackage{float}
\usepackage{lipsum}
\usepackage{balance}

\usepackage{graphicx}
\usepackage{epstopdf}
\usepackage{midfloat}
\usepackage{amsfonts}
\usepackage{amssymb}
\usepackage{amsthm}
\usepackage{times,color}
\usepackage{algorithm}
\usepackage{amsmath}
\newtheorem{definition}{Definition}
\usepackage{flushend}
\newtheorem{theorem}{Theorem}
\newtheorem{lemma}{Lemma}
\newtheorem{corollary}{Corollary}

\newtheorem{remark}{Remark}
\newtheorem{example}{Example}
\usepackage[T1]{fontenc}
\usepackage[latin9]{inputenc}
\usepackage{amsthm}
\usepackage{amsmath}
\usepackage{graphicx}
\usepackage{color}
\usepackage{cite}
\usepackage[T1]{fontenc} 
\usepackage[latin9]{inputenc} 
\usepackage{amsthm}
\usepackage{amsmath} 
\usepackage{graphicx} 
\usepackage{color}
\usepackage{cite}
\usepackage{multirow}
\usepackage{algpseudocode}
\usepackage{algorithm}
\usepackage{colortbl}
\usepackage{amssymb}
\usepackage{subfigure}
\usepackage{esint}
\usepackage{verbatim}

\usepackage{amssymb}
\usepackage{subfigure}
\usepackage[export]{adjustbox}
\usepackage{esint}
\usepackage{algpseudocode}

\theoremstyle{plain}
\theoremstyle{plain}

\definecolor{brown}{rgb}{0.6, 0.2, 0.0}

\normalsize

\begin{document}

\title{A Combinatorial Methodology for Optimizing Non-Binary Graph-Based Codes: Theoretical Analysis and Applications in Data Storage}

\author{Ahmed~Hareedy,~\IEEEmembership{Student~Member,~IEEE,} Chinmayi~Lanka,~\IEEEmembership{Student~Member,~IEEE,} Nian~Guo,~\IEEEmembership{Student~Member,~IEEE,} and~Lara~Dolecek,~\IEEEmembership{Senior~Member,~IEEE}

\thanks{A. Hareedy and L. Dolecek are with the Department of Electrical and Computer Engineering, University of California, Los Angeles (UCLA), Los Angeles, CA 90095 USA. E-mails: ahareedy@ucla.edu; dolecek@ee.ucla.edu. C. Lanka (currently with MathWorks Inc.) and N. Guo (currently with California Institute of Technology) were with UCLA when this research was performed. E-mails: chinmayi.lanka@mathworks.com; nguo@caltech.edu.
}
}

\maketitle

\setlength{\skip\footins}{1.7em}

%%%%%%%%%%%%%%%%%%%%%%%%%%%%%%
\begin{abstract}

Non-binary (NB) low-density parity-check (LDPC) codes are graph-based codes that are increasingly being considered as a powerful error correction tool for modern dense storage devices. Optimizing NB-LDPC codes to overcome their error floor is one of the main code design challenges facing storage engineers upon deploying such codes in practice. Furthermore, the increasing levels of asymmetry incorporated by the channels underlying modern dense storage systems, e.g., multi-level Flash systems, exacerbates the error floor problem by widening the spectrum of problematic objects that contributes to the error floor of an NB-LDPC code. In a recent research, the weight consistency matrix (WCM) framework was introduced as an effective combinatorial NB-LDPC code optimization methodology that is suitable for modern Flash memory and magnetic recording (MR) systems. The WCM framework was used to optimize codes for asymmetric Flash channels, MR channels that have intrinsic memory, in addition to canonical symmetric additive white Gaussian noise channels. In this paper, we provide an in-depth theoretical analysis needed to understand and properly apply the WCM framework. We focus on general absorbing sets of type two (GASTs) as the detrimental objects of interest. In particular, we introduce a novel tree representation of a GAST called the unlabeled GAST tree, using which we prove that the WCM framework is optimal in the sense that it operates on the minimum number of matrices, which are the WCMs, to remove a GAST. Then, we enumerate WCMs and demonstrate the significance of the savings achieved by the WCM framework in the number of matrices processed to remove a GAST. Moreover, we provide a linear-algebraic analysis of the null spaces of WCMs associated with a GAST. We derive the minimum number of edge weight changes needed to remove a GAST via its WCMs, along with how to choose these changes. Additionally, we propose a new set of problematic objects, namely oscillating sets of type two (OSTs), which contribute to the error floor of NB-LDPC codes with even column weights on asymmetric channels, and we show how to customize the WCM framework to remove OSTs. We also extend the domain of the WCM framework applications by demonstrating its benefits in optimizing column weight 5 codes, codes used over Flash channels with additional soft information, and spatially-coupled codes. The performance gains achieved via the WCM framework range between 1 and nearly 2.5 orders of magnitude in the error floor region over interesting channels.

\end{abstract}


%%%%%%%%%%%%%%%%%%%%%%%%%%%%%%
\section{Introduction}\label{sec_intro}

Modern dense storage devices, e.g., multi-level Flash and magnetic recording (MR) devices, operate at very low frame error rate (FER) values, motivating the need for strong error correction techniques. Because of their capacity approaching performance, low-density parity-check (LDPC) codes \cite{gal_th, rich_urb1, rich_urb2, ahh_smm} are becoming the first choice for many of the modern storage systems \cite{ahh_jsac, aslm_fl, maeda_fl, jia_sym, kin_sym, ahh_bas, shafa, yuval1, yuval2}. Under iterative quantized decoding, LDPC codes suffer from the error floor problem, which is a change in the FER slope that undermines the chances of reaching desirable very low FER levels \cite{lara_floor, bani_est, b_ryan, hu_sim, tom_ferr}. It was demonstrated in the literature that absorbing sets (ASs), which are detrimental subgraphs in the Tanner graph of the LDPC code, are the principal cause of the error floor problem \cite{lara_as, behzad_elem}. There are other works that studied different classes of detrimental objects, specifically, stopping sets \cite{decl_nb} and trapping sets \cite{olgica1}, \cite{vas_trap1}. Research works investigating the error floor problem of LDPC codes include \cite{b_ryan, lara_as, olgica1, vas_trap1, jia_cyc, olgica2, siegel_flr, bani_trap, vas_trap2, shu_flr1, shu_flr2}.

Particularly for non-binary LDPC (NB-LDPC) codes, the authors in \cite{behzad_elem} used concepts from \cite{decl_nb} to study non-binary elementary absorbing sets (EASs), and showed that EASs are the detrimental objects which contribute the most to the error floor of NB-LDPC codes over the canonical additive white Gaussian noise (AWGN) channel. The observation that the combinatorial structure of the dominant detrimental objects critically depends on the characteristics of the channel of interest was first introduced in \cite{ahh_glc} and then discussed in \cite{ahh_bas}; we introduced balanced absorbing sets (BASs) and demonstrated their dominance in the error floor of NB-LDPC codes over partial-response (PR) channels, which exemplify 1-D MR channels \cite{vas_prc, cola_pr}. Motivated by the asymmetry possessed by practical Flash channels \cite{mit_nl, cai_fl}, in a recent research \cite{ahh_jsac}, we introduced general absorbing sets (GASs) and general absorbing sets of type two (GASTs) to capture the dominant problematic objects over realistic Flash channels. GASs and GASTs subsume previously introduced AS subclasses (namely EASs and BASs).

In \cite{behzad_elem} and \cite{ahh_bas}, NB-LDPC code optimization algorithms tailored to AWGN and PR channels, respectively, were proposed. While the weight consistency matrix (WCM) framework introduced in \cite{ahh_jsac} was originally motivated by the need to optimize NB-LDPC codes for asymmetric Flash channels, we customized this methodology to be suitable for channels with memory (e.g., PR channels), canonical symmetric channels (e.g., AWGN channels), as well as practical Flash channels, achieving at least $1$ order of magnitude performance gain over all these channels. The principal idea of the WCM framework is representing a problematic object, e.g., a GAST, using a small set of matrices, called WCMs. Since problematic objects in an NB-LDPC code are described in terms of both their weight conditions as well as their topological conditions, there are explicit weight properties associated with the WCMs of an object. By changing the null spaces of the WCMs associated with an object such that the weight conditions of all these WCMs are broken \cite{ahh_jsac}, this problematic object is removed from the Tanner graph of the code. A key feature of the WCM framework is that the GASTs removal process is performed solely via manipulating the edge weights of the Tanner graph of the NB-LDPC code, which consequently preserves all the structural topological properties of the code being optimized.

For NB-LDPC codes with fixed column weights (fixed variable node degrees), our contributions in this paper are:
\begin{enumerate}
\item We characterize GASTs via their WCMs. In particular, we define the unlabeled GAST tree to describe the underlying topology of a GAST, where the leaves of this tree represent the WCMs of the GAST. Using this tree, we prove the optimality of the WCM framework by demonstrating that the framework indeed operates on the minimum possible number of matrices to remove the detrimental object. We also deploy concepts from graph theory and combinatorics to compute the exact number of WCMs associated with a GAST in different cases. We further compare the number of matrices the WCM framework operates on with the number of matrices a suboptimal idea works with, showing the significant reduction (up to about $90\%$) achieved by the WCM framework in the cases of interest.

\item Based on tools from graph theory and linear algebra, we propose a comprehensive analysis of the removal process of GASTs. We start off with discussing the dimensions of the null spaces of WCMs; these null spaces play the central role in the identification and removal of a GAST. Then, we derive the best that can be done to process a short WCM (a WCM that has fewer rows than columns) during the GAST removal process. Finally, we provide the minimum number of edge weight changes\footnote{In the WCM framework, a GAST is removed via careful processing of the weights of its edges (the original and the new weights are not zeros). Throughout this paper, the edge weight changes are always with respect to the original configuration.} needed to remove a GAST, along with how to select the edges and the new weights to guarantee the removal of the GAST through its WCMs.

\item We introduce new combinatorial objects that capture the majority of the non-GAST detrimental objects in the error floor region of NB-LDPC codes that have even column weights over asymmetric Flash channels. We define oscillating sets (OSs) and oscillating sets of type two (OSTs). Furthermore, we expand the analysis of GASTs in \cite{ahh_jsac} to cover OSTs, describing how the WCM framework can be customized to remove OSTs, after GASTs have been removed, to achieve additional performance gains.

\item We extend the scope of the WCM framework by using it to optimize codes with different properties and for various applications. Specifically, we show that despite the good error floor performance of NB-LDPC codes with column weight $5$ before optimization, more than $1$ order of magnitude gain in the uncorrectable bit error rate (UBER) over practical Flash channels is achievable via the WCM framework. We further apply the theoretical concepts in item 3 for NB-LDPC codes with column weight $4$ over practical Flash channels to achieve overall UBER gains up to nearly $2.5$ orders of magnitude. Additionally, we optimize NB-LDPC codes for practical Flash channels with more soft information ($6$ reads). We also use the WCM framework to optimize NB-LDPC codes with irregular check node (CN) degrees and fixed variable node (VN) degrees; we show that more than $1$ order of magnitude performance gain in the FER is achievable by optimizing spatially-coupled (SC) codes \cite{kud_sc, lent_asy, andr_asy, pus_sc, olm_sc, iye_sc} used over PR and AWGN channels.
\end{enumerate}

The rest of the paper is organized as follows. Section~\ref{sec_sum} summarizes the main concepts of the WCM framework. Then, Section~\ref{sec_cch} discusses the characterization of GASTs through their WCMs, in addition to the optimality proof and the WCMs enumeration. In Section~\ref{sec_rem} we detail our analysis for the process of the GAST removal through WCMs. Afterwards, Section~\ref{sec_os} discusses OSTs and how to customize the WCM framework to remove them. The simulation results are presented in Section~\ref{sec_sim}. Finally, the paper is concluded in Section~\ref{sec_conc}.

%%%%%%%%%%%%%%%%%%%%%%%%%%%%%%
\section{Summary of the WCM Framework}\label{sec_sum}

In this section, along with Appendices~\ref{sec_appa} and \ref{sec_appb}, we provide a brief summary of the main concepts and ideas of the WCM framework for the sake of clarity and completeness. Details of the WCM framework were introduced in \cite{ahh_jsac}.

Consider the Tanner graph of an LDPC code. An $(a, b)$ AS in this graph is defined as a set of $a$ VNs with $b$ unsatisfied CNs connected to it such that each VN is connected to strictly more satisfied than unsatisfied CNs, for some set of VN values (these $a$ VNs have non-zero values, while the remaining VNs are set to zero) \cite{lara_as, behzad_elem}. For canonical channels, e.g., the AWGN channel, it was shown that elementary ASs (EASs) are the objects that dominate the error floor region of NB-LDPC codes \cite{behzad_elem}. EASs have the property that all satisfied CNs are of degree $2$, and all unsatisfied CNs are of degree $1$. The different characteristics of storage channels (compared with the AWGN channel) result in changing the combinatorial properties of detrimental objects in NB-LDPC codes simulated over such channels.

Asymmetry, as in Flash channels \cite{mit_nl, cai_fl}, can result in VN errors having high magnitudes, which is typically not the case for canonical channels. These VN errors with high magnitudes make it very difficult for unsatisfied CNs with degree $2$ to participate in correcting an AS error. Consequently, it becomes more likely to have AS errors with degree-$2$ unsatisfied CNs, which are non-elementary AS errors. This was the motivation behind introducing GASs and GASTs in \cite{ahh_jsac} to capture the objects that dominate the error floor region of NB-LDPC codes over asymmetric channels (e.g., Flash channels).

The intrinsic memory, as in PR channels \cite{ahh_bas, vas_prc}, can also result in VN errors having high magnitudes. Moreover, the global iterations (detector-decoder iterations) help the decoder correct AS errors with higher numbers of unsatisfied CNs. Thus, the objects that dominate the error floor region of NB-LDPC codes simulated over PR channels can also have unsatisfied CNs with degree $2$ (non-elementary), and they have a fewer number of unsatisfied (particularly degree-$1$) CNs, which is the reason why they are called ``balanced''. BASs and BASs of type two (BASTs) were introduced in \cite{ahh_bas} and \cite{ahh_jsac} to capture such detrimental objects.

We start off with the definitions of a GAS and an unlabeled GAS \cite{ahh_jsac}.

\vspace{-0.1em}
\begin{definition}\label{def_gas}
Consider a subgraph induced by a subset $\mathcal{V}$ of VNs in the Tanner graph of an NB-LDPC code. Set all the VNs in $\mathcal{V}$ to values $\in$ GF($q$)$\setminus \{0\}$ and set all other VNs to $0$. The set $\mathcal{V}$ is said to be an $(a, b, b_2, d_1, d_2, d_3)$ \textbf{general absorbing set (GAS)} over GF($q$) if and only if the size of $\mathcal{V}$ is $a$, the number of unsatisfied (resp., degree-$2$ unsatisfied) CNs connected to $\mathcal{V}$ is $b$ (resp., $b_2$), the number of degree-$1$ (resp., $2$ and $> 2$) CNs connected to $\mathcal{V}$ is $d_1$ (resp., $d_2$ and  $d_3$), and each VN in $\mathcal{V}$ is connected to strictly more satisfied than unsatisfied neighboring CNs, for some set of VN values.
\end{definition}

Let $\gamma$ be the column weight (VN degree) of the NB-LDPC code. BASs are GASs with $0 \leq b \leq \left \lfloor \frac{ag}{2} \right \rfloor$, where $g = \left \lfloor \frac{\gamma-1}{2} \right \rfloor$ \cite{ahh_bas}. GF refers to Galois field, and $q$ is the GF size (order). We focus here on the case of $q=2^\lambda$, where $\lambda$ is a positive integer $\geq 2$. Furthermore, when we say in this paper that nodes are ``connected'', we mean they are ``directly connected'' or they are ``neighbors'', unless otherwise stated. The same applies conceptually when we say an edge is ``connected'' to a node or vice versa.

\vspace{-0.1em}
\begin{definition}\label{def_ugas}
Let  $\mathcal{V}$ be a subset of VNs in the unlabeled Tanner graph of an NB-LDPC code. Let $\mathcal{O}$ (resp., $\mathcal{T}$ and $\mathcal{H}$) be the set of degree-$1$ (resp., $2$ and $> 2$) CNs connected to $\mathcal{V}$. This graphical configuration is an \textbf{$(a, d_1, d_2, d_3)$ unlabeled GAS} if it satisfies the following two conditions:
\begin{enumerate}
\item {$|\mathcal{V}| = a$, $\vert{\mathcal{O}}\vert=d_1$, $\vert{\mathcal{T}}\vert=d_2$, and $\vert{\mathcal{H}}\vert=d_3$.}
\item Each VN in $\mathcal{V}$ is connected to more neighbors in $(\mathcal{T} \cup \mathcal{H})$ than in $\mathcal{O}$.
\end{enumerate}
\end{definition}

In this paper, all vectors are column vectors, except the cutting vectors of SC codes and the equalization target of the PR channel (see Subsection \ref{subsec_sc}).

\begin{figure}
\vspace{-0.5em}
\center
\includegraphics[width=2.6in]{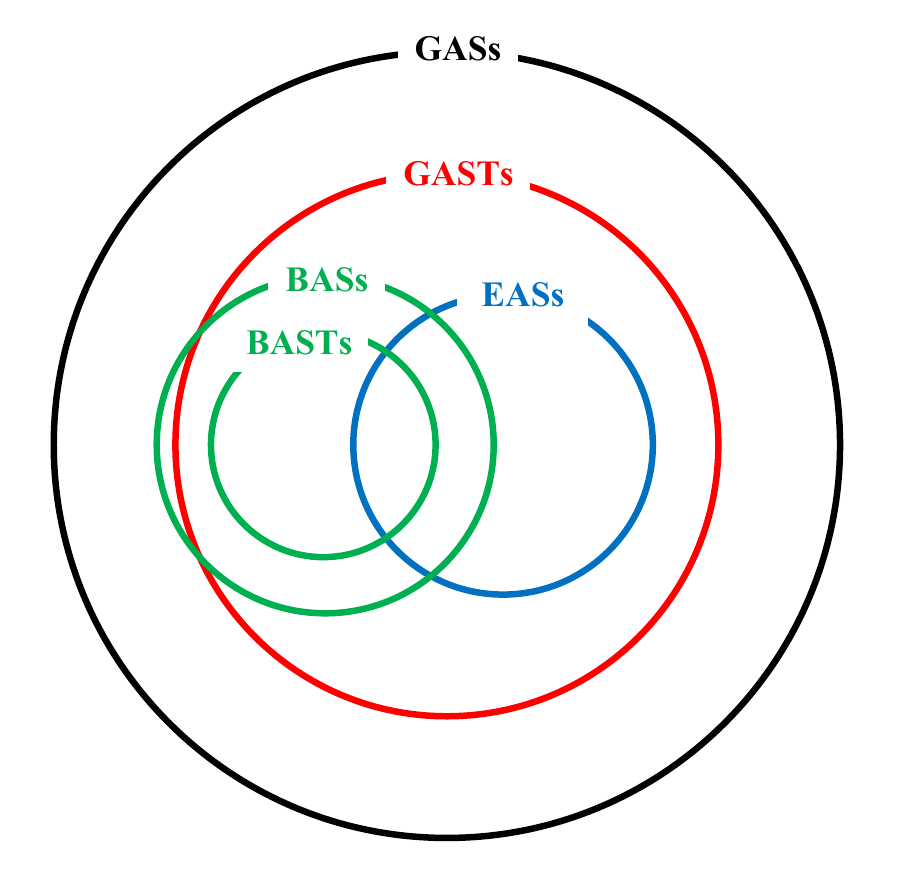}\vspace{-1.2em}
\caption{The relation between different types of absorbing sets demonstrated via a Venn diagram.}
\label{Figure_1}
\vspace{-0.2em}
\end{figure}

Let $\bold{H}$ denote the parity-check matrix of an NB-LDPC code defined over GF($q$). Consider an $(a,b,b_2,d_1,d_2,d_3)$ GAS in the Tanner graph of this code. Let $\bold{A}$ be the $\ell \times a$ submatrix of $\bold{H}$ that consists of $\ell=d_1+d_2+d_3$ rows of  $\bold{H}$, corresponding to the CNs participating in this GAS, and $a$ columns of $\bold{H}$, corresponding to the VNs participating in this GAS. From \cite[Lemma~1]{ahh_jsac}, an $(a, b, b_2, d_1, d_2, d_3)$ GAS must satisfy:
\begin{itemize}
\item \textbf{Topological conditions:} Its unlabeled configuration must satisfy the {unlabeled GAS} conditions stated in Definition~\ref{def_ugas}.
\item \textbf{Weight conditions:} The set is an $(a, b, b_2, d_1, d_2, d_3)$ GAS over GF($q$) if and only if there exists an $(\ell-b) \times a$ submatrix $\bold{W}$ of column rank $\tau_{\bold{W}} < a$, with elements $\psi_{e,f}$, $1 \leq e \leq (\ell-b)$, $1 \leq f \leq a$, of the GAS adjacency matrix $\bold{A}$, that satisfies the following two conditions:
\begin{enumerate}
\item Let $\mathcal{N}(\bold{W})$ be the null space of the submatrix $\bold{W}$, and let $\bold{d}_k^\textup{T}$, $1 \leq k \leq b$, be the $k$th row of the matrix $\bold{D}$ obtained by removing the rows of $\bold{W}$ from $\bold{A}$. Let $\bold{v}$ be a vector of VN values and $\bold{R}$ be an $\ell \times \ell$ permutation matrix. Then,
\begin{align}\label{eq_gas_cond1}
&\exists \text{ } \bold{v}=[v_1 \text{ } v_2 \text{ } \dots \text{ } v_a]^\textup{T} \in \mathcal{N}(\bold{W}) \text{ s.t. } v_f \neq 0, \nonumber \\ &\text{ } \forall f \in \{1, 2, \dots , a\}, \text{ and } \bold{d}_k^\textup{T} \bold{v} = m_k \neq 0, \nonumber \\ &\text{ } \forall k \in \{1, 2, \dots , b\}, \text{ } \bold{m}=[m_1 \text{ } m_2 \text{ } \dots \text{ } m_b]^\textup{T}, \nonumber \\ &\text{i.e.}, \text{ } \bold{RAv}=\begin{bmatrix}\bold{W}_{(\ell-b) \times a}\\ \bold{D}_{b \times a}\end{bmatrix}\bold{v}_{a \times 1}=\begin{bmatrix}\bold{0}_{(\ell-b) \times 1}\\ \bold{m}_{b \times 1}\end{bmatrix}.
\end{align}
\item Let $\theta_{k,f}$, $1 \leq k \leq b$, $1 \leq f \leq a$, be the elements of the matrix $\bold{D}$. Then, $\forall f \in \{1, 2, \dots , a\}$,
\begin{align}\label{eq_gas_cond2}
\left ( \sum\limits_{e=1}^{\ell-b}F\left ( \psi_{e,f} \right ) \right ) > \left ( \sum\limits_{k=1}^{b}F\left ( \theta_{k,f} \right ) \right ),
\end{align}
where $F(\beta)=0$ if $\beta=0$, and $F(\beta)=1$ otherwise.
\end{enumerate}
Computations are performed over GF($q$).
\end{itemize}

In words, $\bold{W}$ is the submatrix of satisfied CNs, and $\bold{D}$ is the submatrix of unsatisfied CNs. Next, we define an important subclass of GASs, which are GASTs.

\begin{definition}\label{def_gast}
A GAS that has $d_2 > d_3$ and all the unsatisfied CNs connected to it (if any) belong to $(\mathcal{O} \cup \mathcal{T})$ (i.e., having either degree $1$ or degree $2$) is defined as an $(a, b, d_1, d_2, d_3)$ \textbf{general absorbing set of type two (GAST)}. Similar to the {unlabeled GAS} definition (Definition~\ref{def_ugas}), we also define the \textbf{$(a, d_1, d_2, d_3)$ unlabeled GAST}.
\end{definition}

A BAST is also a BAS with all the unsatisfied CNs connected to it (if any) having either degree $1$ or degree $2$.

As demonstrated by the Venn diagram in Fig.~\ref{Figure_1}, the set of GASs (resp., GASTs) subsumes the sets of EASs and BASs (resp., BASTs). Moreover, while the WCM framework was originally introduced to remove GASTs from the Tanner graph of an NB-LDPC code, it can be easily customized to efficiently remove EASs and BASTs \cite{ahh_jsac} depending on the application, as we shall discuss shortly in brief.

The three theorems essential for understanding the WCM framework are in \cite{ahh_jsac}. We recall from \cite[Theorem~2]{ahh_jsac} that, given an $(a, d_1, d_2, d_3)$ unlabeled GAST, the maximum number of unsatisfied CNs, $b_{\textup{max}}$, in the resulting GAST after edge labeling is upper bounded by:
\begin{equation}\label{eq_bmax}
b_{\textup{max}} \leq d_1 + b_{\textup{ut}}, \text{ where}
\end{equation}
\begin{equation}\label{eq_but}
b_{\textup{ut}} = \left \lfloor \frac{1}{2} \left ( a\left \lfloor \frac{\gamma-1}{2} \right \rfloor - d_1 \right ) \right \rfloor.
\end{equation}
\vspace{-0.2em}

Here, $b_{\textup{ut}}$ is the upper bound on the maximum number of degree-$2$ unsatisfied CNs the resulting GAST can have. Because of the structure of the underlying unlabeled configuration, sometimes the exact maximum (obtained by \cite[Algorithm~1]{ahh_jsac}, see Appendix~\ref{sec_appa}) is a quantity smaller than $b_{\textup{ut}}$. We refer to this exact maximum as $b_{\textup{et}}$. Thus,
\begin{equation}\label{eq_bmax_exact}
b_{\textup{max}} = d_1 + b_{\textup{et}}.
\end{equation}

Throughout this paper, the notation ``ut'' (resp., ``et'') in the subscript of $b$ refers to the \textit{upper bound on the} (resp., \textit{exact}) maximum number of \textit{degree-$2$} unsatisfied CNs.

For a given $(a, b, d_1, d_2, d_3)$ {GAST}, let  $\mathcal{Z}$ be the set of all $(a, b', d_1, d_2, d_3)$ GASTs with $d_1 \leq b' \leq b_{\textup{max}}$, which have the same {unlabeled GAST} as the original $(a, b, d_1, d_2, d_3)$ {GAST}. Here, $b_{\textup{max}}$ is the largest allowable number of unsatisfied CNs for these configurations.

\begin{figure}
\vspace{-0.5em}
\center
\includegraphics[width=3.6in]{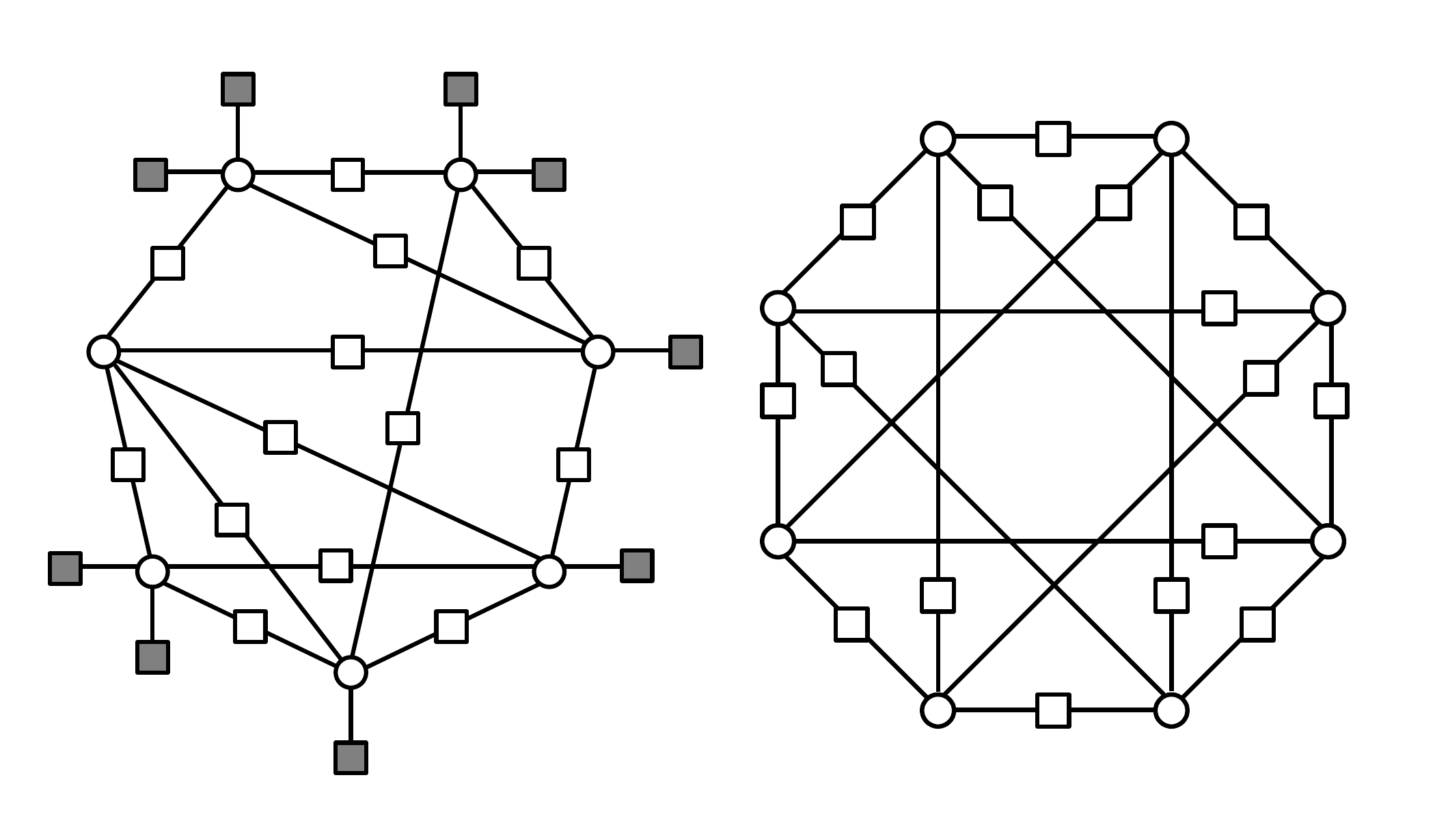}\vspace{-0.7em}
\text{\hspace{1.5em} \footnotesize{(a) \hspace{14em} (b) \hspace{1em}}}
\caption{(a) A $(7, 9, 13, 0)$ unlabeled GAST for $\gamma=5$. (b) An $(8, 0, 16, 0)$ unlabeled GAST for $\gamma=4$.}
\label{Figure_2}
\vspace{-0.6em}
\end{figure}

\begin{definition}\label{def_drem_gast}
An $(a, b, d_1, d_2, d_3)$ \textbf{GAST} is said to be \textbf{removed} from the Tanner graph of an NB-LDPC code if and only if the resulting object (after edge weight processing) $\notin \mathcal{Z}$.
\end{definition}

In all GAST, unlabeled GAST, and OST figures, circles represent VNs. In all GAST and OST figures, grey (resp., white) squares represent unsatisfied (resp., satisfied) CNs. In all unlabeled GAST figures, grey (resp., white) squares represent degree-$1$ (resp., $> 1$) CNs.

\begin{example}\label{ex_prime}
Fig.~\ref{Figure_2}(a) shows a $(7, 9, 13, 0)$ unlabeled GAST for $\gamma=5$. For this unlabeled GAST, $d_1=9$, and from (\ref{eq_but}), $b_{\textup{ut}}=\left \lfloor \frac{1}{2} \left ( 7 \left \lfloor \frac{5-1}{2} \right \rfloor -9 \right ) \right \rfloor=2=b_{\textup{et}}$, which means the resulting GAST after edge labeling can have up to $2$ degree-$2$ unsatisfied CNs. Thus, $b_{\textup{max}}=9+2=11$, and $9 \leq b' \leq 11$. Consequently, $\mathcal{Z}=\{(7, 9, 9, 13, 0), (7, 10, 9, 13, 0), (7, 11, 9, 13, 0)\}$.

Fig.~\ref{Figure_2}(b) shows an $(8, 0, 16, 0)$ unlabeled GAST for $\gamma=4$. For this unlabeled GAST, $d_1=0$, and from (\ref{eq_but}), $b_{\textup{ut}}=\left \lfloor \frac{1}{2} \left ( 8 \left \lfloor \frac{4-1}{2} \right \rfloor -0 \right ) \right \rfloor=4=b_{\textup{et}}$, which means the resulting GAST after edge labeling can have up to $4$ degree-$2$ unsatisfied CNs. Thus, $b_{\textup{max}}=0+4=4$, and $0 \leq b' \leq 4$. Consequently, $\mathcal{Z}=\{(8, 0, 0, 16, 0), (8, 1, 0, 16, 0), (8, 2, 0, 16, 0), \allowbreak (8, 3, 0, 16, 0), (8, 4, 0, 16, 0)\}$.
\end{example}

For a given GAST, define a matrix $\bold{W}^{\textup{z}}$ to be the matrix obtained by removing $b'$, $d_1 \leq b' \leq b_{\textup{max}}$, rows corresponding to CNs $\in (\mathcal{O} \cup \mathcal{T})$ from the matrix $\bold{A}$, the GAST adjacency matrix. These $b'$ CNs can simultaneously be unsatisfied under some edge labeling that produces a GAST which has the same {unlabeled GAST} as the given GAST. Let $\mathcal{U}$ be the set of all such matrices $\bold{W}^{\textup{z}}$. Each element in $\mathcal{Z}$ has one or more matrices in $\mathcal{U}$.

\begin{definition}\label{def_wcms}
For a given $(a,b,d_1,d_2,d_3)$ GAST and its associated adjacency matrix $\bold{A}$ and its associated set $\mathcal{Z}$, we construct a set of $t$ matrices as follows:
\begin{enumerate}
\item Each matrix $\bold{W}_h^{\textup{cm}}$, $1 \leq h \leq t$, in this set is an $(\ell-b^{\textup{cm}}_h)\times a$ submatrix, $d_1 \leq b^{\textup{cm}}_h \leq b_{\textup{max}}$, formed by removing \textbf{different} $b^{\textup{cm}}_h$ rows from the $\ell \times a$ matrix $\bold{A}$ of the GAST. These $b^{\textup{cm}}_h$ rows to be removed correspond to CNs $\in (\mathcal{O} \cup \mathcal{T})$ that can simultaneously be unsatisfied under some edge labeling that produces a GAST which has the same {unlabeled GAST} as the given GAST.
\item Each matrix $\bold{W}^{\textup{z}} \in \mathcal{U}$, for every element $\in \mathcal{Z}$, contains at least one element of the resultant set as its submatrix.
\item This resultant set has the \textbf{smallest cardinality}, which is $t$, among all the sets which satisfy conditions 1 and 2 stated above.
\end{enumerate} 
We refer to the matrices in this set as \textbf{weight consistency matrices (WCMs)}, and to this set itself as $\mathcal{W}$.
\end{definition}
\vspace{-0.3em}

Throughout this paper, the notation ``z'' (resp., ``cm'') in the superscript of a matrix means that the matrix is \textit{associated with an element in the set $\mathcal{Z}$} (resp., \textit{a WCM}).

\begin{figure}
\vspace{-0.5em}
\center
\includegraphics[width=1.8in]{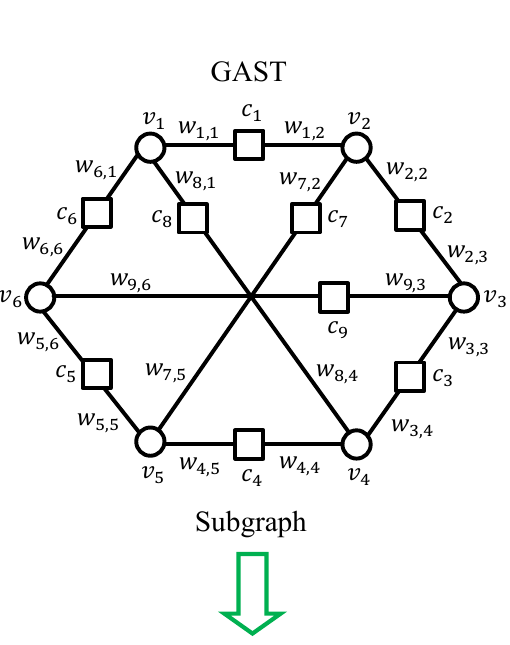}
\includegraphics[width=3.6in]{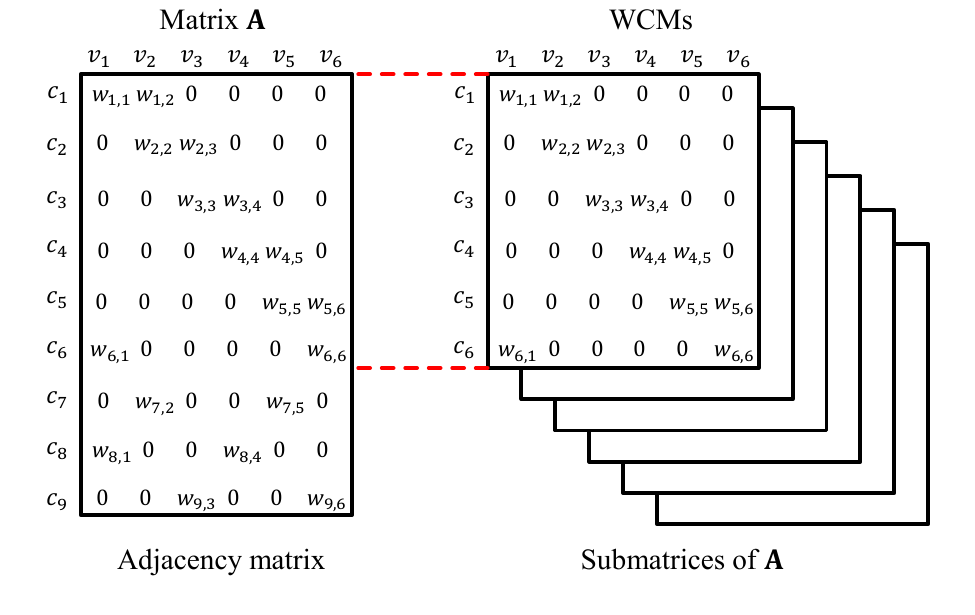}
\vspace{-2.5em}
\caption{An illustrative figure showing the process of extracting the WCMs of a $(6, 0, 0, 9, 0)$ GAST. Appropriate edge weights ($w$'s) $\in$ GF($q$)$\setminus \{0\}$ are assumed.}
\label{Figure_3}
\vspace{-0.7em}
\end{figure}

\begin{definition}\label{def_bet_bst}
Parameter $b_{\textup{et}}$ represents the \textbf{exact maximum number} of rows corresponding to degree-$2$ CNs that can be removed together from $\bold{A}$ to extract a WCM. Similarly, we define $b_{\textup{st}}$ to be the \textbf{exact minimum number} of rows corresponding to degree-$2$ CNs that can be removed together from $\bold{A}$ to extract a WCM. Recall that the rows corresponding to degree-$1$ CNs are always removed while extracting a WCM. Thus, $d_1 \leq d_1+b_{\textup{st}} \leq b^{\textup{cm}}_h \leq d_1+b_{\textup{et}}=b_{\textup{max}}$. Both $b_{\textup{st}}$ and $b_{\textup{et}}$ depend on the unlabeled GAST configuration.
\end{definition}

Fig.~\ref{Figure_3} depicts the relation between a GAST and its associated WCMs, and roughly describes how the WCMs of this GAST are extracted.

The core idea of the WCM framework is in \cite[Theorem~3]{ahh_jsac}. This theorem states that the necessary and sufficient processing needed to remove an $(a, b, d_1, d_2, d_3)$ GAST, according to Definition~\ref{def_drem_gast}, is to change the edge weights such that $\forall h$:
\vspace{-0.1em}\begin{align}\label{eq_rem_cond}
&\text{If } \text{ } \mathcal{N}(\bold{W}^{\textup{cm}}_h)=\textup{span}\{\bold{x}_1, \bold{x}_2, \dots ,\bold{x}_{p_h}\}, \text{ then } \nonumber \\ &\nexists \text{ } \bold{r}=[r_1 \text{ } r_2 \text{ } \dots \text{ } r_{p_h}]^\textup{T} \text{ for} \text{ }
\bold{v}=r_1\bold{x}_1+r_2\bold{x}_2+\dots+r_{p_h}\bold{x}_{p_h} \nonumber \\ &= [v_1 \text{ } v_2 \text{ } \dots v_a]^\textup{T} \text{ s.t. } v_j \neq 0, \text{ } \forall j \in \{1, 2, \dots , a\},
\end{align}
where $p_h$ is the dimension of $\mathcal{N}(\bold{W}^{\textup{cm}}_h)$. Computations are performed over GF($q$).

The WCM framework is easily adjusted to efficiently remove special subclasses of GASTs, namely EASs and BASTs, by customizing the WCM definition. In particular, via replacing $b_{\textup{max}}$ by $b_{\textup{e\_max}}=d_1$ (resp., $b_{\textup{b\_max}}=\left \lfloor \frac{ag}{2} \right \rfloor$ (see \cite{ahh_bas}), where $g=\left \lfloor \frac{\gamma-1}{2} \right \rfloor$), the WCM framework is customized to remove EASs (resp., BASTs), which are the dominant objects in the case of AWGN (resp., PR) channels. More details can be found in \cite{ahh_jsac}.

The two algorithms that constitute the WCM framework are \cite[Algorithm~1]{ahh_jsac}, which is the WCM extraction algorithm, and \cite[Algorithm~2]{ahh_jsac}, which is the code optimization algorithm. The steps of the two algorithms are listed in Appendices~\ref{sec_appa} and \ref{sec_appb} for the reference of the reader.

A WCM that has (\ref{eq_rem_cond}) satisfied is said to be a WCM with \textbf{broken weight conditions}. A GAST is removed if and only if all its WCMs have broken weight conditions.

Note that the complexity of the process of removing a specific GAST using the WCM framework is mainly controlled by the number of WCMs, which is $t$, of that GAST (see the \textbf{for} loop in Step~12 of \cite[Algorithm~2]{ahh_jsac}). Thus, the complexity of the WCM framework depends on the size of the set $\mathcal{G}$ (see Appendix~\ref{sec_appb}) and the numbers of WCMs of the GASTs in $\mathcal{G}$.

%%%%%%%%%%%%%%%%%%%%%%%%%%%%%%
\section{Characterizing GASTs through Their WCMs}\label{sec_cch}

In order to characterize a GAST through its WCMs, we introduce the definition of the GAST tree, which will also be used to derive all the results in this section. Since this tree does not depend on the edge weights of the configuration, we call it the unlabeled GAST tree.

Recall that $\bold{A}$ is the adjacency matrix of the GAST. Both $\bold{W}^{\textup{z}}$ and $\mathcal{U}$ are defined in the paragraph before Definition~\ref{def_wcms}. Recall also that $b_{\textup{et}}$ is the maximum number of degree-$2$ CNs that can be unsatisfied simultaneously while the object remains a GAST. Define $u^0$ as the number of degree-$2$ CNs that can be unsatisfied individually while the object remains a GAST, and $\bold{y}^0$ as the vector in which the indices of such $u^0$ CNs are saved. Note that we always have $b_{\textup{et}} \leq u^0$.

\begin{definition}\label{def_gast_tree}
For a given $(a, d_1, d_2, d_3)$ unlabeled GAST with $b_{\textup{et}} > 0$, we construct the \textbf{unlabeled GAST tree} of $b_{\textup{et}}$ levels (level $0$ is not counted) as follows:
\begin{itemize}
\item Except the root node at level $0$, each tree node represents a degree-$2$ CN in the unlabeled GAST. For any two CNs in the tree, being neighbors means that they can be unsatisfied simultaneously after labeling and the resulting object remains a GAST.
\item Let $i_1, i_2, \dots, i_{b_{\textup{et}}}$ be the running indices used to access nodes at different levels in the tree as follows. The index of a node at level $j$, $1 \leq j \leq b_{\textup{et}}$, is saved in $\bold{y}_{i_1,i_2, \dots, i_{j-1}}^{j-1}$ and given by $y_{i_1,i_2, \dots, i_{j-1}}^{j-1}(i_j)$. CN $c_{y_{i_1,i_2, \dots, i_{j-1}}^{j-1}(i_j)}$ at level $j$ is accessed via the path of nodes ``root node -- $c_{y^{0}(i_1)}$ -- $c_{y_{i_1}^{1}(i_2)}$ -- $c_{y_{i_1,i_2}^{2}(i_3)}$ -- \dots -- $c_{y_{i_1,i_2, \dots, i_{j-1}}^{j-1}(i_j)}$''.
\item At level $0$, a virtual root node is assumed to be connected to the $u^0$ nodes with indices in $\bold{y}^0$ at level $1$. Level $j$ of the tree consists of all the nodes with indices in $\bold{y}_{i_1,i_2, \dots, i_{j-1}}^{j-1},$ $\forall i_1, i_2, \dots, i_{j-1}$. Level $j+1$ of the tree is created as follows. Each CN $c_{y_{i_1,i_2, \dots, i_{j-1}}^{j-1}(i_j)}$ at level $j$ is connected to all the CNs with indices in $\bold{y}_{i_1,i_2, \dots, i_j}^j$ at level $j+1$. These CNs can each be -- simultaneously with the nodes on the path from the root node until $c_{y_{i_1,i_2, \dots, i_{j-1}}^{j-1}(i_j)}$ -- unsatisfied after labeling and the resulting object remains a GAST.
\item The number of nodes at level $j+1$ that are connected to $c_{y_{i_1,i_2, \dots, i_{j-1}}^{j-1}(i_j)}$ at level $j$ is $u_{i_1,i_2, \dots, i_j}^{j}$ (which is the size of the vector $\bold{y}_{i_1,i_2, \dots, i_{j}}^{j}$), with $u_{i_1,i_2, \dots, i_{j}}^{j} < u_{i_1,i_2, \dots, i_{j-1}}^{j-1}, \forall i_1, i_2, \dots, i_j$.
\item The leaves of this tree are linked to the matrices extracted by \cite[Algorithm~1]{ahh_jsac} before removing the repeated matrices (see Appendix~\ref{sec_appa}).
\end{itemize}
\end{definition}

Note that for the parameters $u$ and $\bold{y}$, the superscript refers to the level prior to the level in which the nodes exist, and the subscript refers to the running indices used to access the nodes. Note also that \cite[Algorithm~1]{ahh_jsac} is designed to generate the unlabeled GAST tree.

\begin{figure}
\vspace{-0.5em}
\center
\includegraphics[width=3.1in]{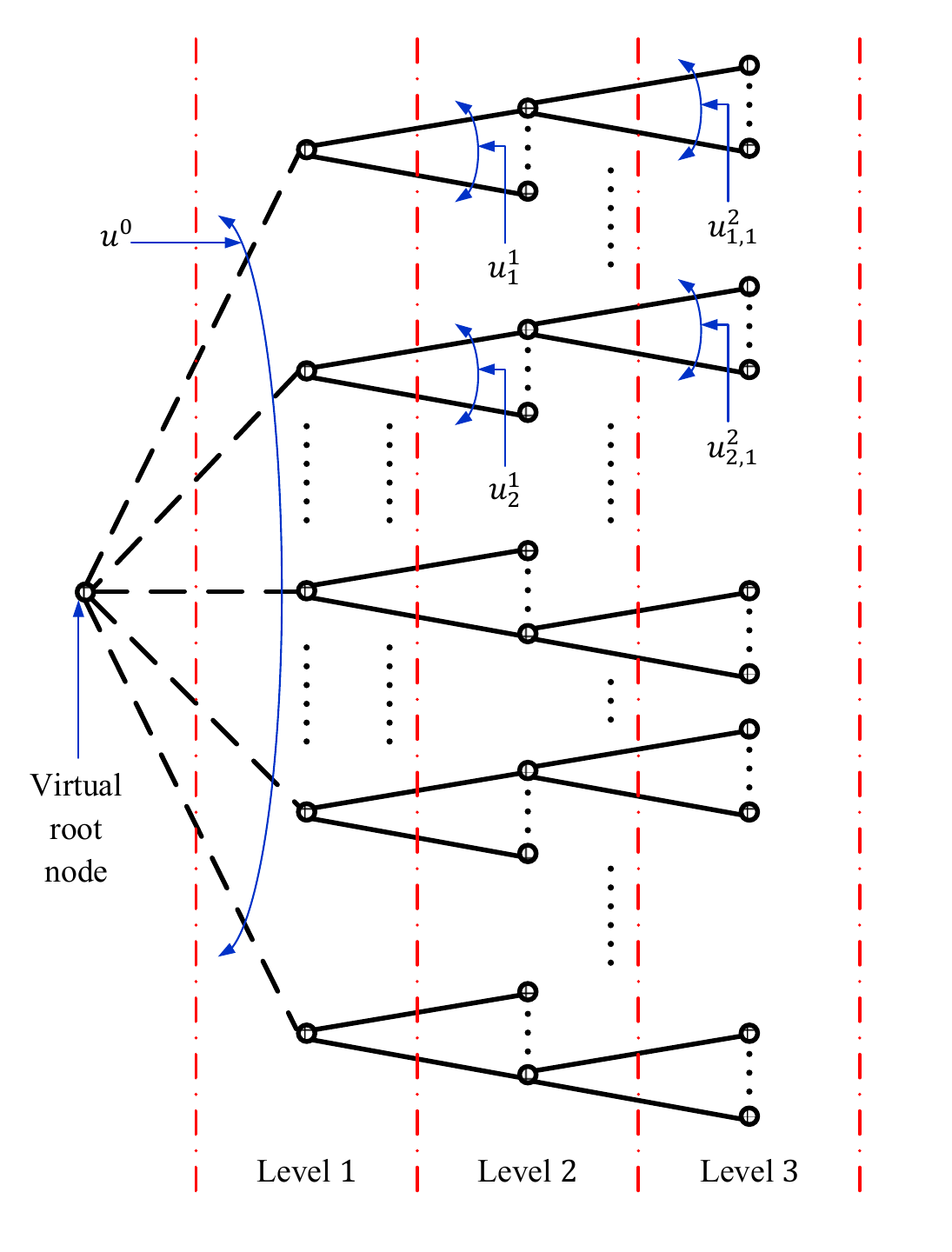}
\vspace{-1.5em}
\caption{An unlabeled GAST tree with $b_{\textup{et}}=3$.}
\label{Figure_4}
\vspace{-0.5em}
\end{figure}

Fig.~\ref{Figure_4} shows an unlabeled GAST tree for a configuration that has $b_{\textup{et}}=3$. The configuration has three levels after the root node. We say that each tree node at level $j$, $j > 0$, in the unlabeled GAST tree is \textit{\textbf{linked to}} a matrix $\bold{W}^{\textup{z}} \in \mathcal{U}$ extracted by removing $(d_1+j)$ rows from the matrix $\bold{A}$. These rows correspond to all the $d_1$ degree-$1$ CNs, and the $j$ degree-$2$ CNs on the path from the virtual root node to this tree node in the configuration. We also say that every valid matrix $\bold{W}^{\textup{z}} \in \mathcal{U}$ is \textit{\textbf{linked to}} one or more tree nodes.

It can be shown that $b_{\textup{et}}$ is the number of levels (nested loops in \cite[Algorithm~1]{ahh_jsac}), after which {$u^{b_{\textup{et}}}_{i_1,i_2, \dots, i_{b_{\textup{et}}}}=0$, $\forall i_1, i_2, \dots, i_{b_{\textup{et}}}$}.  Moreover, because the WCMs do not necessarily have the same row dimension, \cite[Algorithm~1]{ahh_jsac} may stop at $b_{\textup{k}}$ levels, $b_{\textup{k}} \leq b_{\textup{et}}$, starting from some $c_{y^0(i_1)}$, which results in an $(\ell-b^{\textup{cm}}_h)\times a$ WCM with $b^{\textup{cm}}_h = d_1+b_{\textup{k}} \leq b_{\textup{max}} = d_1+b_{\textup{et}}$. The smallest value of $b_{\textup{k}}$ is $b_{\textup{st}}$, i.e., $b_{\textup{st}} \leq b_{\textup{k}} \leq b_{\textup{et}}$.

\begin{remark}\label{rem_1}
Note that the unlabeled GAST tree is unique for a given unlabeled configuration. In other words, two non-isomorphic $(a, d_1, d_2, d_3)$ configurations have two different unlabeled GAST trees even though they have the same $a$, $d_1$, $d_2$, and $d_3$ parameters.
\end{remark}

Repetitions in tree nodes linked to matrices $\bold{W}^{\textup{z}}$ come from the fact that we are addressing the permutations and not the combinations in the tree. In other words, if we have a path from the root node at level $0$ to a tree node at level $2$ that has $c_1$ at level $1$ then $c_4$ at level $2$ on it, there must exist another path from the root node at level $0$ to another tree node at level $2$ that has $c_4$ at level $1$ then $c_1$ at level $2$ on it. Obviously, removing the row of $c_1$ then the row of $c_4$, or first $c_4$ then $c_1$ (in addition to the rows of degree-$1$ CNs) from $\bold{A}$ to extract a matrix produces the same end result.

%%%%%%%%%%%%%%%%%%%%%%%%%%%%%%
\subsection{Proving the Optimality of the WCM Framework}\label{subsec_optim}

The numbers of matrices needed to operate on for different GASTs control the complexity of the code optimization process. In this subsection, we prove that the WCM framework is optimal in the sense that it works on the minimum possible number of matrices to remove a GAST. Our optimization problem is formulated as follows:

\textbf{The optimization problem:} We seek to find the set $\mathcal{W}$ of matrices that has the minimum cardinality, with the matrices in $\mathcal{W}$ representing submatrices of $\bold{A}$ that can be used to remove the problematic GAST, without the need to work on other submatrices.

\textbf{The optimization constraint:} Each matrix in $\mathcal{W}$ has to be a valid $\bold{W}^{\textup{z}}$ matrix in $\mathcal{U}$.

The optimization constraint is set to ensure that we are performing not only sufficient, but also necessary processing to remove the object. Note that, by definition, the set of WCMs is the solution of this optimization problem. Thus, the problem of proving the optimality of the WCM framework reduces to proving that the matrices we extract by \cite[Algorithm~1]{ahh_jsac}, and operate on in \cite[Algorithm~2]{ahh_jsac} to remove the GAST, are indeed the WCMs.

Now, we are ready to present the optimality theorem and its proof.

\begin{theorem}\label{th_opt_frm}
Consider an $(a, b, d_1, d_2, d_3)$ GAST with $b_{\textup{et}} > 0$. After eliminating repetitions, the set of matrices which are linked to the leaves of the unlabeled GAST tree characterized by Definition~\ref{def_gast_tree} is the set of WCMs, i.e., the set $\mathcal{W}$ of minimum cardinality.
\end{theorem}

\begin{IEEEproof}
According to Definition~\ref{def_wcms}{, \cite[Theorem~3]{ahh_jsac}}, and its proof, each matrix $\bold{W}^{\textup{z}} \in \mathcal{U}$ must have at least one matrix $\in \mathcal{W}$ as its submatrix (as $\mathcal{W}$ is the set of WCMs). The relation between a matrix $\bold{W}^{\textup{z}}_1$ linked to tree node $1$ at level $j$ and a matrix $\bold{W}^{\textup{z}}_2$ linked to tree node $2$ at level $j+1$, provided that tree node $2$ is a child of tree node $1$, is as follows. Matrix $\bold{W}^{\textup{z}}_2$ is a submatrix of matrix $\bold{W}^{\textup{z}}_1$, extracted by removing one more row, that corresponds to a degree-$2$ CN (tree node $2$), from $\bold{W}^{\textup{z}}_1$. Following the same logic, the multiset of matrices, say $\mathcal{W}_{\textup{rep}}$, linked to tree nodes with no children (the leaves of the tree) contains submatrices of every possible matrix $\bold{W}^{\textup{z}}$. We let the set $\mathcal{W}_{\textup{nrep}}$ be $\mathcal{W}_{\textup{rep}}$ after eliminating the repetitions.

Now, we prove the sufficiency and minimality, which imply the optimality of $\mathcal{W}_{\textup{nrep}}$. The sufficiency is proved as follows. Any matrix $\bold{W}^{\textup{z}}$ that is linked to a tree node with a child will be redundant if added to the set $\mathcal{W}_{\textup{nrep}}$ because in $\mathcal{W}_{\textup{nrep}}$ there already exists a submatrix of this $\bold{W}^{\textup{z}}$ (from the analysis above). The minimality is proved as follows. If we eliminate any matrix from $\mathcal{W}_{\textup{nrep}}$, there will be at least one matrix $\bold{W}^{\textup{z}}$ that has no submatrices in $\mathcal{W}_{\textup{nrep}}$ (which is the eliminated matrix itself since it is linked to a node (nodes) with no children). Thus, we cannot further reduce the cardinality of $\mathcal{W}_{\textup{nrep}}$. Hence, the set $\mathcal{W}_{\textup{nrep}}$ is indeed the set $\mathcal{W}$ of WCMs, which proves the optimality of the WCM framework.
\end{IEEEproof}

%%%%%%%%%%%%%%%%%%%%%%%%%%%%%%
\subsection{Enumeration of WCMs Associated with a GAST}\label{subsec_enum}

In this subsection, we provide the exact number of distinct WCMs associated with a GAST. Moreover, we present particular examples where this number reduces to a combinatorial function of the column weight of the code. Since the number of WCMs (and also their sizes) associated with a GAST only depends on the unlabeled configuration and not on the edge weights, we relate the number of distinct WCMs, $t$, to the unlabeled GAST throughout this paper.

We first identify the following two types of unlabeled GAST configurations according to the properties of their unlabeled GAST trees.

\begin{definition}
An $(a, d_1, d_2, d_3)$ \textbf{same-size-WCMs} unlabeled GAST satisfies one of the following two conditions:
\begin{enumerate}
\item It has $b_{\textup{et}}=0$, i.e., $u^0=0$, which results in $\vert{\mathcal{W}}\vert=1$.
\item It has $b_{\textup{et}} > 0$; thus, $u^0 > 0$, and its tree has the property that $u_{i_1,i_2, \dots, i_j}^{j}=0$ only if $j=b_{\textup{et}}$, $\forall i_1, i_2, \dots, i_{b_{\textup{et}}}$, which results in all the WCMs having the same $(\ell-b_{\textup{max}}) \times a$ size, $b_{\textup{max}}=d_1+b_{\textup{et}}$.
\end{enumerate}
\end{definition}

\begin{definition}
An $(a, d_1, d_2, d_3)$ \textbf{u-symmetric} unlabeled GAST is a same-size-WCMs unlabeled GAST which satisfies the following condition. If $u^0 > 0$, its tree has the property that at any level $j$, $u_{i_1,i_2, \dots, i_{j-1}}^{j-1}$ is the same, $\forall i_1, i_2, \dots, i_{j-1}$.
\end{definition}

\begin{figure}
\vspace{-0.5em}
\center
\includegraphics[trim={0.08in 0.0in 0.5in 0.0in},clip,width=3.5in]{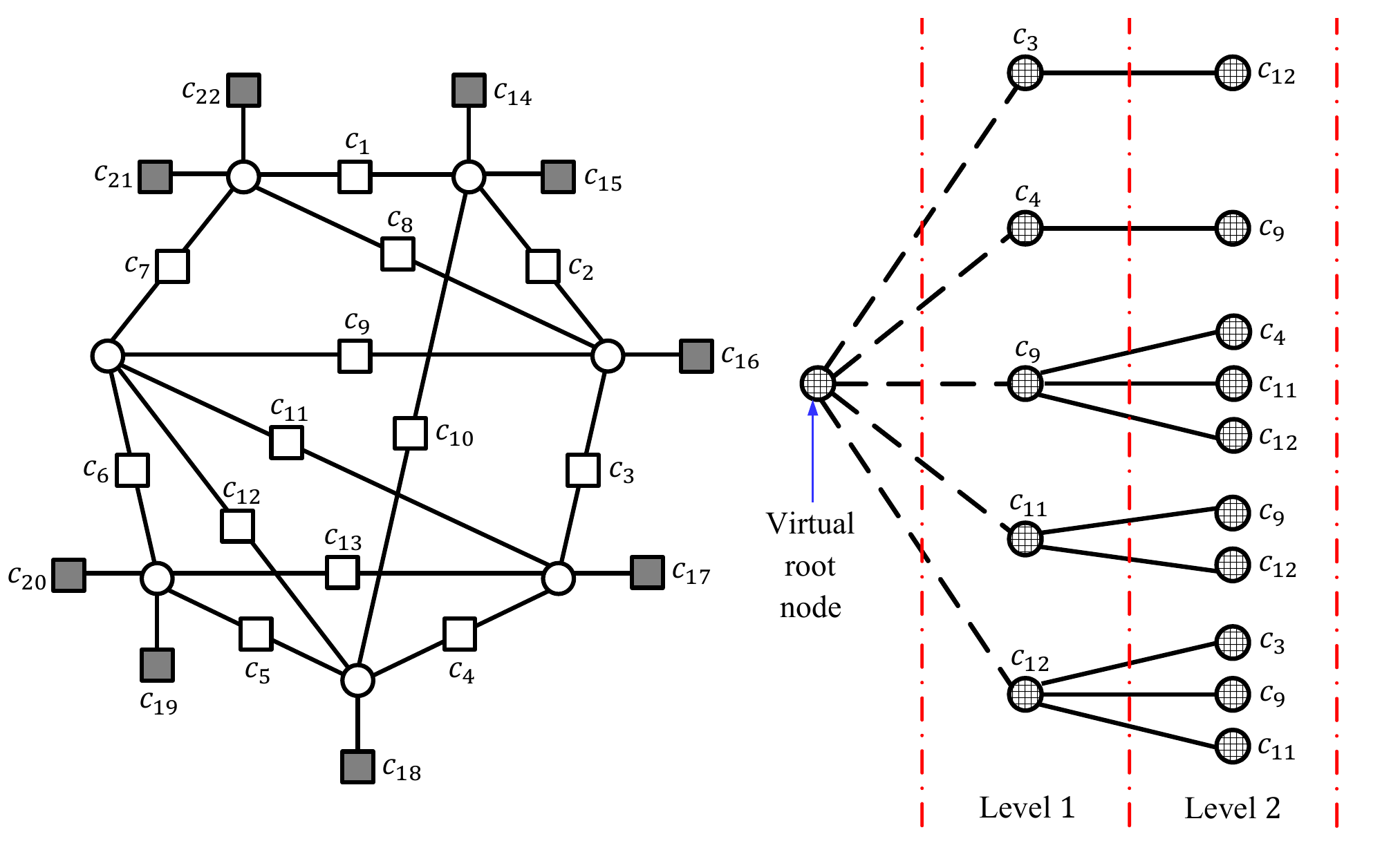}\vspace{-0.5em}
\text{\hspace{1.5em} \footnotesize{(a) \hspace{14em} (b) \hspace{1em}}}
\caption{(a) A $(7, 9, 13, 0)$ unlabeled GAST for $\gamma=5$. (b) The associated unlabeled GAST tree with $b_{\textup{et}}=2$.}
\label{Figure_6}
\vspace{-0.5em}
\end{figure}

\begin{figure}
\vspace{-0.5em}
\center
\includegraphics[trim={0.2in 0.0in 0.35in 0.0in},clip,width=3.5in]{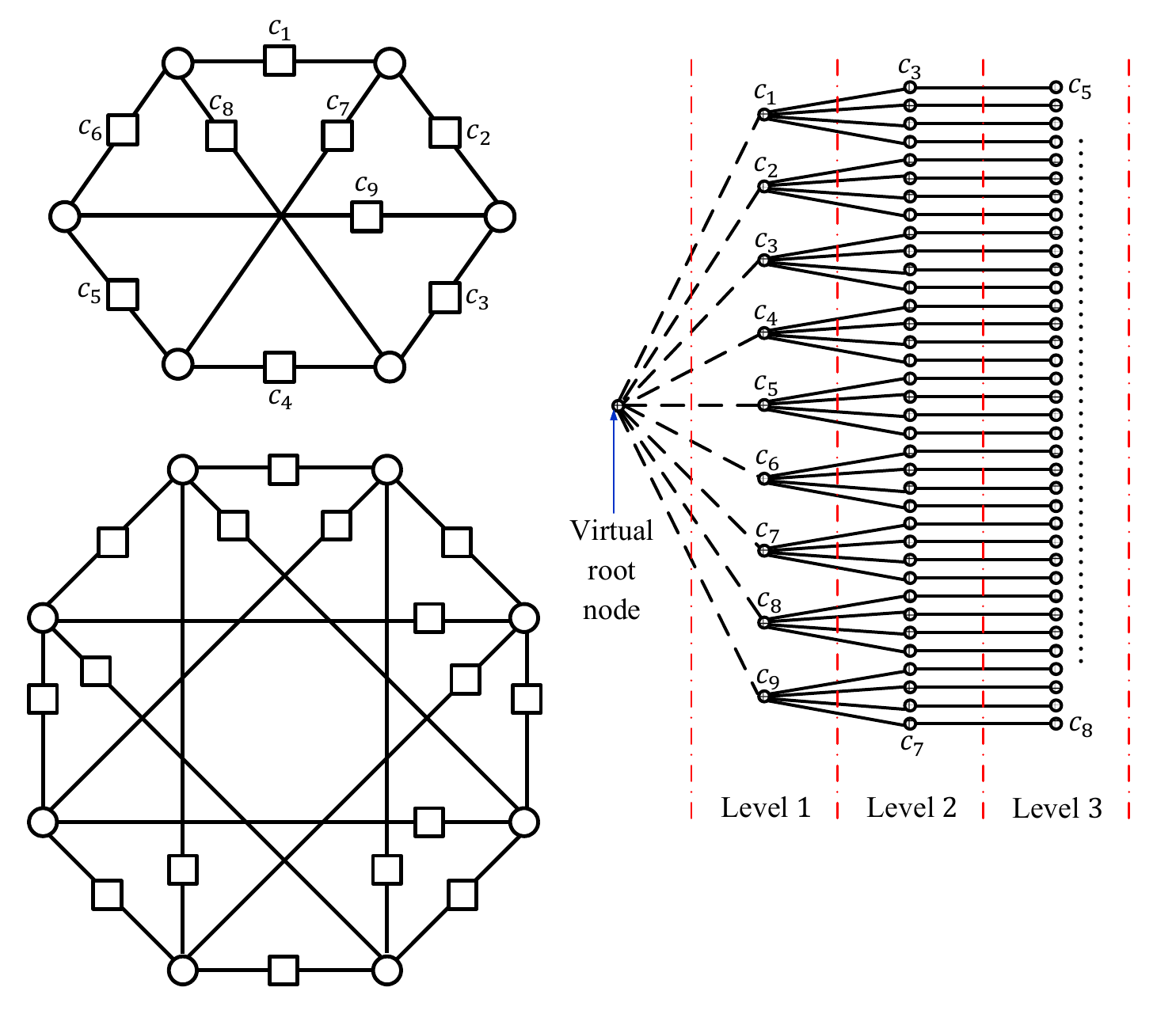}\vspace{-0.5em}
\text{\hspace{0.3em} \footnotesize{(a) \hspace{14em} (b) \hspace{1em}}}
\caption{(a) Upper panel: the u-symmetric $(6, 0, 9, 0)$ unlabeled GAST for $\gamma=3$.  Lower panel: the u-symmetric $(8, 0, 16, 0)$ unlabeled GAST for $\gamma=4$. (b) The associated unlabeled GAST tree for the $(6, 0, 9, 0)$ unlabeled GAST with $b_{\textup{et}}=3$.}
\label{Figure_8}
\vspace{-0.5em}
\end{figure}

An example of a same-size-WCMs unlabeled GAST that is not u-symmetric is the $(7, 9, 13, 0)$ configuration shown in Fig.~\ref{Figure_6}(a). The $(6, 0, 9, 0)$ and the $(8, 0, 16, 0)$ configurations shown in Fig.~\ref{Figure_8}(a) are examples of u-symmetric unlabeled GASTs.

We start off with the count for the general case.

\begin{theorem}\label{th_exact_gen}
Given the unlabeled GAST tree, an $(a, d_1, d_2, d_3)$ unlabeled GAST, with the parameters $b_{\textup{st}} > 0$ and $b_{\textup{et}} > 0$, results in the following number, $t$, of distinct WCMs ($t$ is the size of the set $\mathcal{W}$) for the labeled configuration:
\vspace{-0.1em}\begin{equation}\label{eq_exact_gen}
t = \sum_{b_{\textup{k}}=b_{\textup{st}}}^{b_{\textup{et}}}\frac{1}{b_{\textup{k}}!} \sum_{i_1=1}^{u^0} \sum_{i_2=1}^{u_{i_1}^1} \sum_{i_3=1}^{u_{i_1,i_2}^2} \dots \sum_{i_{b_{\textup{k}}}=1}^{u_{i_1,i_2, \dots, i_{b_{\textup{k}}-1}}^{b_{\textup{k}}-1}} T\left ( u_{i_1,i_2, \dots, i_{b_{\textup{k}}}}^{b_{\textup{k}}} \right ),
\end{equation}
where $b_{\textup{st}} \leq b_{\textup{k}} \leq b_{\textup{et}}$. Here, $T\left ( u_{i_1,i_2, \dots, i_{b_{\textup{k}}}}^{b_{\textup{k}}} \right )=1$ if $u_{i_1,i_2, \dots, i_{b_{\textup{k}}}}^{b_{\textup{k}}}\allowbreak=0$, and $T\left ( u_{i_1,i_2, \dots, i_{b_{\textup{k}}}}^{b_{\textup{k}}} \right )=0$ otherwise.
\end{theorem}

\begin{IEEEproof}
To prove Theorem~\ref{th_exact_gen}, we recall the unlabeled GAST tree. The number of nodes in this tree at any level $b_{\textup{k}} > 0$ is given by:
\vspace{-0.3em}\begin{equation}\label{eq_leaves_bk}
\mu_{b_{\textup{k}}} = \sum_{i_1=1}^{u^0} \sum_{i_2=1}^{u_{i_1}^1} \sum_{i_3=1}^{u_{i_1,i_2}^2} \dots \sum_{i_{b_{\textup{k}}}=1}^{u_{i_1,i_2, \dots, i_{b_{\textup{k}}-1}}^{b_{\textup{k}}-1}} \left ( 1 \right ).
\end{equation}
From the previous subsection, the number, $t_{\textup{rep},b_{\textup{k}}}$, of WCMs (not necessarily distinct) extracted by removing $b^{\textup{cm}}_h=d_1+b_{\textup{k}}$ rows from $\bold{A}$ equals the number of leaves at level $b_{\textup{k}}$. Note that the leaves at level $b_{\textup{k}}$ do not have connections to level $b_{\textup{k}}+1$ (no children) in the tree. As a result, $t_{\textup{rep},b_{\textup{k}}}$ is given by:
\begin{equation}\label{eq_repwcms_bk}
t_{\textup{rep},b_{\textup{k}}} = \sum_{i_1=1}^{u^0} \sum_{i_2=1}^{u_{i_1}^1} \sum_{i_3=1}^{u_{i_1,i_2}^2} \dots \sum_{i_{b_{\textup{k}}}=1}^{u_{i_1,i_2, \dots, i_{b_{\textup{k}}-1}}^{b_{\textup{k}}-1}} T\left ( u_{i_1,i_2, \dots, i_{b_{\textup{k}}}}^{b_{\textup{k}}} \right ).
\end{equation}

To compute the number of distinct WCMs, we need to eliminate repeated WCMs. Since a WCM extracted by removing $(d_1+b_{\textup{k}})$ rows from $\bold{A}$ appears $b_{\textup{k}}!$ times, we compute the number of distinct WCMs that are extracted by removing $(d_1+b_{\textup{k}})$ rows from $\bold{A}$ using (\ref{eq_repwcms_bk}) as follows:
\begin{equation}\label{eq_wcms_bk}
t_{b_{\textup{k}}} = \frac{1}{b_{\textup{k}}!} \sum_{i_1=1}^{u^0} \sum_{i_2=1}^{u_{i_1}^1} \sum_{i_3=1}^{u_{i_1,i_2}^2} \dots \sum_{i_{b_{\textup{k}}}=1}^{u_{i_1,i_2, \dots, i_{b_{\textup{k}}-1}}^{b_{\textup{k}}-1}} T\left ( u_{i_1,i_2, \dots, i_{b_{\textup{k}}}}^{b_{\textup{k}}} \right ).
\end{equation}
The total number of distinct WCMs is then obtained by summing $t_{b_{\textup{k}}}$ in (\ref{eq_wcms_bk}) over all values of $b_{\textup{k}}$, $b_{\textup{st}} \leq b_{\textup{k}} \leq b_{\textup{et}}$, to reach $t$ in (\ref{eq_exact_gen}).
\end{IEEEproof}

Recall that $\gamma$ is the column weight (VN degree) of the code.

\begin{figure}
\vspace{-0.5em}
\center
\includegraphics[trim={0.15in 0.0in 0.4in 0.0in},clip,width=3.5in]{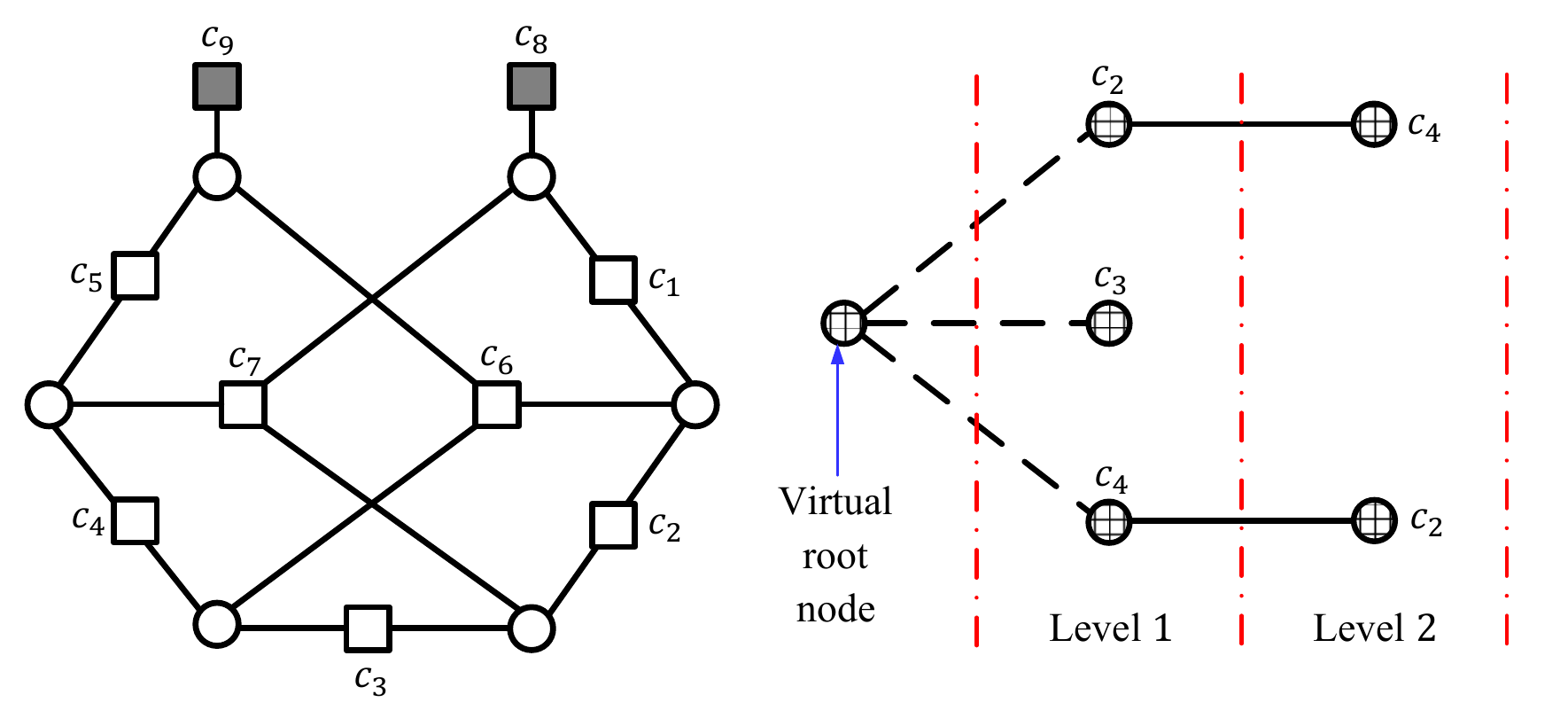}\vspace{-0.3em}
\text{\hspace{0.3em} \footnotesize{(a) \hspace{14em} (b) \hspace{1em}}}
\caption{(a) A $(6, 2, 5, 2)$ unlabeled GAST for $\gamma=3$. (b) The associated unlabeled GAST tree with $b_{\textup{et}}=2$.}
\label{Figure_5}
\vspace{-0.5em}
\end{figure}

\begin{example}\label{ex_1}
Fig.~\ref{Figure_5}(a) shows a $(6, 2, 5, 2)$ unlabeled GAST for $\gamma=3$. As demonstrated by the unlabeled GAST tree in Fig.~\ref{Figure_5}(b), the configuration has WCMs that are not of the same size. Since $b_{\textup{st}}=1$, $b_{\textup{et}}=b_{\textup{ut}}=2$, and $u^0=3$ (that are $c_2$, $c_3$, and $c_4$), (\ref{eq_exact_gen}) reduces to:
\vspace{-0.1em}\begin{align}
t &= \sum_{b_{\textup{k}}=1}^{2}\frac{1}{b_{\textup{k}}!} \sum_{i_1=1}^{3} \sum_{i_2=1}^{u_{i_1}^1} T\left ( u_{i_1, \dots, i_{b_{\textup{k}}}}^{b_{\textup{k}}} \right ) \nonumber \\ &=\frac{1}{1!}(0+1+0)+\frac{1}{2!}(1+0+1)=2. \nonumber
\end{align}
Thus, the configuration has only $2$ WCMs, extracted by removing the rows of the following groups of CNs from $\bold{A}$: $\{(c_3,\mathcal{O}_\textup{sg}), (c_2, c_4,\mathcal{O}_\textup{sg})\}$, where $\mathcal{O}_\textup{sg}$ is $(c_8, c_9)$. {We explicitly list the subgroup $\mathcal{O}_\textup{sg}$ of degree-$1$ CNs to highlight the fact that the rows of these CNs are always removed, irrespective of the action on the remaining rows in $\bold{A}$.}
\end{example}
\vspace{-0.3em}

Now, we analyze the important special case of same-size-WCMs configurations.

\begin{lemma}\label{lem_exact_samelen}
Given the unlabeled GAST tree, a same-size-WCMs $(a, d_1, d_2, d_3)$ unlabeled GAST, with the parameter $b_{\textup{et}} > 0$, results in the following number, $t$, of distinct WCMs ($t$ is the size of the set $\mathcal{W}$) for the labeled configuration:
\begin{equation}\label{eq_exact_samelen}
t = \frac{1}{b_{\textup{et}}!} \sum_{i_1=1}^{u^0} \sum_{i_2=1}^{u_{i_1}^1} \sum_{i_3=1}^{u_{i_1,i_2}^2} \dots \sum_{i_{b_{\textup{et}}}=1}^{u_{i_1,i_2, \dots, i_{b_{\textup{et}}-1}}^{b_{\textup{et}}-1}} \left ( 1 \right ).
\end{equation}
\end{lemma}

\begin{IEEEproof}
We prove Lemma~\ref{lem_exact_samelen} by substituting $b_{\textup{k}}=b_{\textup{st}}=b_{\textup{et}}$ in (\ref{eq_exact_gen}):
\vspace{-0.2em}\begin{align}\label{eq_lem1_pr}
t &= \frac{1}{b_{\textup{et}}!} \sum_{i_1=1}^{u^0} \sum_{i_2=1}^{u_{i_1}^1} \sum_{i_3=1}^{u_{i_1,i_2}^2} \dots \sum_{i_{b_{\textup{et}}}=1}^{u_{i_1,i_2, \dots, i_{b_{\textup{et}}-1}}^{b_{\textup{et}}-1}} T\left ( u_{i_1,i_2, \dots, i_{b_{\textup{et}}}}^{b_{\textup{et}}} \right ) \nonumber \\ &= \frac{1}{b_{\textup{et}}!} \sum_{i_1=1}^{u^0} \sum_{i_2=1}^{u_{i_1}^1} \sum_{i_3=1}^{u_{i_1,i_2}^2} \dots \sum_{i_{b_{\textup{et}}}=1}^{u_{i_1,i_2, \dots, i_{b_{\textup{et}}-1}}^{b_{\textup{et}}-1}} \left ( 1 \right ).
\end{align}
The second equality in (\ref{eq_lem1_pr}) follows from the fact that $T\left ( u_{i_1,i_2, \dots, i_{b_{\textup{et}}}}^{b_{\textup{et}}} \right )=1$ since $u_{i_1,i_2, \dots, i_{b_{\textup{et}}}}^{b_{\textup{et}}}=0$, $\forall i_1, i_2, \dots, i_{b_{\textup{et}}}$, by definition of $b_{\textup{et}}$.
\end{IEEEproof}

\begin{example}\label{ex_2}
Fig.~\ref{Figure_6}(a) shows a $(7, 9, 13, 0)$ unlabeled GAST for $\gamma=5$. As demonstrated by the unlabeled GAST tree in Fig.~\ref{Figure_6}(b), this is a same-size-WCMs configuration. Since $b_{\textup{et}}=b_{\textup{ut}}=2$ and $u^0=5$ (that are $c_3$, $c_4$, $c_9$, $c_{11}$, and $c_{12}$), (\ref{eq_exact_samelen}) reduces to:
\vspace{-0.1em}
\begin{equation}
t = \frac{1}{2!} \sum_{i_1=1}^{5} \sum_{i_2=1}^{u_{i_1}^1} \left ( 1 \right )=\frac{1}{2}(1+1+3+2+3)=5. \nonumber
\end{equation}
Thus, the configuration has $5$ WCMs, all of the same size ($11 \times 7$), extracted by removing the rows of the following groups of CNs from the matrix $\bold{A}$: $\{(c_3, c_{12},\mathcal{O}_\textup{sg}), (c_4, c_9,\mathcal{O}_\textup{sg}), \allowbreak (c_9, c_{11},\mathcal{O}_\textup{sg}), (c_9, c_{12},\mathcal{O}_\textup{sg}), (c_{11}, c_{12},\mathcal{O}_\textup{sg})\}$, where $\mathcal{O}_\textup{sg}$ is $(c_{14}, c_{15}, c_{16}, c_{17}, c_{18}, c_{19}, c_{20}, c_{21}, c_{22})$.
\end{example}

Another important special case to study is the case of u-symmetric configurations.

\begin{corollary}\label{lem_rel_symm}
Given the unlabeled GAST tree, a u-symmetric $(a, d_1, d_2, d_3)$ unlabeled GAST, with the parameter $b_{\textup{et}} > 0$, results in the following number, $t$, of distinct WCMs ($t$ is the size of the set $\mathcal{W}$) for the labeled configuration:
\vspace{-0.2em}\begin{equation}\label{eq_rel_symm}
t = \frac{1}{b_{\textup{et}}!} \prod_{j=1}^{b_{\textup{et}}} u^{j-1}.
\end{equation}
\end{corollary}

\begin{IEEEproof}
Since the u-symmetric case is a special case of the same-size-WCMs case, we use (\ref{eq_exact_samelen}) to conclude:
\begin{align}
t &= \frac{1}{b_{\textup{et}}!} \sum_{i_1=1}^{u^0} \sum_{i_2=1}^{u^1} \sum_{i_3=1}^{u^2} \dots \sum_{i_{b_{\textup{et}}}=1}^{u^{b_{\textup{et}}-1}} \left ( 1 \right ) = \frac{1}{b_{\textup{et}}!} \prod_{j=1}^{b_{\textup{et}}} u^{j-1}. \label{eq_lem2_pr2}
\end{align}
Equation (\ref{eq_lem2_pr2}) follows from the fact that for a u-symmetric configuration, at any level $j$, $u_{i_1,i_2, \dots, i_{j-1}}^{j-1}$ is the same, $\forall i_1, i_2, \dots, i_{j-1}$. Thus, we can express $u_{i_1,i_2, \dots, i_{j-1}}^{j-1}$ in (\ref{eq_exact_samelen}) as $u^{j-1}$, which is independent of $i_1, i_2, \dots, i_{b_{\textup{et}}-1}$, $\forall j \in \{1, 2, \dots, b_{\textup{et}}\}$.
\end{IEEEproof}

\begin{figure}
\vspace{-0.5em}
\center
\includegraphics[trim={0.3in 0.0in 0.4in 0.0in},clip,width=3.5in]{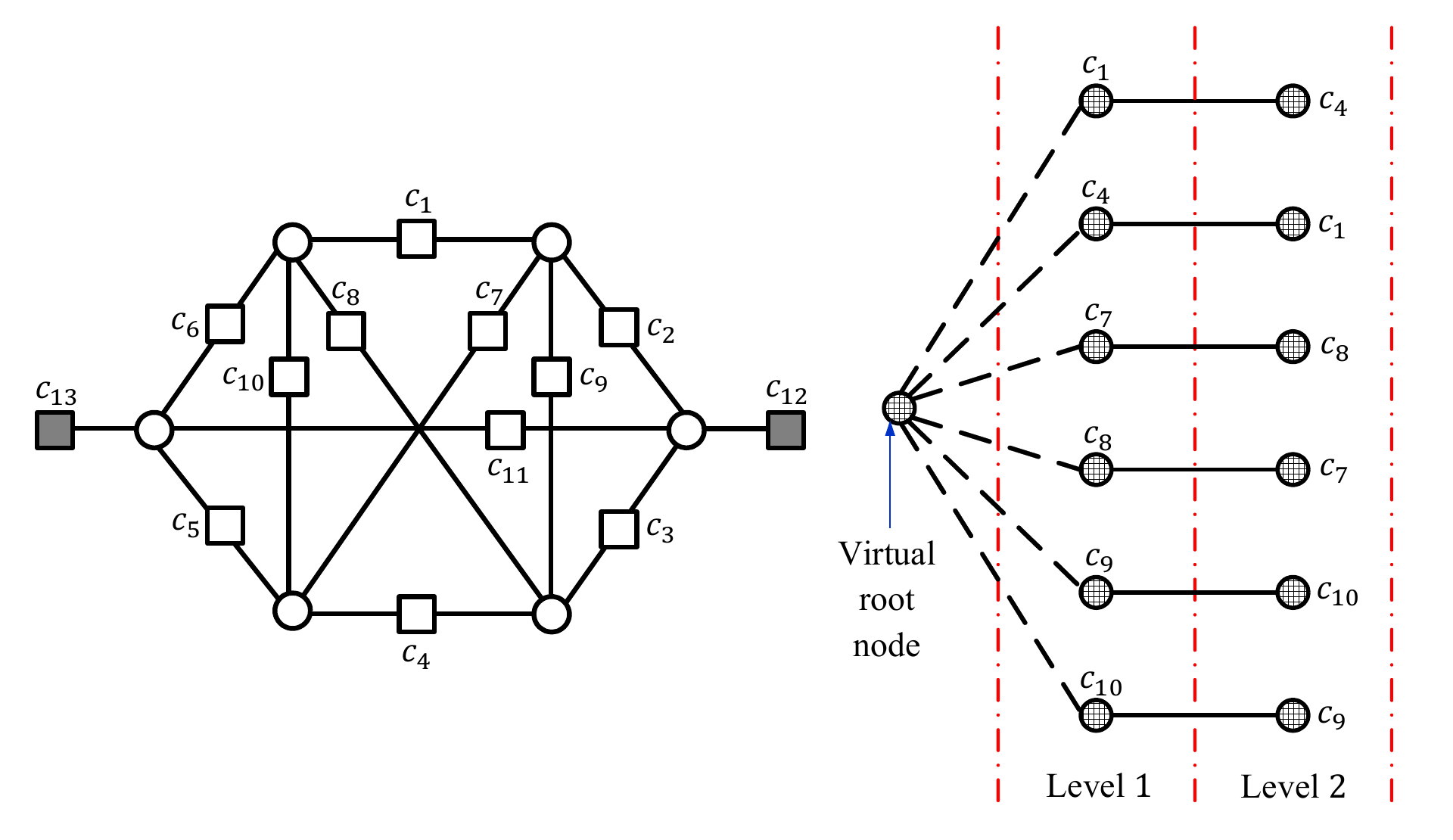}\vspace{-0.5em}
\text{\hspace{2.7em} \footnotesize{(a) \hspace{13em} (b) \hspace{1em}}}
\caption{(a) A $(6, 2, 11, 0)$ unlabeled GAST for $\gamma=4$. (b) The associated unlabeled GAST tree with $b_{\textup{et}}=2$.}
\label{Figure_7}
\vspace{-0.5em}
\end{figure}

\begin{example}\label{ex_3}
Fig.~\ref{Figure_7}(a) shows a $(6, 2, 11, 0)$ unlabeled GAST for $\gamma=4$. As demonstrated by the unlabeled GAST tree in Fig.~\ref{Figure_7}(b), the configuration is u-symmetric. Since $b_{\textup{et}}=b_{\textup{ut}}=2$,  $u^0=6$, and $u^1=1$, (\ref{eq_rel_symm}) reduces to:
\begin{equation}
t = \frac{1}{2!} \prod_{j=1}^{2} u^{j-1}=\frac{1}{2}(6)(1)=3. \nonumber
\end{equation}
Thus, the configuration has $3$ WCMs, all of the same size ($9 \times 6$), extracted by removing the rows of the following groups of CNs from the matrix $\bold{A}$: $\{(c_1, c_4,\mathcal{O}_\textup{sg}), (c_7, c_8,\mathcal{O}_\textup{sg}), \allowbreak (c_9, c_{10},\mathcal{O}_\textup{sg})\}$, where $\mathcal{O}_\textup{sg}$ is $(c_{12}, c_{13})$.
\end{example}

After providing the exact number of WCMs for different cases, we now study examples where the number of distinct WCMs associated with a configuration is proved to be a function only of the column weight $\gamma$ (the VN degree). We study the u-symmetric version of the $(2\gamma, 0, \gamma^2, 0)$ unlabeled GASTs with $g=\left \lfloor \frac{\gamma-1}{2} \right \rfloor=1$ (i.e., for $\gamma=3$ or $\gamma=4$). Studying these configurations is important because they are unlabeled low weight codewords, and their multiplicity in the Tanner graph of a code typically strongly affects the error floor (and also the waterfall) performance of this code.

\begin{lemma}\label{lem_cw_g1}
A u-symmetric $(2 \gamma, 0, \gamma^2, 0)$ unlabeled GAST, with $\gamma \in \{3, 4\}$ (see Fig.~\ref{Figure_8}(a)), results in $t = \gamma!$ distinct WCMs for the labeled configuration.
\end{lemma}

\begin{IEEEproof}
From (\ref{eq_but}), for a u-symmetric $(2 \gamma, 0, \gamma^2, 0)$ unlabeled GAST\footnote{Unlike the superscript of $u$, the superscript of $\gamma$ and all linear expressions of $\gamma$ refers to the mathematical power (if exists).}, we have:
\begin{equation}\label{eq_bet_cwg1}
b_{\textup{et}}=b_{\textup{ut}}=\left \lfloor \frac{1}{2}\left ( 2 \gamma \left \lfloor \frac{\gamma-1}{2} \right \rfloor-0 \right ) \right \rfloor=\gamma.
\end{equation}
Notice that $\left \lfloor \frac{\gamma-1}{2} \right \rfloor=1$ for $\gamma \in \{3, 4\}$. Substituting (\ref{eq_bet_cwg1}) in (\ref{eq_rel_symm}) gives:
\begin{equation}\label{eq_temp_cwg1}
t = \frac{1}{\gamma!} \prod_{j=1}^{\gamma} u^{j-1} = \frac{\gamma^2}{\gamma!} \prod_{j=2}^{\gamma} u^{j-1},
\end{equation}
where the second equality in (\ref{eq_temp_cwg1}) follows from the property that for a $(2 \gamma, 0, \gamma^2, 0)$ unlabeled GAST, $u^0=\gamma^2$.

Next, we compute $u^{j-1}$, $2 \leq j \leq b_{\textup{et}} = \gamma$. At level $1$, a degree-$2$ CN that has its index in $\bold{y}^0$ will be marked as unsatisfied resulting in:
\begin{equation}
u^1 = u^0-1-2 \left (\gamma -1 \right ) = \gamma^2 - 2 \gamma +1 = \left (\gamma -1 \right )^2. \label{eq_u1_2}
\end{equation}
Equation (\ref{eq_u1_2}) follows from the fact that after such a degree-$2$ CN is selected to be marked as unsatisfied at level $1$, all the remaining $\left ( \gamma-1 \right )$ CNs connected to each of the two VNs sharing this CN cannot be selected at level $2$ (because $g=1$ for $\gamma \in \{3, 4\}$). Thus, $u^1 = u^0-(1+2 \left ( \gamma-1 \right ))$, where the additional $1$ represents the already selected CN itself. Furthermore:
\begin{equation}
u^2 = u^1-1-2 \left (\gamma -2 \right ) = \left (\gamma -1 \right )^2-2\gamma+3 = \left (\gamma -2 \right )^2. \label{eq_u2_2}
\end{equation}
Note that the $2 \left ( \gamma -1 \right )$ CNs that cannot be selected at level $2$ are connected to all the remaining $(2\gamma -2)$ VNs in the configuration (after excluding the two VNs sharing the CN selected at level $1$). Thus, any CN to be selected at level $2$ results in $2 \left ( \gamma-2 \right )$ extra CNs that cannot be selected\footnote{The reason why it is $2 \left ( \gamma-2 \right )$ and not $2 \left ( \gamma-1 \right )$ is that two CNs from the group that cannot be selected at level $3$ were already accounted for while computing $u^1$ as they could not be selected at level $2$ (recall that the configuration is u-symmetric).} at level $3$. As a result, $u^2 = u^1-(1+2 \left ( \gamma-2 \right ))$, which is equation (\ref{eq_u2_2}).  This analysis also applies for $u^{j-1}$ with $j > 3$. By means of induction, we conclude the following for every $u^{j-1}$ with $1 \leq j \leq b_{\textup{et}} = \gamma$:
\begin{equation}\label{eq_uj_gen}
u^{j-1} = \left ( \gamma - (j-1) \right )^2.
\end{equation}

Substituting (\ref{eq_uj_gen}) into (\ref{eq_temp_cwg1}) gives:
\begin{equation}\label{eq_final}
t = \frac{1}{\gamma!} \gamma^2 \left (\gamma -1 \right )^2 \left (\gamma -2 \right )^2 \cdots 1^2 = \gamma!.
\end{equation}
As a result, $t=\gamma!$, which completes the proof.
\end{IEEEproof}

\begin{example}\label{ex_4}
Fig.~\ref{Figure_8}(a), upper panel, shows the u-symmetric $(6, 0, 9, 0)$ unlabeled GAST for $\gamma=3$. Fig.~\ref{Figure_8}(b) confirms that the configuration is u-symmetric with $b_{\textup{et}}=b_{\textup{ut}}=3$,  $u^0=9$, $u^1=4$, and $u^2=1$. Thus, (\ref{eq_rel_symm}) reduces to (\ref{eq_final}), implying:
\begin{equation}
t = \frac{1}{3!} \prod_{j=1}^{3} u^{j-1}=3!=6. \nonumber
\end{equation}
The configuration has $6$ WCMs (size $6 \times 6$), extracted by removing the rows of the following groups of CNs from $\bold{A}$: $\{(c_1, c_3, c_5),$ $(c_1, c_4, c_9),$ $(c_2, c_4, c_6),$ $(c_2, c_5, c_8),$ $(c_3, c_6, c_7),$ $(c_7, c_8, c_9)\}$.

Fig.~\ref{Figure_8}(a), lower panel, shows the u-symmetric $(8, 0, 16, 0)$ unlabeled GAST for $\gamma=4$. We omit the tree of this unlabeled GAST for brevity. Following the same logic we used for the u-symmetric $(6, 0, 9, 0)$ unlabeled GAST ($\gamma=3$), we conclude that this configuration has $b_{\textup{et}}=4$,  $u^0=16$, $u^1=9$, $u^2=4$, and $u^3=1$. Thus, from (\ref{eq_final}):
\vspace{-0.3em}\begin{equation}
t = \frac{1}{4!} \prod_{j=1}^{4} u^{j-1}=4!=24. \nonumber
\end{equation}
\end{example}

We conclude Subsection~\ref{subsec_enum} with Table~\ref{Table1}. Table~\ref{Table1} lists the number of distinct WCMs for different types of unlabeled GASTs.

\begin{table*}
\caption{Number of distinct WCMs for different types of unlabeled GASTs.}
\vspace{-0.5em}
\centering
\scalebox{1.00}
{
\begin{tabular}{|c|c|}
\hline
Unlabeled GAST type & Number of distinct WCMs ($t$) \\
\hline
General & $t=\sum\limits_{b_{\textup{k}}=b_{\textup{st}}}^{b_{\textup{et}}}\frac{1}{b_{\textup{k}}!} \sum\limits_{i_1=1}^{u^0} \sum\limits_{i_2=1}^{u_{i_1}^1} \sum\limits_{i_3=1}^{u_{i_1,i_2}^2} \dots \sum\limits_{i_{b_{\textup{k}}}=1}^{u_{i_1,i_2, \dots, i_{b_{\textup{k}}-1}}^{b_{\textup{k}}-1}} T\left ( u_{i_1,i_2, \dots, i_{b_{\textup{k}}}}^{b_{\textup{k}}} \right )$ \\
\hline
Same-size-WCMs & $t=\frac{1}{b_{\textup{et}}!} \sum\limits_{i_1=1}^{u^0} \sum\limits_{i_2=1}^{u_{i_1}^1} \sum\limits_{i_3=1}^{u_{i_1,i_2}^2} \dots \sum\limits_{i_{b_{\textup{et}}}=1}^{u_{i_1,i_2, \dots, i_{b_{\textup{et}}-1}}^{b_{\textup{et}}-1}} \left ( 1 \right )$ \\
\hline
U-symmetric & $t=\frac{1}{b_{\textup{et}}!} \prod\limits_{j=1}^{b_{\textup{et}}} u^{j-1}$ \\
\hline
U-symmetric, $(2 \gamma, 0, \gamma^2, 0)$, with $\gamma \in \{3, 4\}$ & $t=\gamma!$ \\
\hline
\end{tabular}}
\label{Table1}
\vspace{+1.0em}
\end{table*}

%%%%%%%%%%%%%%%%%%%%%%%%%%%%%%
\subsection{Complexity Comparison with a Suboptimal Idea}\label{subsec_comp}

We have already proved the optimality of the WCM framework in Subsection~\ref{subsec_optim}. In this subsection, we demonstrate the complexity reduction we gain by focusing only on the set of WCMs, $\mathcal{W}$, to remove a GAST. We compute the total number of distinct matrices to operate on in an alternative idea (a suboptimal idea), and compare it with the number of distinct WCMs we operate on, which is $t$ derived in Subsection~\ref{subsec_enum}. The suboptimal idea we compare with is operating on the set of all distinct matrices $\bold{W}^{\textup{z}}$.

Computational savings of the WCM framework relative to the suboptimal idea mentioned above on a prototypical example of an NB-LDPC code are quite apparent; it takes roughly only four days to optimize this code using the WCM framework (via operating on the WCMs of each GAST to be removed), compared with roughly a month of computations using the suboptimal approach (via operating on all distinct matrices $\bold{W}^{\textup{z}}$ of each GAST to be removed). In this subsection, we justify this observation.

Here, we seek to compare the number of distinct WCMs, which is the size of the set $\mathcal{W}$, with the number of distinct $\bold{W}^{\textup{z}}$ matrices, which is the size of the set $\mathcal{U}$. For convenience, we assume for this comparison that $b_{\textup{et}} > 0$ and $u^0 > 0$.

\begin{theorem}\label{th_diff_gen}
Given the unlabeled GAST tree, the difference between the cardinalities of the sets $\mathcal{U}$ and $\mathcal{W}$ (the reduction in the number of matrices to operate on) for an $(a, d_1, d_2, d_3)$ unlabeled GAST, with the parameters $b_{\textup{st}} > 0$ and $b_{\textup{et}} > 0$, is:
\begin{equation}\label{eq_diff_gen}
t'-t=1+\sum_{j=1}^{b_{\textup{et}}-1}\frac{1}{j!} \sum_{i_1=1}^{u^0} \sum_{i_2=1}^{u_{i_1}^1} \sum_{i_3=1}^{u_{i_1,i_2}^2} \dots \sum_{i_j=1}^{u_{i_1,i_2, \dots, i_{j-1}}^{j-1}} \hspace{-1.0em} T_{\textup{c}}\left ( u_{i_1,i_2, \dots, i_{j}}^{j} \right ),
\end{equation}
where $t'=\vert{\mathcal{U}}\vert$ and $t=\vert{\mathcal{W}}\vert$. Here, $T_{\textup{c}}\left ( u_{i_1,i_2, \dots, i_{j}}^{j} \right )=1$ if $u_{i_1,i_2, \dots, i_{j}}^{j} \neq 0$, and $T_{\textup{c}}\left ( u_{i_1,i_2, \dots, i_{j}}^{j} \right )=0$ otherwise.
\end{theorem}

\begin{IEEEproof}
Given that $t$ (which is $\vert{\mathcal{W}}\vert$) is known from Subsection~\ref{subsec_enum}, we need to derive $t'$ (which is $\vert{\mathcal{U}}\vert$). Since $\mathcal{U}$ is the set of all distinct matrices $\bold{W}^{\textup{z}}$, it follows that the cardinality of $\mathcal{U}$ is a function of the total number of nodes in the unlabeled GAST tree. Note that each node at level $j$ in the unlabeled GAST tree is linked to a matrix $\bold{W}^{\textup{z}}$ (see the proof of Theorem~\ref{th_opt_frm}). The total number of these tree nodes is:
\begin{equation}\label{eq_nodes_wrep}
\eta_{\textup{rep}}=\sum_{j=1}^{b_{\textup{et}}} \sum_{i_1=1}^{u^0} \sum_{i_2=1}^{u_{i_1}^1} \sum_{i_3=1}^{u_{i_1,i_2}^2} \dots \sum_{i_j=1}^{u_{i_1,i_2, \dots, i_{j-1}}^{j-1}} \left ( 1 \right ).
\end{equation}
To remove the repeated $\bold{W}^{\textup{z}}$ matrices from that count, we need to divide the number of tree nodes at each level $j$ by $j!$. Thus, we reach:
\begin{equation}\label{eq_nodes_norep}
\eta=\sum_{j=1}^{b_{\textup{et}}} \frac{1}{j!} \sum_{i_1=1}^{u^0} \sum_{i_2=1}^{u_{i_1}^1} \sum_{i_3=1}^{u_{i_1,i_2}^2} \dots \sum_{i_j=1}^{u_{i_1,i_2, \dots, i_{j-1}}^{j-1}} \left ( 1 \right ).
\end{equation}
The cardinality of the set $\vert{\mathcal{U}}\vert$, which is $t'$, is then:
\begin{equation}\label{eq_wz_count}
t'=1+\eta=1+\sum_{j=1}^{b_{\textup{et}}} \frac{1}{j!} \sum_{i_1=1}^{u^0} \sum_{i_2=1}^{u_{i_1}^1} \sum_{i_3=1}^{u_{i_1,i_2}^2} \dots \sum_{i_j=1}^{u_{i_1,i_2, \dots, i_{j-1}}^{j-1}} \left ( 1 \right ),
\end{equation}
where the additional $1$ is for the particular matrix $\bold{W}^{\textup{z}}$ extracted by removing $d_1$ rows from $\bold{A}$ corresponding to all degree-$1$ CNs in the configuration. Note that we can consider the virtual root node as the node linked to this particular $\bold{W}^{\textup{z}}$ matrix in the tree.

To compute $t'-t$, we subtract (\ref{eq_exact_gen}) from (\ref{eq_wz_count}). Consequently,
\vspace{-0.1em}\begin{align}
&t'-t=1+\sum_{j=1}^{b_{\textup{st}}-1}\frac{1}{j!} \sum_{i_1=1}^{u^0} \sum_{i_2=1}^{u_{i_1}^1} \sum_{i_3=1}^{u_{i_1,i_2}^2} \dots \sum_{i_j=1}^{u_{i_1,i_2, \dots, i_{j-1}}^{j-1}} \left ( 1 \right ) + \nonumber \\ & \sum_{j=b_{\textup{st}}}^{b_{\textup{et}}}\frac{1}{j!} \sum_{i_1=1}^{u^0} \sum_{i_2=1}^{u_{i_1}^1} \sum_{i_3=1}^{u_{i_1,i_2}^2} \dots \sum_{i_j=1}^{u_{i_1,i_2, \dots, i_{j-1}}^{j-1}} \left [ 1-T\left ( u_{i_1,i_2, \dots, i_{j}}^{j} \right ) \right ]. \label{eq_th3_pr1}
\end{align}
Thus, we complete the proof as follows:
\vspace{-0.1em}\begin{align}\label{eq_th3_pr2}
&t'-t = 1+ \nonumber \\ &\sum_{j=1}^{b_{\textup{et}}}\frac{1}{j!} \sum_{i_1=1}^{u^0} \sum_{i_2=1}^{u_{i_1}^1} \sum_{i_3=1}^{u_{i_1,i_2}^2} \dots \sum_{i_j=1}^{u_{i_1,i_2, \dots, i_{j-1}}^{j-1}} \left [ 1-T\left ( u_{i_1,i_2, \dots, i_{j}}^{j} \right ) \right ] \nonumber \\
&=1+\sum_{j=1}^{b_{\textup{et}}-1}\frac{1}{j!} \sum_{i_1=1}^{u^0} \sum_{i_2=1}^{u_{i_1}^1} \sum_{i_3=1}^{u_{i_1,i_2}^2} \dots \sum_{i_j=1}^{u_{i_1,i_2, \dots, i_{j-1}}^{j-1}} \hspace{-1.0em} T_{\textup{c}}\left ( u_{i_1,i_2, \dots, i_{j}}^{j} \right ).
\end{align}
The first equality in (\ref{eq_th3_pr2}) is derived by observing that $\left [ 1-T\left ( u_{i_1,i_2, \dots, i_{j}}^{j} \right ) \right ]=1$ for $j \in \{1, 2, \dots, b_{\textup{st}}-1\}$. The second equality in (\ref{eq_th3_pr2}) follows from that $T\left ( u_{i_1,i_2, \dots, i_{j}}^{j} \right )=1$ for $j=b_{\textup{et}}$, and $\left [ 1-T\left ( u_{i_1,i_2, \dots, i_{j}}^{j} \right ) \right ] =T_{\textup{c}}\left ( u_{i_1,i_2, \dots, i_{j}}^{j} \right )$~(from the definitions of both $T$ and $T_{\textup{c}}$). We can simply consider $T_{\textup{c}}\left ( u_{i_1,i_2, \dots, i_{j}}^{j} \right )$ as the complement function (binary inversion) of $T\left ( u_{i_1,i_2, \dots, i_{j}}^{j} \right )$.
\end{IEEEproof}

\begin{table*}
\caption{Reduction in the number of matrices to operate on for different types of unlabeled GASTs.}
\vspace{-0.5em}
\centering
\scalebox{1.00}
{
\begin{tabular}{|c|c|}
\hline
Unlabeled GAST type & Reduction in the number of matrices ($t'-t$) \\
\hline
General & $t'-t=1+\sum\limits_{j=1}^{b_{\textup{et}}-1}\frac{1}{j!} \sum\limits_{i_1=1}^{u^0} \sum\limits_{i_2=1}^{u_{i_1}^1} \sum\limits_{i_3=1}^{u_{i_1,i_2}^2} \dots \sum\limits_{i_j=1}^{u_{i_1,i_2, \dots, i_{j-1}}^{j-1}} T_{\textup{c}}\left ( u_{i_1,i_2, \dots, i_{j}}^{j} \right )$ \\
\hline
Same-size-WCMs & $t'-t=1+\sum\limits_{j=1}^{b_{\textup{et}}-1}\frac{1}{j!} \sum\limits_{i_1=1}^{u^0} \sum\limits_{i_2=1}^{u_{i_1}^1} \sum\limits_{i_3=1}^{u_{i_1,i_2}^2} \dots \sum\limits_{i_j=1}^{u_{i_1,i_2, \dots, i_{j-1}}^{j-1}} \left ( 1 \right )$ \\
\hline
U-symmetric & $t'-t=1+\sum\limits_{j=1}^{b_{\textup{et}}-1}\frac{1}{j!} \prod\limits_{i=1}^{j} u^{i-1}$ \\
\hline
U-symmetric, $(2 \gamma, 0, \gamma^2, 0)$, with $\gamma \in \{3, 4\}$ & $t'-t=1+\sum\limits_{j=1}^{\gamma-1}\frac{1}{j!} \prod\limits_{i=1}^{j} \left ( \gamma - (i-1) \right )^2$ \\
\hline
\end{tabular}}
\label{Table2}
\end{table*}

\begin{example}\label{ex_6}
Consider the $(6, 2, 5, 2)$ unlabeled GAST ($\gamma=3$) shown in Fig.~\ref{Figure_5}(a). Since $b_{\textup{st}}=1$, $b_{\textup{et}}=b_{\textup{ut}}=2$, and $u^0=3$, and aided by the unlabeled GAST tree in Fig.~\ref{Figure_5}(b), the complexity reduction (the reduction in the number of matrices to operate on) is (see (\ref{eq_diff_gen})):
\begin{equation}
t'-t=1+\frac{1}{1!} \sum_{i_1=1}^{3} T_{\textup{c}}\left ( u_{i_1}^{1} \right )=1+(1+0+1)=3. \nonumber
\end{equation}
In other words, the cardinality of the set $\mathcal{U}$ is $t'=5$, while from Example~\ref{ex_1}, the cardinality of the set $\mathcal{W}$ (the number of distinct WCMs) is $t=2$. Thus, the complexity reduction is $60\%$ for this configuration.
\end{example}
Now, we study the case of same-size-WCMs unlabeled GASTs.

\begin{lemma}\label{lem_diff_samelen}
Given the unlabeled GAST tree, the difference between the cardinalities of the sets $\mathcal{U}$ and $\mathcal{W}$ (the reduction in the number of matrices to operate on) for a same-size-WCMs $(a, d_1, d_2, d_3)$ unlabeled GAST, with the parameter $b_{\textup{et}} > 0$, is:
\begin{equation}\label{eq_diff_samelen}
t'-t=1+\sum_{j=1}^{b_{\textup{et}}-1}\frac{1}{j!} \sum_{i_1=1}^{u^0} \sum_{i_2=1}^{u_{i_1}^1} \sum_{i_3=1}^{u_{i_1,i_2}^2} \dots \sum_{i_j=1}^{u_{i_1,i_2, \dots, i_{j-1}}^{j-1}} \left ( 1 \right ),
\end{equation}
where $t'=\vert{\mathcal{U}}\vert$ and $t=\vert{\mathcal{W}}\vert$.
\end{lemma}

\begin{IEEEproof}
Knowing that the configuration is a same-size-WCMs unlabeled GAST does not simplify the expression of $t'$ in (\ref{eq_wz_count}). Thus, to compute $(t'-t)$, we subtract (\ref{eq_exact_samelen}) from (\ref{eq_wz_count}). The result of this subtraction is (\ref{eq_diff_samelen}).
\end{IEEEproof}

\begin{example}\label{ex_7}
Consider the $(7, 9, 13, 0)$ unlabeled GAST ($\gamma=5$) shown in Fig.~\ref{Figure_6}(a). Since $b_{\textup{et}}=b_{\textup{ut}}=2$ and $u^0=5$, and aided by the unlabeled GAST tree in Fig.~\ref{Figure_6}(b), the complexity reduction (the reduction in the number of matrices to operate on) is (see (\ref{eq_diff_samelen})):
\begin{equation}
t'-t=1+\frac{1}{1!} \sum_{i_1=1}^{5} \left ( 1 \right )=1+5=6. \nonumber
\end{equation}
In other words, the cardinality of the set $\mathcal{U}$ is $t'=11$, while from Example~\ref{ex_2}, the cardinality of the set $\mathcal{W}$ (the number of distinct WCMs) is $t=5$. Thus, the complexity reduction is over $50\%$.
\end{example}

\begin{corollary}\label{cor_diff_symm}
Given the unlabeled GAST tree, the difference between the cardinalities of the sets $\mathcal{U}$ and $\mathcal{W}$ (the reduction in the number of matrices to operate on) for a u-symmetric $(a, d_1, d_2, d_3)$ unlabeled GAST, with the parameter $b_{\textup{et}} > 0$, is:
\begin{equation}\label{eq_diff_symm}
t'-t=1+\sum_{j=1}^{b_{\textup{et}}-1}\frac{1}{j!} \prod_{i=1}^{j} u^{i-1},
\end{equation}
where $t'=\vert{\mathcal{U}}\vert$ and $t=\vert{\mathcal{W}}\vert$.
\end{corollary}

\begin{IEEEproof}
We recall again that the u-symmetric configuration is a special case of the same-size-WCMs configuration. Consequently, we can use the same idea from the proof of Corollary~\ref{lem_rel_symm} in (\ref{eq_diff_samelen}) to reach (\ref{eq_diff_symm}).
\end{IEEEproof}

\begin{example}\label{ex_8}
Consider the u-symmetric $(2\gamma, 0, \gamma^2, 0)$ unlabeled GAST. From (\ref{eq_uj_gen}), we know that $u^{j-1} = \left ( \gamma - (j-1) \right )^2$. Thus, from Corollary~\ref{cor_diff_symm}, the complexity reduction (the reduction in the number of matrices to operate on) is:
\vspace{-0.1em}\begin{equation}
t'-t=1+\sum_{j=1}^{b_{\textup{et}}-1}\frac{1}{j!} \prod_{i=1}^{j} u^{i-1}=1+\sum_{j=1}^{\gamma-1}\frac{1}{j!} \prod_{i=1}^{j} \left ( \gamma - (i-1) \right )^2.
\end{equation}
For $\gamma=3$ (corresponding to the u-symmetric $(6, 0, 9, 0)$ unlabeled GAST), the complexity reduction is $1+\frac{1}{1!}(9)+\frac{1}{2!}(9)(4)=28$, which is over $80\%$ (i.e., $t'=34$ while $t=6$). For $\gamma=4$ (corresponding to the u-symmetric $(8, 0, 16, 0)$ unlabeled GAST), the complexity reduction is $1+\frac{1}{1!}(16)+\frac{1}{2!}(16)(9)+\frac{1}{3!}(16)(9)(4)=185$, which is about $90\%$ (i.e., $t'=209$ while $t=24$).
\end{example}

We conclude Subsection~\ref{subsec_comp} with Table~\ref{Table2}. Table~\ref{Table2} lists the reduction, $t'-t$, in the number of matrices to operate on for different types of unlabeled GASTs.

\begin{remark}\label{rem_4}
The analysis in Section~\ref{sec_cch} is focusing on the case where $b_{\textup{et}} > 0$ (thus, $u^0 > 0$) because if $b_{\textup{et}} = 0$ (i.e., $u^0 = 0$), $t=1$ always. In other words, there exists only one matrix $\bold{W}^{\textup{z}}$. As a result, there exists only one WCM of size $(\ell-d_1) \times a$, which is the single matrix $\bold{W}^{\textup{z}}$ itself. Note that if $b_{\textup{et}} > 0$, the matrix $\bold{W}^{\textup{z}}$ of size $(\ell-d_1) \times a$ cannot be a WCM (this is the reason why we do not add $1$ in (\ref{eq_exact_gen}) as we do in (\ref{eq_wz_count})).
\end{remark}

\begin{remark}\label{rem_bas}
An analysis similar to what we presented in Section~\ref{sec_cch} can be done for BASTs.
\end{remark}

%%%%%%%%%%%%%%%%%%%%%%%%%%%%%%
\section{More on How GASTs Are Removed}\label{sec_rem}

After demonstrating the complexity reduction achieved by operating only on the set of WCMs to remove a GAST, in this section, we provide more details on the removal of GASTs via their WCMs. We first investigate the dimension of the null space of a WCM. Then, we discuss the best that can be done to break the weight conditions of a short WCM. Finally, we discuss the exact minimum number of edge weight changes needed by the WCM framework to remove a GAST from the Tanner graph of an NB-LDPC code, and we provide a useful topological upper bound on that minimum. A further discussion about the null spaces of WCMs that belong to GASTs having $b=d_1$ is provided in Appendix~\ref{sec_appc}.

%%%%%%%%%%%%%%%%%%%%%%%%%%%%%%
\subsection{The Dimension of the Null Space of a WCM}\label{subsec_dim}

A GAST is removed via breaking the weight conditions of all its WCMs, i.e., via satisfying (\ref{eq_rem_cond}) for all its WCMs. This breaking is performed by forcing the null spaces of these WCMs to have a particular property. Thus, studying the dimension of the null space of a WCM is critical to understand how GASTs are removed.

Consider a WCM $\bold{W}^{\textup{cm}}_h$, $1 \leq h \leq t$, of a GAST. Recall that $\mathcal{N}(\bold{M})$ is the null space of a matrix $\bold{M}$, and let $\textup{dim}(\mathcal{N}(\bold{M}))$ denote the dimension of the null space of a matrix $\bold{M}$. Moreover, let $G^{\textup{cm}}_h$ be the subgraph created by removing $b^{\textup{cm}}_h$ degree $\leq 2$ CNs from the GAST subgraph. The $b^{\textup{cm}}_h$ rows that are removed from $\bold{A}$ to reach $\bold{W}^{\textup{cm}}_h$ correspond to these $b^{\textup{cm}}_h$ CNs. Note that these CNs are the ones on the path from the root node until the tree node linked to $\bold{W}^{\textup{cm}}_h$ in the unlabeled GAST tree (the CNs marked as unsatisfied by \cite[Algorithm~1]{ahh_jsac} to extract $\bold{W}^{\textup{cm}}_h$). Moreover, let $\bold{M}(G)$ denote the adjacency matrix of a graph $G$. 

\begin{theorem}\label{th_dim_null}
The dimension $p_h$ of the null space of a WCM $\bold{W}^{\textup{cm}}_h$, $1 \leq h \leq t$, of an $(a, b, d_1, d_2, d_3)$ GAST, given that this WCM has unbroken weight conditions, is given by:
\vspace{-0.1em}\begin{equation}\label{eq_dim_null}
p_h = \textup{dim}\left (\mathcal{N}(\bold{W}^{\textup{cm}}_h) \right ) = \sum_{k=1}^{\delta_h}\textup{dim}\left (\mathcal{N}\left (\bold{M}(G^{\textup{disc}}_{h,k}) \right ) \right ) \geq \delta_h,
\end{equation}
where $\delta_h$ is the number of disconnected components in $G^{\textup{cm}}_h$, and $G^{\textup{disc}}_{h,k}$ is the $k$th disconnected component in $G^{\textup{cm}}_h$, with $1 \leq k \leq \delta_h$.
\end{theorem}

\begin{IEEEproof}
It is known from graph theory that if graph $G^{\textup{cm}}_h$ has $\delta_h$ disconnected components defined as $G^{\textup{disc}}_{h,k}$, $1 \leq k \leq \delta_h$, then:
\vspace{-0.4em}\begin{align}\label{eq_th4_pr1}
p_h &= \textup{dim}\left (\mathcal{N}(\bold{W}^{\textup{cm}}_h) \right ) = \textup{dim}\left (\mathcal{N}\left (\bold{M}(G^{\textup{cm}}_h) \right ) \right ) \nonumber \\ &= \sum_{k=1}^{\delta_h}\textup{dim}\left (\mathcal{N}\left (\bold{M}(G^{\textup{disc}}_{h,k}) \right ) \right ).
\end{align}
Note that by definition of $G^{\textup{cm}}_h$, $\bold{W}^{\textup{cm}}_h=\bold{M}(G^{\textup{cm}}_h)$.

Then, we prove the inequality $p_h \geq \delta_h$. If $\exists \text{ } G^{\textup{disc}}_{h,k}$ s.t. $\textup{dim}\left (\mathcal{N}\left (\bold{M}(G^{\textup{disc}}_{h,k}) \right ) \right ) = 0$ (which means $\mathcal{N}\left (\bold{W}^{\textup{cm}}_h \right ) = \{\bold{0}\}$), then it is impossible to have a vector $\bold{v}=[v_1 \text{ } v_2 \text{ } \dots \text{  } v_a]^\textup{T} \in \mathcal{N}\left (\bold{M}(G^{\textup{cm}}_h) \right )=\mathcal{N}(\bold{W}^{\textup{cm}}_h)$ s.t. $v_f \neq 0$, $\forall f \in \{1, 2, \dots, a\}$, where $a$ is the size of the GAST. Thus, in order to have a WCM that has unbroken weight conditions, we must have $\textup{dim}\left (\mathcal{N}\left (\bold{M}(G^{\textup{disc}}_{h,k}) \right ) \right ) > 0$, $\forall k \in \{1, 2, \dots, \delta_h\}$. Noting that if $\textup{dim}\left (\mathcal{N}\left (\bold{M}(G^{\textup{disc}}_{h,k}) \right ) \right ) > 0$, $\forall k \in \{1, 2, \dots, \delta_h\}$, then it has to be the case that $p_h=\sum_{k=1}^{\delta_h}\textup{dim}\left (\mathcal{N}\left (\bold{M}(G^{\textup{disc}}_{h,k}) \right ) \right ) \geq \delta_h$ completes the proof of Theorem~\ref{th_dim_null}.
\end{IEEEproof}

\begin{remark}\label{rem_5}
Consider a WCM $\bold{W}^{\textup{cm}}_h$ that has unbroken weight conditions. In the majority of the GASTs we have studied, if $G^{\textup{cm}}_h$ (the graph corresponding to $\bold{W}^{\textup{cm}}_h$) has $\delta_h=1$ (the graph is fully connected), then $\textup{dim}\left (\mathcal{N}\left (\bold{W}^{\textup{cm}}_h \right ) \right ) =1$. Similarly, we have typically observed that if $\delta_h > 1$, then $\forall G^{\textup{disc}}_{h,k}$, $1 \leq k \leq \delta_h$, $\textup{dim}\left (\mathcal{N}\left (\bold{M}(G^{\textup{disc}}_{h,k}) \right ) \right ) = 1$. In other words, in most of the cases we have seen, $p_h = \textup{dim}\left (\mathcal{N}(\bold{W}^{\textup{cm}}_h) \right ) = \delta_h$. Having said that, we have already encountered few examples where $p_h = \textup{dim}\left (\mathcal{N}(\bold{W}^{\textup{cm}}_h) \right ) > \delta_h$ (see the next subsection).
\end{remark}

Typically, if $\delta_h = 1$, breaking the weight conditions of a WCM $\bold{W}^{\textup{cm}}_h$ yields $\textup{dim}\left (\mathcal{N}\left (\bold{W}^{\textup{cm}}_h \right ) \right ) = 0$ (there are few exceptions to that). Contrarily, it is important to note that if $\delta_h > 1$ (which again means the graph corresponding to $\bold{W}^{\textup{cm}}_h$ has more than one disconnected components), the weight conditions of this WCM can be broken while $\textup{dim}\left (\mathcal{N}\left (\bold{W}^{\textup{cm}}_h \right ) \right ) > 0$. This situation occurs if $\exists$ $G^{\textup{disc}}_{h,k_1}$ and~$G^{\textup{disc}}_{h,k_2}$, $1 \leq k_1, k_2 \leq \delta_h$, s.t. $\textup{dim}\left (\mathcal{N}\left (\bold{M}(G^{\textup{disc}}_{h,k_1}) \right ) \right ) = 0$ while $\textup{dim}\left (\mathcal{N}\left (\bold{M}(G^{\textup{disc}}_{h,k_2}) \right ) \right ) > 0$. Thus, breaking the weight conditions of such a WCM by making $\textup{dim}\left (\mathcal{N}\left (\bold{W}^{\textup{cm}}_h \right ) \right ) = 0$ (if possible) is, albeit sufficient, not necessary. A conceptually-similar observation will be presented in the next subsection. We present Example~\ref{ex_9} to illustrate Theorem~\ref{th_dim_null} as well as this discussion.

\begin{figure}
\vspace{-0.5em}
\center
\includegraphics[trim={0.06in 0.0in 0.07in 0.0in},clip,width=3.5in]{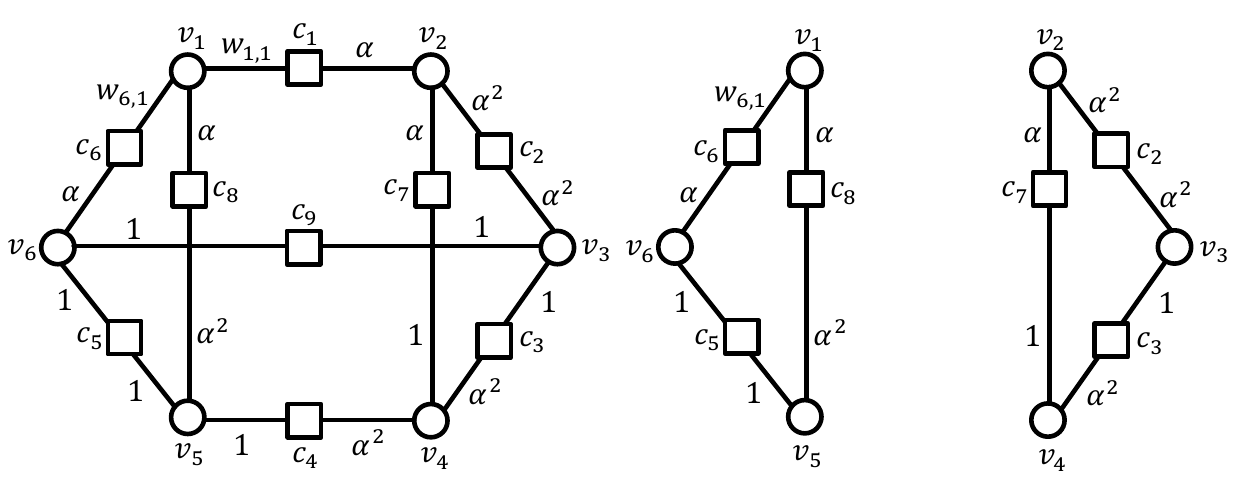}\vspace{-0.5em}
\text{\hspace{0.6em} \footnotesize{(a) \hspace{14em} (b) \hspace{1em}}}
\caption{(a) A $(6, 0, 0, 9, 0)$ GAST for $\gamma=3$. (b) The graph created by removing $(c_1, c_4, c_9)$ from the GAST graph. GF($4$) is assumed.}
\label{Figure_9}
\vspace{-0.5em}
\end{figure}

\begin{example}\label{ex_9}
Consider the $(6, 0, 0, 9, 0)$ GAST ($\gamma=3$) over GF($4$), where GF($4$) $=\{0, 1 , \alpha, \alpha^2\}$ and $\alpha$ is a primitive element (a root of the primitive polynomial $\rho(x)=x^2+x+1$, i.e., $\alpha^2=\alpha+1$), that is shown in Fig.~\ref{Figure_9}(a). The matrix $\bold{A}$ of this configuration is:
\vspace{-1.0em}
\begin{gather*}\label{ex_wcms}
\begin{matrix}
\text{  }\text{  }\text{  }\text{  }\text{  }\text{  }\text{   }\text{   } v_1 \text{   } & v_2 \text{   } & v_3 \text{   } & v_4 & v_5 & v_6 \vspace{-0.3em}
\end{matrix}
\\
\bold{A}=
\begin{matrix}
c_1 \vspace{-0.0em}\\ 
c_2 \vspace{-0.0em}\\
c_3 \vspace{-0.0em}\\
c_4 \vspace{-0.0em}\\
c_5 \vspace{-0.0em}\\
c_6 \vspace{-0.0em}\\
c_7 \vspace{-0.0em}\\
c_8 \vspace{-0.0em}\\
c_9 \\
\end{matrix}
\begin{bmatrix}
w_{1,1} & \alpha & 0 & 0 & 0 & 0 \vspace{-0.0em}\\
0 & \alpha^2 & \alpha^2 & 0 & 0 & 0 \vspace{-0.0em}\\
0 & 0 & 1 & \alpha^2 & 0 & 0 \vspace{-0.0em}\\
0 & 0 & 0 & \alpha^2 & 1 & 0 \vspace{-0.0em}\\
0 & 0 & 0 & 0 & 1 & 1 \vspace{-0.0em}\\
w_{6,1} & 0 & 0 & 0 & 0 & \alpha \vspace{-0.0em}\\
0 & \alpha & 0 & 1 & 0 & 0 \vspace{-0.0em}\\
\alpha & 0 & 0 & 0 & \alpha^2 & 0 \vspace{-0.0em}\\
0 & 0 & 1 & 0 & 0 & 1
\end{bmatrix}.
\end{gather*}
For the original configuration, we assume that $w_{1,1}=w_{6,1} \allowbreak =1$. The unlabeled GAST tree of this configuration reveals that it is neither u-symmetric nor same-size-WCMs. The configuration has $10$ WCMs (of different sizes), extracted by removing the rows of the following groups of CNs from $\bold{A}$: $\{(c_1, c_3, c_5), (c_1, c_4, c_9), (c_2, c_4, c_6), (c_2, c_5), (c_2, c_8), (c_3, c_6), \allowbreak (c_3, c_8), (c_5, c_7), (c_6, c_7), (c_7, c_8, c_9)\}$. We index these groups of CNs (and consequently, the resulting WCMs) by $h$, $1 \leq h \leq t=10$. The WCM of interest in this example is $\bold{W}^{\textup{cm}}_2$, which is extracted by removing the rows of $(c_1, c_4, c_9)$ from $\bold{A}$. The graph corresponding to $\bold{W}^{\textup{cm}}_2$, which is $G^{\textup{cm}}_2$, is shown in Fig.~\ref{Figure_9}(b). Note that this graph has $\delta_2=2$ disconnected components. For the given edge weight assignment, $\bold{W}^{\textup{cm}}_2$ (as well as all the remaining $9$ WCMs) has unbroken weight conditions. Thus, according to Theorem~\ref{th_dim_null}, $\textup{dim}\left (\mathcal{N}(\bold{W}^{\textup{cm}}_2) \right ) = \sum_{k=1}^{2}\textup{dim}\left (\mathcal{N}\left (\bold{M}(G^{\textup{disc}}_{2,k}) \right ) \right ) \geq 2$ must be satisfied. Solving for the null space of $\bold{W}^{\textup{cm}}_2$ yields:
\begin{equation}\label{eq_null_od2}
\mathcal{N}(\bold{W}^{\textup{cm}}_2)=\textup{span}\{[\alpha \text{ } 0 \text{ } 0 \text{ } 0 \text{ } 1 \text{ } 1]^\textup{T}, [0 \text{ } 1 \text{ } 1 \text{ } \alpha \text{ } 0 \text{ } 0]^\textup{T}\}, 
\end{equation}
which means that $\textup{dim}\left (\mathcal{N}(\bold{W}^{\textup{cm}}_2) \right ) = 2$, and the reason is that $\textup{dim}\left (\mathcal{N}\left (\bold{M}(G^{\textup{disc}}_{2,k}) \right ) \right ) = 1$, $\forall k \in \{1, 2\}$.~Note that $\mathcal{N}\left (\bold{M}(G^{\textup{disc}}_{2,1}) \right ) = \textup{span}\{[\alpha \text{ } 1 \text{ } 1]^\textup{T}\}$, where $G^{\textup{disc}}_{2,1}$ is the subgraph grouping $\{v_1, v_5, v_6\}$ in Fig.~\ref{Figure_9}(b), while $\mathcal{N}\left (\bold{M}(G^{\textup{disc}}_{2,1}) \right ) = \textup{span}\{[1 \text{ } 1 \text{ } \alpha]^\textup{T}\}$, where $G^{\textup{disc}}_{2,2}$ is the subgraph grouping $\{v_2, v_3, v_4\}$ in Fig.~\ref{Figure_9}(b). Observe that the existance of the vector:
\begin{align}
\bold{v}&=[\alpha \text{ } 0 \text{ } 0 \text{ } 0 \text{ } 1 \text{ } 1]^\textup{T} + [0 \text{ } 1 \text{ } 1 \text{ } \alpha \text{ } 0 \text{ } 0]^\textup{T} \nonumber \\ &=[\alpha \text{ } 1 \text{ } 1 \text{ } \alpha \text{ } 1 \text{ } 1]^\textup{T} \in \mathcal{N}(\bold{W}^{\textup{cm}}_2)
\end{align}
verifies that the weight conditions of $\bold{W}^{\textup{cm}}_2$ are unbroken.

Now, assume that in the process of removing the GAST, we break the weight conditions of $\bold{W}^{\textup{cm}}_2$ via the following set of a single edge weight change: $\{w_{6,1}:1 \rightarrow \alpha^2\}$. This change results in breaking the weight conditions of $\bold{W}^{\textup{cm}}_2$, i.e., $\nexists \text{ } \bold{v}=[v_1 \text{ } v_2 \text{ } \dots \text{  } v_6] \in \mathcal{N}(\bold{W}^{\textup{cm}}_2)$ s.t. $v_f \neq 0$, $\forall f \in \{1, 2, \dots, 6\}$. However, $\mathcal{N}(\bold{W}^{\textup{cm}}_2) = \textup{span}\{[0 \text{ } 1 \text{ } 1 \text{ } \alpha \text{ } 0 \text{ } 0]^\textup{T}\} \neq \{\bold{0}\}$, i.e., $\textup{dim}\left (\mathcal{N}(\bold{W}^{\textup{cm}}_2) \right ) = 1$ (it was originally $2$). This is an example of how the weight conditions of a WCM that has a corresponding graph with $\delta_h > 1$ can be broken while $\textup{dim}\left (\mathcal{N}(\bold{W}^{\textup{cm}}_h) \right ) > 0$ (for $h=2$ here). Obviously, it is possible to make another edge weight change for an edge in $G^{\textup{disc}}_{2,2}$ to make $\mathcal{N}(\bold{W}^{\textup{cm}}_2) = \{\bold{0}\}$. However, this is by no means necessary for the GAST removal process.
\end{example}

%%%%%%%%%%%%%%%%%%%%%%%%%%%%%%
\subsection{Breaking the Weight Conditions of Short WCMs}\label{subsec_break}

In this subsection, we discuss the best that can be done to break the weight conditions of a short WCM. The following lemma states this result.

\begin{lemma}\label{th_short_wcm}
The null space of a short WCM $\bold{W}^{\textup{cm}}_h$ (a WCM that has fewer rows than columns) after a successful removal process for the $(a, b, d_1, d_2, d_3)$ GAST to which this WCM belongs satisfies the following two conditions\footnote{Note that Lemma~\ref{th_short_wcm} also applies to any WCM $\bold{W}^{\textup{cm}}_h$ that has $G^{\textup{cm}}_h$ with $\delta_h > 1$ and at least one of the disconnected components having a short adjacency matrix (a very rare case).}:
\begin{enumerate}
\item Its dimension is strictly more than $0$, i.e., $p_h = \textup{dim}\left (\mathcal{N}(\bold{W}^{\textup{cm}}_h) \right ) > 0$.
\item For any $\bold{v}=[v_1 \text{ } v_2 \text{ } \dots \text{  } v_a]^\textup{T} \in \mathcal{N}(\bold{W}^{\textup{cm}}_h)$, where $a$ is the size of the GAST, the number of non-zero elements in $\bold{v}$ is strictly less than $a$, i.e., $\left \| \bold{v} \right \|_0 < a$.
\end{enumerate}
\end{lemma}

\begin{IEEEproof}
Since the number of rows is less than the number of columns in this WCM, the WCM cannot have a full column rank. Thus, $\mathcal{N}\left (\bold{W}^{\textup{cm}}_h \right ) \neq \{\bold{0}\}$, which means $\textup{dim}\left (\mathcal{N}\left (\bold{W}^{\textup{cm}}_h \right ) \right ) > 0$, and proves the first condition in the lemma. Moreover, if a GAST is removed successfully, this implies that each WCM in the set $\mathcal{W}$ associated with that GAST has broken weight conditions. Thus, the second condition in the lemma is satisfied for the short WCM.
\end{IEEEproof}

Lemma~\ref{th_short_wcm} further emphasizes on the fact that the weight conditions of a WCM $\bold{W}^{\textup{cm}}_h$ can be broken while $\textup{dim}\left (\mathcal{N}\left (\bold{W}^{\textup{cm}}_h \right ) \right ) > 0$ (i.e., $\mathcal{N}\left (\bold{W}^{\textup{cm}}_h \right ) \neq \{\bold{0}\}$). One way this case can happen is if $\delta_h > 1$ (which is discussed in the previous subsection). Another way is if the WCM is short, even with $\delta_h=1$. For many short WCMs, the reason why this case occurs is that before breaking the weight conditions of a short WCM, it typically has $p_h=\textup{dim}\left (\mathcal{N}\left (\bold{W}^{\textup{cm}}_h \right ) \right ) > \delta_h$. Here, we are more interested in short WCMs with $\delta_h = 1$. The difference between the two ways is that if $\delta_h > 1$ and there does not exist any disconnected component having a short adjacency matrix, we can still break the weight conditions of the WCM by making $\textup{dim}\left (\mathcal{N}\left (\bold{W}^{\textup{cm}}_h \right ) \right ) = 0$. However, such processing is not necessary, and it would require more edge weight changes than the minimum needed. Contrarily, if the WCM is short, it is impossible to break the weight conditions by making $\textup{dim}\left (\mathcal{N}\left (\bold{W}^{\textup{cm}}_h \right ) \right ) = 0$, and the best we can do is what is described in Lemma~\ref{th_short_wcm}. The following example demonstrates Lemma~\ref{th_short_wcm}.

\begin{figure}
\vspace{-0.7em}
\center
\includegraphics[trim={0.08in 0.0in 0.07in 0.0in},clip,width=3.5in]{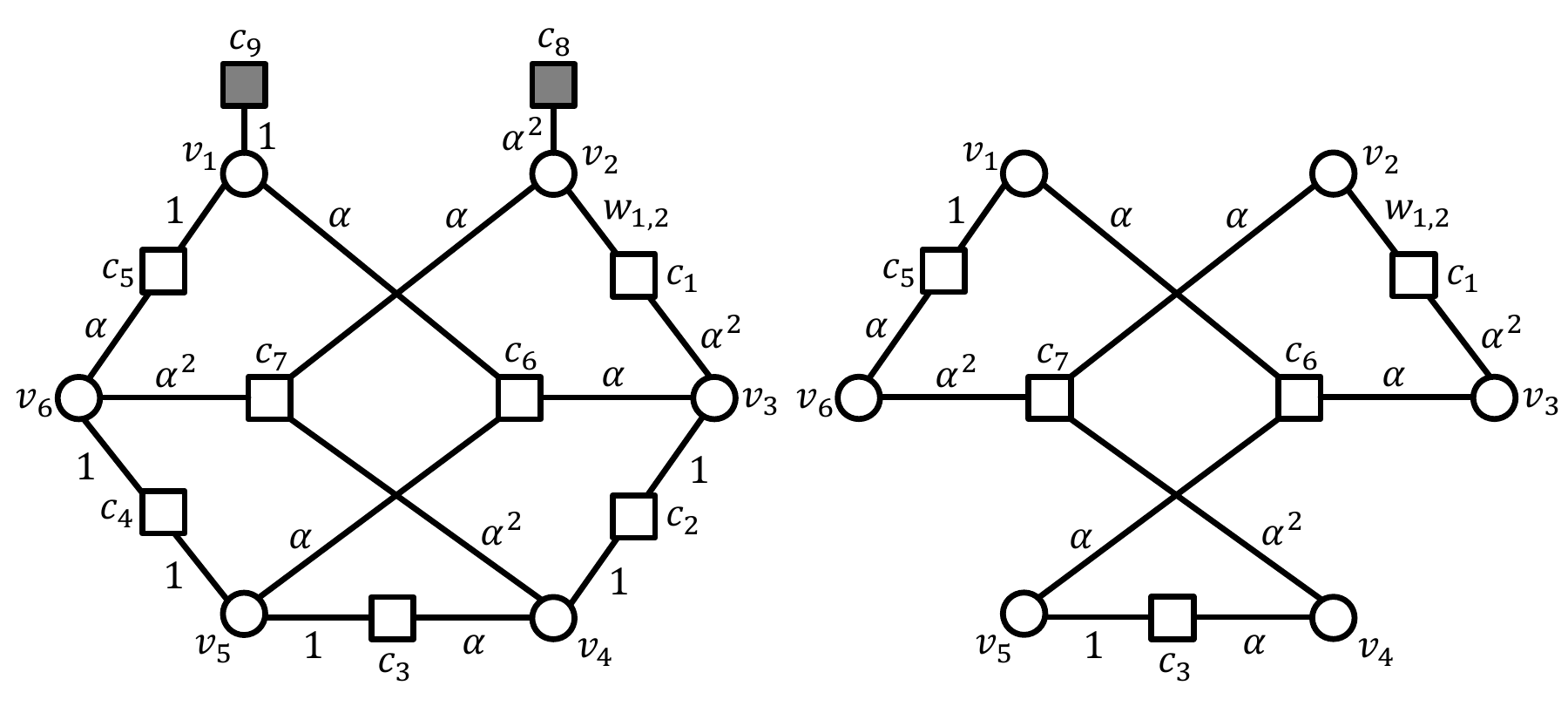}\vspace{-0.5em}
\text{\hspace{0.6em} \footnotesize{(a) \hspace{14em} (b) \hspace{1em}}}
\caption{(a) A $(6, 2, 2, 5, 2)$ GAST for $\gamma=3$. (b) The subgraph created by removing $(c_2, c_4, c_8, c_9)$ from the GAST subgraph. GF($4$) is assumed.}
\label{Figure_10}
\vspace{-0.5em}
\end{figure}

\vspace{-0.1em}
\begin{example}\label{ex_10}
Consider the $(6, 2, 2, 5, 2)$ GAST ($\gamma=3$) over GF($4$), where GF($4$) $=\{0, 1 , \alpha, \alpha^2\}$ and $\alpha$ is a primitive element, that is shown in Fig.~\ref{Figure_10}(a). The matrix $\bold{A}$ of this configuration is:
\vspace{-0.5em}
\begin{gather*}\label{ex_wcms}
\begin{matrix}
\text{  }\text{  }\text{  }\text{  }\text{  }\text{  }\text{   } v_1 & v_2 \text{   } & v_3 \text{   } & v_4 & v_5 & v_6 \vspace{-0.3em}
\end{matrix}
\\
\bold{A}=
\begin{matrix}
c_1 \vspace{-0.0em}\\ 
c_2 \vspace{-0.0em}\\
c_3 \vspace{-0.0em}\\
c_4 \vspace{-0.0em}\\
c_5 \vspace{-0.0em}\\
c_6 \vspace{-0.0em}\\
c_7 \vspace{-0.0em}\\
c_8 \vspace{-0.0em}\\
c_9 \\
\end{matrix}
\begin{bmatrix}
0 & w_{1,2} & \alpha^2 & 0 & 0 & 0 \vspace{-0.0em}\\
0 & 0 & 1 & 1 & 0 & 0 \vspace{-0.0em}\\
0 & 0 & 0 & \alpha & 1 & 0 \vspace{-0.0em}\\
0 & 0 & 0 & 0 & 1 & 1 \vspace{-0.0em}\\
1 & 0 & 0 & 0 & 0 & \alpha \vspace{-0.0em}\\
\alpha & 0 & \alpha & 0 & \alpha & 0 \vspace{-0.0em}\\
0 & \alpha & 0 & \alpha^2 & 0 & \alpha^2 \vspace{-0.0em}\\
0 & \alpha^2 & 0 & 0 & 0 & 0 \vspace{-0.0em}\\
1 & 0 & 0 & 0 & 0 & 0
\end{bmatrix}.
\end{gather*}
For the original configuration, we assume that $w_{1,2}=\alpha^2$. From Example~\ref{ex_1}, this configuration has $2$ WCMs, extracted by removing the rows of the following groups of CNs from $\bold{A}$: $\{(c_3,\mathcal{O}_\textup{sg}), (c_2, c_4,\mathcal{O}_\textup{sg})\}$, where $\mathcal{O}_\textup{sg}$ is $(c_8, c_9)$. We index these groups of CNs (and consequently, the resulting WCMs) by $h$, $1 \leq h \leq t=2$. {The WCM of interest in this example is $\bold{W}^{\textup{cm}}_2$, which is extracted by removing the rows of $(c_2, c_4, c_8, c_9)$ from $\bold{A}$.} The graph corresponding to $\bold{W}^{\textup{cm}}_2$, which is $G^{\textup{cm}}_2$, is shown in Fig.~\ref{Figure_10}(b). Note that $\bold{W}^{\textup{cm}}_2$ is of size $5 \times 6$ (a short matrix). For the given edge weight assignment, $\bold{W}^{\textup{cm}}_2$ (as well as $\bold{W}^{\textup{cm}}_1$) has unbroken weight conditions. Solving for the null space of $\bold{W}^{\textup{cm}}_2$ yields the following:
\begin{equation}\label{eq_null_od2}
\mathcal{N}(\bold{W}^{\textup{cm}}_2)=\textup{span}\{[0 \text{ } 1 \text{ } 1 \text{ } \alpha^2 \text{ } 1 \text{ } 0]^\textup{T}, [1 \text{ } 1 \text{ } 1 \text{ } 0 \text{ } 0 \text{ } \alpha^2]^\textup{T}\}, 
\end{equation}
This is one of the cases where we have $\textup{dim}\left (\mathcal{N}\left (\bold{W}^{\textup{cm}}_h \right ) \right ) > 1$ (for $h=2$) with $\delta_h = 1$ (the corresponding graph to $\bold{W}^{\textup{cm}}_2$ is fully connected). Observe also that the existance of the vector:
\begin{align}
\bold{v}&=[0 \text{ } 1 \text{ } 1 \text{ } \alpha^2 \text{ } 1 \text{ } 0]^\textup{T} + \alpha[1 \text{ } 1 \text{ } 1 \text{ } 0 \text{ } 0 \text{ } \alpha^2]^\textup{T} \nonumber \\ &= [\alpha \text{ } \alpha^2 \text{ } \alpha^2 \text{ } \alpha^2 \text{ } 1 \text{ } 1]^\textup{T} \in \mathcal{N}(\bold{W}^{\textup{cm}}_2)
\end{align}
verifies that the weight conditions of $\bold{W}^{\textup{cm}}_2$ are unbroken.

Now, assume that in the process of removing the GAST, we break the weight conditions of $\bold{W}^{\textup{cm}}_2$ via the following set of a single edge weight change: $\{w_{1,2}:\alpha^2 \rightarrow \alpha\}$. This change results in breaking the weight conditions of $\bold{W}^{\textup{cm}}_2$, i.e., $\nexists \text{ } \bold{v}=[v_1 \text{ } v_2 \text{ } \dots \text{  } v_6] \in \mathcal{N}(\bold{W}^{\textup{cm}}_2)$ s.t. $v_f \neq 0$, $\forall f \in \{1, 2, \dots, 6\}$. However, $\mathcal{N}(\bold{W}^{\textup{cm}}_2) = \textup{span}\{[1 \text{ } 0 \text{ } 0 \text{ } \alpha^2 \text{ } 1 \text{ } \alpha^2]^\textup{T}\} \neq \{\bold{0}\}$, i.e., $\textup{dim}\left (\mathcal{N}(\bold{W}^{\textup{cm}}_2) \right ) = 1$ (it was originally $2$). This example illustrates that the weight conditions of the short WCM can only be broken with $\textup{dim}\left (\mathcal{N}(\bold{W}^{\textup{cm}}_2) \right ) > 0$ regardless of the edge weight change(s) we perform.
\end{example}

%%%%%%%%%%%%%%%%%%%%%%%%%%%%%%
\vspace{-0.2em}
\subsection{The Number of Edge Weight Changes Needed}\label{subsec_num}

In this subsection, we discuss the minimum number of edge weight changes needed, in addition to how to select these edge weight changes in order to have a successful removal of the problematic object. Recall that we need to break the weight conditions of all the WCMs of a GAST in order to remove the GAST.

\begin{lemma}\label{lem_emin_gas}
The minimum number of edge weight changes (with respect to the original configuration) needed to remove an $(a, b, b_2, d_1, d_2, d_3)$ GAS (convert it into a non-AS, i.e., a non-GAS) is given by:
\begin{equation}\label{eq_emin_gas}
E_{\textup{GAS},\textup{min}}=g-b_{\textup{vn},\textup{max}}+1,
\end{equation}
where $g=\left \lfloor \frac{\gamma-1}{2} \right \rfloor$, and $b_{\textup{vn},\textup{max}}$ is the maximum number of existing unsatisfied CNs per VN in the GAS. A topological upper bound on that minimum is given by:
\begin{equation}\label{eq_emin_bnd}
E_{\textup{GAS},\textup{min}} \leq g-d_{1,\textup{vn},\textup{max}}+1,
\end{equation}
where $d_{1,\textup{vn},\textup{max}}$ is the maximum number of existing degree-$1$ CNs per VN in the GAS. 
\end{lemma}

\begin{IEEEproof}
The set of GASs is simply the set of absorbing sets (ASs). Thus, the proof of (\ref{eq_emin_gas}) is exactly the same as the proof of $E_{\textup{AS},\textup{min}}$ in \cite[Lemma~2]{ahh_bas}.

Now, we prove the upper bound in (\ref{eq_emin_bnd}). Recall that degree-$1$ CNs are always unsatisfied. Thus, irrespective of whether the VN that has the maximum number of existing unsatisfied CNs is the same as the VN that has the maximum number of existing degree-$1$ CNs or not, the following inequality is always satisfied:
\begin{equation}\label{eq_lem6_pr}
b_{\textup{vn},\textup{max}} \geq d_{1,\textup{vn},\textup{max}}.
\end{equation}
Substituting (\ref{eq_lem6_pr}) in (\ref{eq_emin_gas}) gives (\ref{eq_emin_bnd}), and completes the proof.
\end{IEEEproof}

While theoretically a GAST can be removed by forcing a single degree $> 2$ CN (if any) to be unsatisfied via a single edge weight change, this is not the strategy we follow. The reason is that the described process can result in another GAS with a degree $> 2$ unsatisfied CN ($b > d_1+b_2$). Despite that GASs with $b > d_1+b_2$ are generally less harmful than GASTs, it is not preferred to remove GASTs by converting some of them into other types of GASs. Thus, we remove a GAST by performing $E_{\textup{GAST},\textup{min}}=E_{\textup{GAS},\textup{min}}$ edge weight changes for edges connected to only degree-$2$ CNs (all the weights of edges connected to degree $> 2$ CNs remain untouched). In other words, (\ref{eq_emin_gas}) and (\ref{eq_emin_bnd}) are applicable to both GASTs and GASs. Furthermore, $E_{\textup{GAST},\textup{min}}$ we use in \cite[Algorithm~2]{ahh_jsac} (see Appendix~\ref{sec_appb}) is given by (\ref{eq_emin_gas}) and bounded by (\ref{eq_emin_bnd}).

The upper bound in (\ref{eq_emin_bnd}) is purely topological (determined from the unlabeled GAST), i.e., it does not require any knowledge of $b$ of the GAST being processed (nor $b_{\textup{vn},\textup{max}}$, consequently). The importance of this topological bound will be illustrated shortly.

A useful definition and a subsequent corollary, which are used to simplify the process of selecting $E_{\textup{GAST},\textup{min}}$ edge weights to change, are proposed below.

\begin{definition}
A \textbf{borderline VN} in an $(a, b, d_1, d_2, d_3)$ GAST in a code with column weight $\gamma$ is a VN that is connected to exactly $g=\left \lfloor \frac{\gamma-1}{2} \right \rfloor$ degree-$1$ CNs.
\end{definition}

\begin{corollary}\label{cor_emin_1}
An $(a, b, d_1, d_2, d_3)$ GAST that has at least one borderline VN has $E_{\textup{GAST},\textup{min}}=1$, and the upper bound on $E_{\textup{GAST},\textup{min}}$ is also $1$.
\end{corollary}

\begin{IEEEproof}
A borderline VN already has the maximum number of unsatisfied CNs a VN can have in a GAST (or in an AS in general), which is $g=\left \lfloor \frac{\gamma-1}{2} \right \rfloor$. Consequently, a GAST with at least one borderline VN has:
\begin{equation}\label{eq_cor6_pr}
b_{\textup{vn},\textup{max}} = d_{1,\textup{vn},\textup{max}} = g = \left \lfloor \frac{\gamma-1}{2} \right \rfloor.
\end{equation}
Substituting (\ref{eq_cor6_pr}) in (\ref{eq_emin_gas}) gives $E_{\textup{GAST},\textup{min}}=1$. Noting that $b_{\textup{vn},\textup{max}} = d_{1,\textup{vn},\textup{max}}$ proves that the upper bound is also $1$ (by substituting in (\ref{eq_emin_bnd})).
\end{IEEEproof}

\begin{remark}\label{rem_7}
Lemma~\ref{lem_emin_gas}, Corollary~\ref{cor_emin_1}, and the discussion below give the minimum number of edge weight changes in addition to the specification of which edge weights need to be changed in order to remove a GAST. However, they do not determine what these particular changes should be, i.e., they do not specify the new values of the edge weights to be changed. Specifying the new values of the edge weights is performed by the WCM framework via checking the null spaces of all WCMs of the GAST being processed, and making sure that all the WCMs have broken weight conditions after the edge weight changes (see also \cite[Theorem~3]{ahh_jsac}, \cite[Algorithm~2]{ahh_jsac}, and Example~\ref{ex_12}). This justification is the reason why the word ``\textbf{properly}'' is used to describe the edge weight changes in this subsection.
\end{remark}

It can be concluded from Corollary~\ref{cor_emin_1} that any degree-$2$ unsatisfied CN connected to a borderline VN results in an object that is not a GAST. Thus, assuming that the object being processed is identified to be a GAST via the WCM framework (at least one of its WCMs has unbroken weight conditions), we select the edge weights to be changed based on the following two cases\footnote{Note that these two items are for a stand-alone GAST. It happens in few cases that we need more than the minimum number of edge weight changes to remove a GAST because of previously removed GASTs that share edges with the GAST being processed (or other reasons).
}:

\begin{enumerate}
\item If the GAST has at least one borderline VN ($E_{\textup{GAST},\textup{min}}=1$), then we \textbf{properly} change the weight of an edge connected to a degree-$2$ CN connected to any of the borderline VNs. If every VN in the GAST is borderline, then we change the weight of an edge connected to any degree-$2$ CN.
\item If the GAST does not have any borderline VNs ($E_{\textup{GAST},\textup{min}} \geq 1$), then we determine the VN(s) that has (have) the maximum number, $d_{1,\textup{vn},\textup{max}}$, of degree-$1$ CNs connected to it (them). Then we \textbf{properly} change the weights of a maximum of $(g-d_{1,\textup{vn},\textup{max}}+1)$ edges connected to different degree-$2$ CNs connected to a particular VN of those having $d_{1,\textup{vn},\textup{max}}$ neighboring degree-$1$ CNs.
\end{enumerate}

To relate the above analysis to the WCMs, recall that every CN in a GAST has a corresponding row in the matrix $\bold{A}$ of this GAST. The GAST is removed by breaking the weight conditions of all its WCMs. To achieve this, we operate on a set of rows in $\bold{A}$, that has the minimum cardinality and its rows correspond to degree-$2$ CNs, with the property that every WCM has at least one row in that set. Any set of $(g-d_{1,f}+1)$ rows in $\bold{A}$ satisfies the stated property if they correspond to degree-$2$ CNs connected to the same VN, $v_f$, $f \in \{1, 2, \dots, a\}$, where $d_{1,f}$ is the number of degree-$1$ CNs connected to VN $v_f$. The reason is that we cannot together remove $(g-d_{1,f}+1)$ rows of degree-$2$ CNs connected to VN $v_f$ from $\bold{A}$ to extract a WCM since the resulting matrix then will not be a valid $\bold{W}^{\textup{z}}$. Thus, a set of $(g-d_{1,\textup{vn},\textup{max}}+1)$ rows of degree-$2$ CNs connected to the same VN that achieves $d_{1,\textup{vn},\textup{max}}$ is indeed a set of minimum cardinality with the property that every WCM has at least one row in that set. Consequently, the topological upper bound in (\ref{eq_emin_bnd}) provides the cardinality of that set of rows satisfying the stated property. Properly operating on a maximum of $(g-d_{1,\textup{vn},\textup{max}}+1)$ weights in these rows (only one weight per row) is what is needed to remove the GAST. Examples~\ref{ex_12} and \ref{ex_13} illustrate the process performed by the WCM framework to remove a stand-alone GAST.

\begin{remark}\label{rem_8}
Typically, we only need to perform $(g-b_{\textup{vn},\textup{max}}+1) \leq (g-d_{1,\textup{vn},\textup{max}}+1)$ edge weight changes to remove the GAST. When $b_{\textup{vn},\textup{max}} \neq d_{1,\textup{vn},\textup{max}}$, the number of WCMs with unbroken weight conditions becomes strictly less than $t$, and only $(g-b_{\textup{vn},\textup{max}}+1)$ rows are enough to establish a set of minimum cardinality with the property that every WCM with unbroken weight conditions has at least one row in that set.
\end{remark}

\begin{example}\label{ex_12}
We again discuss the $(6, 0, 0, 9, 0)$ GAST ($\gamma=3$) in Fig.~\ref{Figure_9}(a), with $w_{1,1}=w_{6,1}=1$. The null spaces of the $10$ WCMs of that GAST are given in Example~\ref{ex_11}, (\ref{eq_ex11_1}) (see Example~\ref{ex_9} for how the WCMs are extracted). Since $b_{\textup{vn},\textup{max}}=d_{1,\textup{vn},\textup{max}}=0$, (\ref{eq_emin_gas}) gives $E_{\textup{GAST},\textup{min}}=\left \lfloor \frac{3-1}{2} \right \rfloor-0+1=2$ (same as the upper bound). Given that all VNs have $0$ degree-$1$ neighboring CNs, any VN can be selected. Suppose that $v_1$ is selected. Each of the $10$ WCMs has at least one of the two rows corresponding to $c_1$ and $c_6$ (both are connected to $v_1$) in $\bold{A}$. Thus, we consider the following two sets of edge weight changes for $w_{1,1}$ and $w_{6,1}$ (cardinality $2$). The first set is $\{w_{1,1}:1 \rightarrow \alpha, w_{6,1}:1 \rightarrow \alpha\}$. Using this set of edge weight changes, the null spaces of the $10$ WCMs become:
\begin{align}\label{eq_ex12_1}
\mathcal{N}(\bold{W}^{\textup{cm}}_1) &= \mathcal{N}(\bold{W}^{\textup{cm}}_3) = \mathcal{N}(\bold{W}^{\textup{cm}}_4) \nonumber \\ &= \mathcal{N}(\bold{W}^{\textup{cm}}_6) = \mathcal{N}(\bold{W}^{\textup{cm}}_8) = \mathcal{N}(\bold{W}^{\textup{cm}}_9) = \{\bold{0}\}, \nonumber \\
\mathcal{N}(\bold{W}^{\textup{cm}}_2)&=\textup{span}\{[0 \text{ } 1 \text{ } 1 \text{ } \alpha \text{ } 0 \text{ } 0]^\textup{T}\}, \text{ and} \nonumber \\
\mathcal{N}(\bold{W}^{\textup{cm}}_5)&= \mathcal{N}(\bold{W}^{\textup{cm}}_7) = \mathcal{N}(\bold{W}^{\textup{cm}}_{10}) = \textup{span}\{[1 \text{ } 1 \text{ } 1 \text{ } \alpha \text{ } 1 \text{ } 1]^\textup{T}\}.
\end{align}
Clearly, the GAST is not removed as there are $3$ WCMs with unbroken weight conditions: $\bold{W}^{\textup{cm}}_5$, $\bold{W}^{\textup{cm}}_7$, and $\bold{W}^{\textup{cm}}_{10}$. The second set is $\{w_{1,1}:1 \rightarrow \alpha, w_{6,1}:1 \rightarrow \alpha^2\}$. Using this set of edge weight changes, the null spaces of the $10$ WCMs become:
\vspace{-0.1em}\begin{align}\label{eq_ex12_2}
\mathcal{N}(\bold{W}^{\textup{cm}}_1) &= \mathcal{N}(\bold{W}^{\textup{cm}}_3) = \mathcal{N}(\bold{W}^{\textup{cm}}_4) \nonumber \\ &= \mathcal{N}(\bold{W}^{\textup{cm}}_5) = \mathcal{N}(\bold{W}^{\textup{cm}}_6) = \mathcal{N}(\bold{W}^{\textup{cm}}_7) \nonumber \\ &= \mathcal{N}(\bold{W}^{\textup{cm}}_8) = \mathcal{N}(\bold{W}^{\textup{cm}}_9) = \mathcal{N}(\bold{W}^{\textup{cm}}_{10}) = \{\bold{0}\} \text{ and} \nonumber \\
\mathcal{N}(\bold{W}^{\textup{cm}}_2)&=\textup{span}\{[0 \text{ } 1 \text{ } 1 \text{ } \alpha \text{ } 0 \text{ } 0]^\textup{T}\},
\end{align}
which means that the GAST is successfully removed as the $10$ WCMs have broken weight conditions. As a result, it can be concluded that properly identifying which edge weights to change is important but not enough. Checking the null spaces of all WCMs is what determines which set of edge weight changes is sufficient for a successful GAST removal.

Now, consider the case of $w_{1,1}=\alpha$ and $w_{6,1}=1$ for the original configuration (i.e., before any removal attempt). The configuration in this case is a $(6, 1, 0, 9, 0)$ GAST with $b_{\textup{vn},\textup{max}}=1$ and $d_{1,\textup{vn},\textup{max}}=0$. Thus, (\ref{eq_emin_gas}) gives $E_{\textup{GAST},\textup{min}}=1$, while the upper bound is $2$ from (\ref{eq_emin_bnd}). The null spaces of the $10$ WCMs are:
\begin{align}\label{eq_ex12_3}
\mathcal{N}(\bold{W}^{\textup{cm}}_1) &=\textup{span}\{[\alpha \text{ } 1 \text{ } 1 \text{ } \alpha \text{ } 1 \text{ } 1]^\textup{T}\}, \nonumber \\ \mathcal{N}(\bold{W}^{\textup{cm}}_2) &=\textup{span}\{[\alpha \text{ } 0 \text{ } 0 \text{ } 0 \text{ } 1 \text{ } 1]^\textup{T}, [0 \text{ } 1 \text{ } 1 \text{ } \alpha \text{ } 0 \text{ } 0]^\textup{T}\}, \text{ and} \nonumber \\
\mathcal{N}(\bold{W}^{\textup{cm}}_3) &= \mathcal{N}(\bold{W}^{\textup{cm}}_4) = \mathcal{N}(\bold{W}^{\textup{cm}}_5) = \mathcal{N}(\bold{W}^{\textup{cm}}_6) = \mathcal{N}(\bold{W}^{\textup{cm}}_7) \nonumber \\ &= \mathcal{N}(\bold{W}^{\textup{cm}}_8) = \mathcal{N}(\bold{W}^{\textup{cm}}_9) = \mathcal{N}(\bold{W}^{\textup{cm}}_{10}) = \{\bold{0}\}.
\end{align}
Only $2$ WCMs, $\bold{W}^{\textup{cm}}_1$ and $\bold{W}^{\textup{cm}}_2$, have unbroken weight conditions. Both of them share the row corresponding to $c_6$. Consequently, only one edge weight change is needed to break the weight conditions of the $2$ WCMs and remove the object ($E_{\textup{GAST},\textup{min}}$ is achieved). A set of a single edge weight change, e.g., $\{w_{6,1}:1 \rightarrow \alpha^2\}$, is sufficient to perform the removal (see also Remark~\ref{rem_8}).
\end{example}

\begin{example}\label{ex_13}
We discuss the $(6, 2, 2, 5, 2)$ GAST ($\gamma=3$) in Fig.~\ref{Figure_10}(a), with $w_{1,2}=\alpha^2$. The null spaces of the $2$ WCMs of that GAST are given in Example~\ref{ex_11}, (\ref{eq_ex11_2}) (see Example~\ref{ex_10} for how the WCMs are extracted). Since $b_{\textup{vn},\textup{max}}=d_{1,\textup{vn},\textup{max}}=1$, (\ref{eq_emin_gas}) gives $E_{\textup{GAST},\textup{min}}=1$ (same as the upper bound). Either $v_1$ or $v_2$ can be selected as both are borderline VNs. Suppose that $v_2$ is selected. Each of the $2$ WCMs has the row of $c_1$ (see Example~\ref{ex_10}). Applying the set of a single edge weight change, $\{w_{1,2}:\alpha^2 \rightarrow \alpha\}$, yields the following null spaces:
\vspace{-0.1em}\begin{align}\label{eq_ex12_4}
\mathcal{N}(\bold{W}^{\textup{cm}}_1)&= \{\bold{0}\} \text{ and } \mathcal{N}(\bold{W}^{\textup{cm}}_2)=\textup{span}\{[1 \text{ } 0 \text{ } 0 \text{ } \alpha^2 \text{ } 1 \text{ } \alpha^2]^\textup{T}\},
\end{align}
which means that the GAST is successfully removed.
\end{example}

%%%%%%%%%%%%%%%%%%%%%%%%%%%%%%
\vspace{-0.2em}
\section{Removing Oscillating Sets to Achieve More Gain}\label{sec_os}

Now that we have presented the in-depth analysis of the baseline WCM framework, we are ready to introduce an extension to the framework. In particular, in this section, we discuss a new set of detrimental objects, namely oscillating sets of type two (OSTs), that are the second-order cause of the error floor of NB-LDPC codes with even column weights over asymmetric channels. We show how to remove OSTs using the WCM framework. In the simulation results section, we will show that performing another optimization phase that addresses OSTs, after the GASTs removal phase, secures up to nearly $2.5$ orders of magnitude overall performance gain in the error floor region over practical (asymmetric) Flash channels.

%%%%%%%%%%%%%%%%%%%%%%%%%%%%%%
\subsection{Defining OSs and OSTs}\label{subsec_ost}

Before we introduce an oscillating set (OS), we define an oscillating VN.

\begin{definition}\label{def_osc_vn}
Consider a subgraph induced by a subset $\mathcal{V}$ of VNs in the Tanner graph of an NB-LDPC code. Set all the VNs in $\mathcal{V}$ to values $\in$ GF($q$)$\setminus \{0\}$ and set all other VNs to $0$. A VN in $\mathcal{V}$ is said to be an \textbf{oscillating VN} if the number of its neighboring satisfied CNs equals the number of its neighboring unsatisfied CNs for some set of VN values. The set of all oscillating VNs in $\mathcal{V}$ is referred to as $\mathcal{S}$.
\end{definition}

It is clear that for codes with fixed column weights (fixed VN degrees), there can exist an oscillating VN only under the condition that the column weight $\gamma$ is even. Based on Definition~\ref{def_osc_vn}, we define the oscillating set.

\begin{definition}\label{def_os}
Consider a subgraph induced by a subset $\mathcal{V}$ of VNs in the Tanner graph of an NB-LDPC code. Set all the VNs in $\mathcal{V}$ to values $\in$ GF($q$)$\setminus \{0\}$ and set all other VNs to $0$. The set $\mathcal{V}$ is said to be an $(a, b, b_2, d_1, d_2, d_3)$ \textbf{oscillating set (OS)} over GF($q$) if and only if the size of $\mathcal{V}$ is $a$, the number of unsatisfied (resp., degree-$2$ unsatisfied) CNs connected to $\mathcal{V}$ is $b$ (resp., $b_2$), the number of degree-$1$ (resp., $2$ and $> 2$) CNs connected to $\mathcal{V}$ is $d_1$ (resp., $d_2$ and  $d_3$), the set of oscillating VNs $\mathcal{S} \subseteq \mathcal{V}$ is not empty, and each VN (if any) in $\mathcal{V} \setminus \mathcal{S}$ is connected to strictly more satisfied than unsatisfied neighboring CNs, for some set of VN values.
\end{definition}

Recall that $\mathcal{O}$ (resp., $\mathcal{T}$ and $\mathcal{H}$) is the set of degree-$1$ (resp., $2$ and $> 2$) CNs connected to $\mathcal{V}$.

The unlabeled OS is defined in a way similar to the unlabeled GAS except for that Condition 2 in Definition~\ref{def_ugas} is replaced by ``\textit{Each VN in $\mathcal{V}$ has a number of neighbors in $(\mathcal{T} \cup \mathcal{H})$ that is at least the same as its number of neighbors in $\mathcal{O}$.}'' Moreover, \cite[Lemma~1]{ahh_jsac} can be changed to suit OSs by referring to the unlabeled OS instead of the unlabeled GAS in the topological conditions, and by using the following equation instead of (\ref{eq_gas_cond2}) in the weight conditions:
\begin{align}\label{eq_os_cond2}
\left ( \sum\limits_{e=1}^{\ell-b}F\left ( \psi_{e,f} \right ) \right ) \geq \left ( \sum\limits_{k=1}^{b}F\left ( \theta_{k,f} \right ) \right ).
\end{align}
Note that the equality in (\ref{eq_os_cond2}) must hold for at least one VN in $\mathcal{V}$. We also define an oscillating set of type two (OST) as follows.

\begin{definition}\label{def_ost}
An OS that has $d_2 > d_3$ and all the unsatisfied CNs connected to it $\in (\mathcal{O} \cup \mathcal{T})$ (having either degree $1$ or degree $2$), is defined as an $(a, b, d_1, d_2, d_3)$ \textbf{oscillating set of type two (OST)}. Similar to the {unlabeled OS} definition, we also define the \textbf{$(a, d_1, d_2, d_3)$ unlabeled OST}.
\end{definition}

If hard decision, majority rule decoding, e.g., Gallager B decoding \cite{gal_th}, is assumed, an oscillating VN in error receives the exact same number of ``stay'' and ``change'' messages. This observation makes it harder for the decoder to correct it compared with a VN with more neighboring unsatisfied than satisfied CNs. Over aggressively asymmetric channels, oscillating VNs in error are even less likely to be corrected in many cases under soft decision decoding because of the high error magnitudes. Based on our extensive simulations, OSTs typically result in between $5\%$ and $10\%$ of the errors of NB-LDPC codes with even $\gamma$ in the error floor region over practical (asymmetric) Flash channels, making OSTs the second-order cause, after GASTs, of the error floor in such channels. As we shall see in Section~\ref{sec_sim}, removing OSTs from the Tanner graphs of NB-LDPC codes offers about $0.5$ of an order of magnitude or more additional performance gain.

\begin{figure}
\vspace{-0.5em}
\center
\includegraphics[trim={0.07in 0.0in 0.2in 0.0in},clip,width=3.5in]{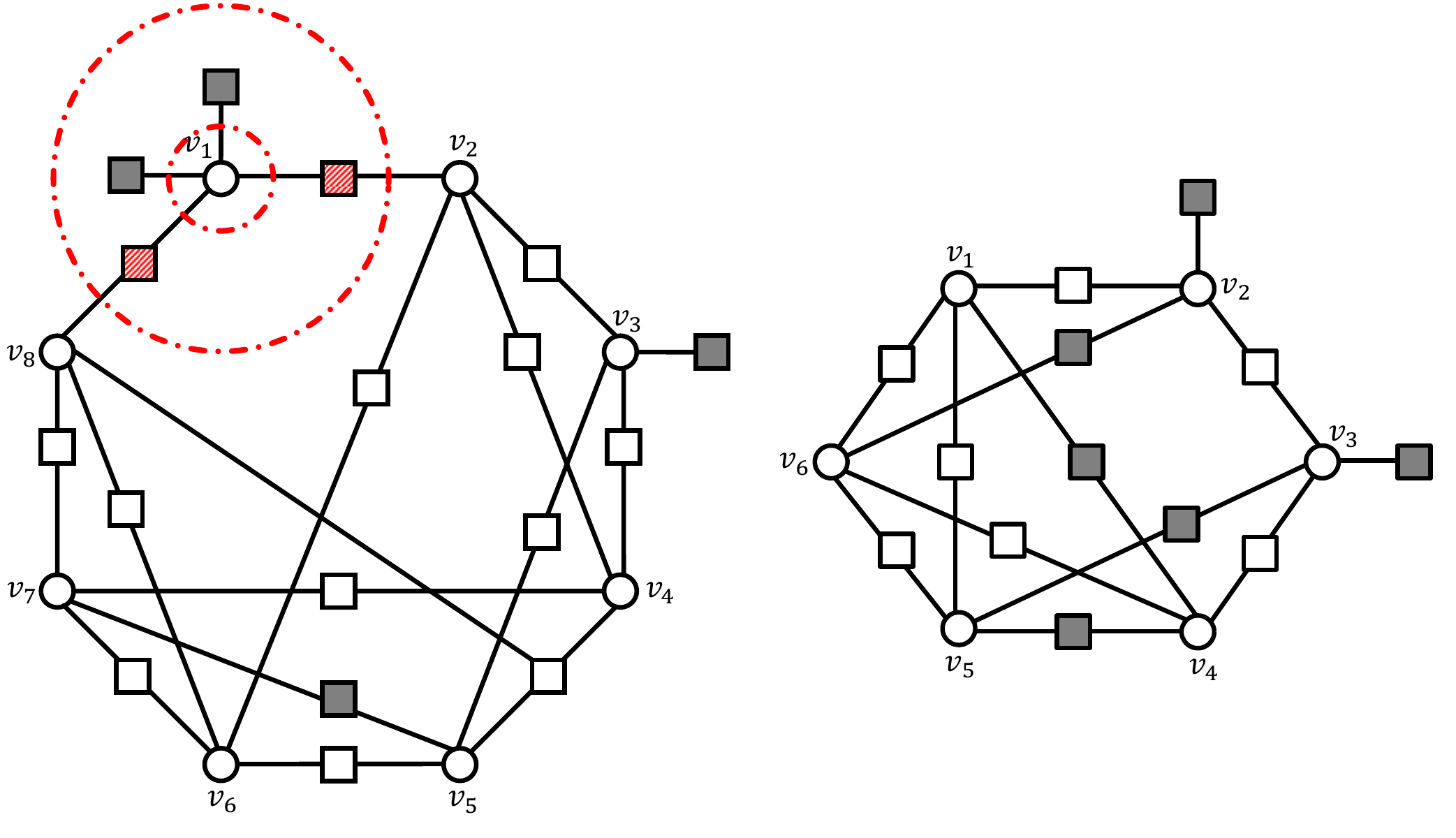}\vspace{-0.5em}
\text{\hspace{0.3em} \footnotesize{(a) \hspace{14.5em} (b) \hspace{1em}}}
\caption{(a) An $(8, 4, 3, 13, 1)$ OST for $\gamma=4$. (b) A $(6, 6, 2, 11, 0)$ OST for $\gamma=4$. Appropriate non-binary edge weights are assumed. Unlabeled OSTs are reached by setting all the weights in the configurations
to $1$.}
\label{Figure_11}
\vspace{-0.5em}
\end{figure}

Fig.~\ref{Figure_11}(a) shows an $(8, 4, 3, 13, 1)$ OST that has $\mathcal{S}=\{v_1\}$. Fig.~\ref{Figure_11}(b) shows a $(6, 6, 2, 11, 0)$ OST that has $\mathcal{S}=\{v_2, v_3, v_4, v_5\}$. Some OSTs have underlying GASTs as subgraphs, while others do not. For example, if the VN $v_1$ is eliminated from the $(8, 4, 3, 13, 1)$ OST in Fig.~\ref{Figure_11}(a), the underlying object is a $(7, 4, 3, 11, 1)$ GAST (the two CNs shaded in red will be degree-$1$ unsatisfied CNs as a result of the elimination of $v_1$). Contrarily, the $(6, 6, 2, 11, 0)$ OST in Fig.~\ref{Figure_11}(b) does not have an underlying GAST.

%%%%%%%%%%%%%%%%%%%%%%%%%%%%%%
\subsection{How to Remove OSTs Using WCMs}\label{subsec_orem}

Before we propose the lemma that discusses the removal of OSTs, we need to state several auxiliary results.

\begin{lemma}\label{lem_odeg2_unsat}
Consider an $(a,d_1,d_2,d_3)$ {unlabeled OST} with its sets $\mathcal{T}$ and $\mathcal{H}$. A CN $c \in \mathcal{T}$ can be unsatisfied in the resulting OST (with proper edge labeling) resulting in $b > d_1$ if and only if the two neighboring VNs of $c$ (with respect to this {unlabeled OST}) each has the property that strictly more than $\frac{\gamma}{2}$ of its neighboring CNs belong to $(\mathcal{T} \cup \mathcal{H})$.
\end{lemma}

\begin{IEEEproof}
The proof follows the same logic as the proof of \cite[Theorem~1]{ahh_jsac}.
\end{IEEEproof}

\begin{lemma}\label{lem_obmax}
Given an $(a, d_1, d_2, d_3)$ {unlabeled OST}, the maximum number of unsatisfied CNs, $b_{\textup{o\_max}}$, in the resulting OST after edge labeling is upper bounded by:
\begin{equation}\label{eq_obmax}
b_{\textup{o\_max}} \leq d_1 + b_{\textup{o\_ut}}, \text{ where}
\end{equation}
\begin{equation}\label{eq_obut}
b_{\textup{o\_ut}} = \left \lfloor \frac{1}{2} \left ( a\left( \frac{\gamma}{2} \right ) - d_1 \right ) \right \rfloor.
\end{equation}
\end{lemma}

\begin{IEEEproof}
The proof follows the same logic as the proof of \cite[Theorem~2]{ahh_jsac}. The main equation in the proof is:
\vspace{-0.1em}\begin{align}\label{eq_lem9_pr}
b_{\textup{o\_ut}}&=\sum_{f=1}^{a}\left [ \frac{\gamma}{2} - b_{\textup{o\_up},f}\right ] = a\left ( \frac{\gamma}{2} \right ) - \left ( d_1 + b_{\textup{o\_ut}} \right ),
\end{align}
where $b_{\textup{o\_ut}}$ is the upper bound on the maximum number of degree-$2$ unsatisfied CNs the resulting OST can have after labeling, {and $b_{\textup{o\_up},f}$ is the number of the already-unsatisfied CNs connected to VN $v_f$, $f \in \{1, 2, \dots, a \}$, updated by what has been done for all the VNs processed prior to VN $v_f$.}
\end{IEEEproof}

The following example illustrates Lemmas~\ref{lem_odeg2_unsat} and \ref{lem_obmax}.

\begin{example}\label{ex_14}
Both configurations in Fig.~\ref{Figure_11} can have degree-$2$ unsatisfied CNs in the resulting OSTs. For the $(8, 3, 13, 1)$ unlabeled OST, $b_{\textup{o\_ut}}=6$ {($1$ of these $6$ CNs is unsatisfied in the $(8, 4, 3, 13, 1)$ OST in Fig.~\ref{Figure_11}(a))}, while for the $(6, 2, 11, 0)$ unlabeled OST, $b_{\textup{o\_ut}}=5$ {($4$ of these $5$ CNs are unsatisfied in the $(6, 6, 2, 11, 0)$ OST in Fig.~\ref{Figure_11}(b))}. The upper bound of $b_{\textup{o\_max}}$ is achieved for both.
\end{example}

For a given $(a, b, d_1, d_2, d_3)$ {OST}, let  $\mathcal{Z}_{\textup{o}}$ be the set of all $(a, b_\textup{o}', d_1, d_2, d_3)$ GASTs/OSTs with $d_1 \leq b_\textup{o}' \leq b_{\textup{o\_max}}$, which have the same {unlabeled GAST/OST} as the original {OST}. Here, $b_{\textup{o\_max}}$ is the largest allowable number of unsatisfied CNs for these configurations.

\begin{definition}\label{def_drem_ost}
An $(a, b, d_1, d_2, d_3)$ \textbf{OST} is said to be \textbf{removed} from the Tanner graph of an NB-LDPC code if and only if the resulting object (after edge weight processing) $\notin \mathcal{Z}_{\textup{o}}$.
\end{definition}

We thus augment the code optimization process for asymmetric channels to consist of two phases. The first phase, as before, focuses on the removal of GASTs, and the second phase focuses on the removal of OSTs. The ordering of the phases is critical because of the following. While it is allowed to remove a GAST by converting it to an OST during the first phase, it is not allowed to remove an OST by converting it into a GAST during the second phase because GASTs, not OSTs, are the principal cause of the error floor. This is the reason why the set $\mathcal{Z}_{\textup{o}}$ is the set of all $(a, b_\textup{o}', d_1, d_2, d_3)$ GASTs/OSTs with $d_1 \leq b_\textup{o}' \leq b_{\textup{o\_max}}$. For example, to remove the $(6, 6, 2, 11, 0)$ OST in Fig.~\ref{Figure_11}(b), the configuration needs to be converted into an object $\notin \{(6, 2, 2, 11, 0) \text{ GAST}, (6, 3, 2, 11, 0) \text{ GAST/OST}, \allowbreak (6, 4, 2, 11, 0) \text{ GAST/OST}, (6, 5, 2, 11, 0) \text{ OST}, (6, 6, 2, 11, 0) \allowbreak \text{ OST}, (6, 7, 2, 11, 0) \text{ OST}\}$ (this is because $d_1=2$ and $b_{\textup{o\_max}}=d_1+b_{\textup{o\_ut}}=7$).

For a given OST, define a matrix $\bold{W}_{\textup{o}}^{\textup{z}}$ to be the matrix obtained by removing $b_\textup{o}'$, $d_1 \leq b_\textup{o}' \leq b_{\textup{o\_max}}$, rows corresponding to CNs $\in (\mathcal{O} \cup \mathcal{T})$ from the matrix $\bold{A}$, which is the OST adjacency matrix. These $b_\textup{o}'$ CNs can simultaneously be unsatisfied under some edge labeling that produces a GAST/an OST which has the same {unlabeled GAST/OST} as the given OST. Let $\mathcal{U}_{\textup{o}}$ be the set of all matrices $\bold{W}_{\textup{o}}^{\textup{z}}$. Each element $\in \mathcal{Z}_{\textup{o}}$ has one or more matrices $\in \mathcal{U}_{\textup{o}}$.

\begin{definition}\label{def_owcms}
For a given $(a,b,d_1,d_2,d_3)$ OST and its associated adjacency matrix $\bold{A}$ and its associated set $\mathcal{Z}_{\textup{o}}$, we construct a set of $t_{\textup{o}}$ matrices as follows:
\begin{enumerate}
\item Each matrix $\bold{W}_h^{\textup{o\_cm}}$, $1 \leq h \leq t_{\textup{o}}$, in this set is an $(\ell-b^{\textup{o\_cm}}_h)\times a$ submatrix, $d_1 \leq b^{\textup{o\_cm}}_h \leq b_{\textup{o\_max}}$, formed by removing \textbf{different} $b^{\textup{o\_cm}}_h$ rows from the $\ell \times a$ matrix $\bold{A}$ of the OST. These $b^{\textup{o\_cm}}_h$ rows to be removed correspond to CNs $\in (\mathcal{O} \cup \mathcal{T})$ that can simultaneously be unsatisfied under some edge labeling that produces a GAST/an OST which has the same {unlabeled GAST/OST} as the given OST.
\item Each matrix $\bold{W}_{\textup{o}}^{\textup{z}} \in \mathcal{U}_{\textup{o}}$, for every element $\in \mathcal{Z}_{\textup{o}}$, contains at least one element of the resultant set as its submatrix.
\item This resultant set has the \textbf{smallest cardinality}, which is $t_{\textup{o}}$, among all the sets which satisfy conditions 1 and 2 stated above.
\end{enumerate} 
We refer to the matrices in this set as \textbf{oscillating weight consistency matrices (OWCMs)}, and to this set itself as $\mathcal{W}_{\textup{o}}$.
\end{definition}

Similar to the definition of $b_{\textup{et}}$ in GASTs, we also define $b_{\textup{o\_et}} \leq b_{\textup{o\_ut}}$ for OSTs such that $b_{\textup{o\_max}} = d_1 + b_{\textup{o\_et}}$. The following lemma addresses the removal of OSTs via their OWCMs. In other words, the lemma shows how the WCM framework can be customized to remove OSTs.

\begin{lemma}\label{lem_rem_ost}
The necessary and sufficient processing needed to remove an $(a, b, d_1, d_2, d_3)$ OST, according to Definition~\ref{def_drem_ost}, is to change the edge weights such that {for every OWCM $\bold{W}^{\textup{o\_cm}}_h \in \mathcal{W}_{\textup{o}}$, there does not exist any vector with all its entries $\neq 0$ in the null space of that OWCM.} Mathematically, $\forall h$:
\begin{align}\label{eq_orem_cond}
&\text{If } \text{ } \mathcal{N}(\bold{W}^{\textup{o\_cm}}_h)=\textup{span}\{\bold{x}_1, \bold{x}_2, \dots ,\bold{x}_{p_h}\}, \text{ then } \nonumber \\ &\nexists \text{ } \bold{r}=[r_1 \text{ } r_2 \text{ } \dots \text{ } r_{p_h}]^\textup{T} \text{ for} \text{ }
\bold{v}=r_1\bold{x}_1+r_2\bold{x}_2+\dots+r_{p_h}\bold{x}_{p_h} \nonumber \\ &= [v_1 \text{ } v_2 \text{ } \dots v_a]^\textup{T} \text{ s.t. } v_j \neq 0, \text{ } \forall j \in \{1, 2, \dots , a\},
\end{align}
{where $p_h$ is the dimension of $\mathcal{N}(\bold{W}^{\textup{o\_cm}}_h)$}. Computations are performed over GF($q$).
\end{lemma}

\begin{IEEEproof}
The proof follows the same logic as the proof of \cite[Theorem~3]{ahh_jsac}.
\end{IEEEproof}

A similar analysis to the one in Section~\ref{sec_cch} can be performed to compute the number of OWCMs in the set $\mathcal{W}_{\textup{o}}$, with few changes (e.g., $b_{\textup{o\_et}}$ should be used instead of $b_{\textup{et}}$). Now, we propose the minimum number of edge weight changes needed to remove an OST from the Tanner graph of an NB-LDPC code with even column weight.

\begin{corollary}\label{cor_emin_os}
The minimum number of edge weight changes (with respect to the original configuration) needed to remove an $(a, b, b_2, d_1, d_2, d_3)$ OS (convert it into a non-OS/non-AS) is given by:
\begin{equation}\label{eq_emin_os}
E_{\textup{OS},\textup{min}}=1 \leq \frac{\gamma}{2}-d_{1,\textup{vn},\textup{max}}+1,
\end{equation}
where $d_{1,\textup{vn},\textup{max}}$ is the maximum number of existing degree-$1$ CNs per VN in the OS. 
\end{corollary}

\begin{IEEEproof}
The proof follows the same logic as the proof of Lemma~\ref{lem_emin_gas} (see also \cite{ahh_bas}), with $\frac{\gamma}{2}$ replacing $g$. Note that by definition of an OS, at least one of its VNs has exactly $\frac{\gamma}{2}$ neighboring unsatisfied CNs. Thus,
\vspace{-0.1em}\begin{equation}
E_{\textup{OS},\textup{min}}=\frac{\gamma}{2}-b_{\textup{vn},\textup{max}}+1=\frac{\gamma}{2}-\frac{\gamma}{2}+1=1,
\end{equation}
where $b_{\textup{vn},\textup{max}}$ is the maximum number of existing unsatisfied CNs per VN in the OS, which equals $\frac{\gamma}{2}$ for any OS.
\end{IEEEproof}

Similar to the case of GASTs, since we only change~the weights of edges connected to degree-$2$ CNs to remove OSTs, (\ref{eq_emin_os}) also holds for $E_{\textup{OST},\textup{min}}$. Moreover, a \textbf{topologically-oscillating VN} in an OST in a code with even column weight $\gamma$ is connected to exactly $\frac{\gamma}{2}$ degree-$1$ CNs. An OST with such a VN has the upper bound on $E_{\textup{OST},\textup{min}}$ equal to $1$.

The following simple algorithm illustrates the procedure we follow to optimize NB-LDPC codes with even column weights for usage over asymmetric channels.

\begin{algorithm}[H]\label{alg_code_opt}
\caption{Optimizing NB-LDPC Codes with Even Column Weights}
\begin{algorithmic}[1]
\State Apply \cite[Algorithm~2]{ahh_jsac} to optimize the NB-LDPC code by removing the detrimental GASTs.
\State Using initial simulations and combinatorial techniques (e.g., \cite{bani_cycle}) for the output code of Step~1, determine the set of OSTs to be removed.
\State Apply a customized version of \cite[Algorithm~2]{ahh_jsac} to further optimize the NB-LDPC code generated in Step~1 by removing the detrimental OSTs.
\end{algorithmic}
\end{algorithm}
\vspace{-0.2em}

A crucial check to make while removing an OST is that the edge weight changes to be performed do not undo the removal of any of the already removed GASTs nor OSTs.

%%%%%%%%%%%%%%%%%%%%%%%%%%%%%%
\section{Applications of the WCM Framework}\label{sec_sim}

In this section, we apply the WCM framework to optimize NB-LDPC codes with different structures and for various applications, demonstrating significant performance gains in the error floor region. We used a finite-precision, fast Fourier transform based $q$-ary sum-product algorithm (FFT-QSPA) LDPC decoder \cite{dec_fft}, which performs a maximum of $50$ iterations (except for PR channel simulations), and it stops if a codeword is reached sooner.

In the following items, we provide some details about the performed simulations and about our choices for the simulated NB-LDPC codes:

\begin{itemize}
\item We provide simulation results over practical storage channels, which are the main focus of this work. In particular, we present results over asymmetric Flash channels (with $3$ and $6$ voltage reads) and a 1-D MR channel with intrinsic memory (a PR channel). As an additional example, we also present results over the AWGN channel. These results collectively demonstrate the effectiveness of the WCM framework in optimizing NB-LDPC codes for various channels with different characteristics.

\item All our codes are circulant-based codes. The unoptimized NB block codes are chosen to be protograph-based codes as in \cite{lara_prot} and \cite{baz_qc} (more details about the construction are provided below). The reasons are that NB protograph-based codes enable faster encoding and decoding, and they have good performance \cite{ahh_bas, lara_prot, baz_qc}.

\item We provide results for NB-LDPC codes with various column weights (particularly $\gamma \in \{3, 4, 5\}$). The justification is that we want to demonstrate that the WCM framework typically offers at least $1$ order (and up to almost $2.5$ orders) of magnitude performance gain in the error floor region for NB-LDPC codes with initial average, good, and very good error floor performance (average in the case of $\gamma = 3$, good in the case of $\gamma = 4$, and very good in the case of $\gamma = 5$).

\item Details about the constructions and the parameters of the SC codes we simulate are provided in Subsection~\ref{subsec_sc}.
\end{itemize}

All the unoptimized NB-LDPC codes we are using in Subsections~\ref{subsec_five}, \ref{subsec_osc}, and \ref{subsec_soft} are regular non-binary protograph-based LDPC (NB-PB-LDPC) codes. These codes are constructed as follows. First, a binary protograph matrix $\bold{H}_{\textup{p}}$ is designed. Then, (the Tanner graph of) $\bold{H}_{\textup{p}}$ is lifted via a lifting parameter $\zeta$ to create (the Tanner graph of) the binary image of $\bold{H}$, which is $\bold{H}_{\textup{b}}$. The lifting process means that every $1$ in $\bold{H}_{\textup{p}}$ is replaced by a $\zeta \times \zeta$ circulant matrix, while every $0$ (if any) in $\bold{H}_{\textup{p}}$ is replaced by a $\zeta \times \zeta$ all-zero matrix. The circulant powers are adjusted such that the unlabeled Tanner graph of the resulting code does not have cycles of length $4$. Then, the $1$'s in $\bold{H}_{\textup{b}}$ are replaced by non-zero values $\in$ GF($q$) to generate $\bold{H}$. These unoptimized codes are high performance NB-PB-LDPC codes (see also \cite{lara_prot} and \cite{baz_qc}, in addition to \cite{ahh_jsac}). Note that the WCM framework works for any regular, or even irregular with fixed column weight, NB-LDPC codes. Moreover, the WCM framework also works for any GF size, $q$, and for any code rate.

\begin{remark}
While the WCM framework works for NB-LDPC codes defined over any GF size, $q$, the performance gains achieved are expected to be relatively smaller for higher GF sizes, e.g., $q \geq 32$ (over all channels). The reason is that the fraction of edge weight assignments under which a configuration is a detrimental GAST becomes smaller as $q$ increases (see also \cite{behzad_elem}), which means NB-LDPC codes with higher GF sizes naturally have improved error floor performance. However, increasing the GF size dramatically increases the decoding complexity, as the decoding complexity either has $O(q^2)$ or $O(q \log_2 (q))$ \cite{dec_fft}. Consequently, the approach of solely increasing the GF size in order to mitigate the error floor is not advised for storage systems because of complexity constraints. Note that NB-LDPC codes with $\gamma=2$ can only provide good error floor performance if the GF size is large. This discussion is the reason why in our simulations, we work with various column weights ($\gamma \in \{3, 4, 5\}$), but we keep the GF size relatively small ($q \in \{4, 8\}$). In summary, we provide NB-LDPC codes with superior error floor performance, achieved via the WCM framework, without a dramatic increase in the decoding complexity. An NB-LDPC decoder implementation customized for storage application, which uses a GF size $q=8$, is provided in \cite{dec_yuta}.
\end{remark}

In this paper, RBER is the raw bit error rate, which is the number of raw (uncoded) data bits in error divided by the total number of raw (uncoded) data bits read \cite{cai_defn}. UBER is the uncorrectable bit error rate, which is a metric for the fraction of bits in error out of all bits read after the error correction is applied via encoding/decoding \cite{cai_defn}. One formulation of UBER, as recommended by industry, is the frame error rate (FER) divided by the sector size in bits.

%%%%%%%%%%%%%%%%%%%%%%%%%%%%%%
\subsection{Optimizing Column Weight $5$ Codes}\label{subsec_five}

In this subsection, we use the WCM framework to optimize NB-LDPC codes with column weight $5$ for the first time. Column weight $5$ codes generally guarantee better performance compared with column weight $3$ and $4$ codes in the error floor region. We show in this subsection, that more than $1$ order of magnitude performance gain is still achievable via the WCM framework for such codes despite their improved error floor performance. The channel used in this subsection is a practical Flash channel: the normal-Laplace mixture (NLM) Flash channel \cite{mit_nl}. Here, we use $3$ reads, and the sector size is $512$ bytes.

In the NLM channel, the threshold voltage distribution of sub-$20$nm multi-level cell (MLC) Flash memories is carefully modeled. The four levels are modeled as different NLM distributions, incorporating several sources of error due to wear-out effects, e.g., programming errors, thereby resulting in significant asymmetry \cite{mit_nl}. Furthermore, the authors provided accurate fitting results of their model for program/erase (P/E) cycles up to $10$ times the manufacturer's endurance specification. We implemented the NLM channel based on the parameters described in \cite{mit_nl}.

\begin{figure}
\vspace{-0.5em}
\center
\includegraphics[trim={0.0in 0.0in 0.0in 0.2in},clip,width=3.5in]{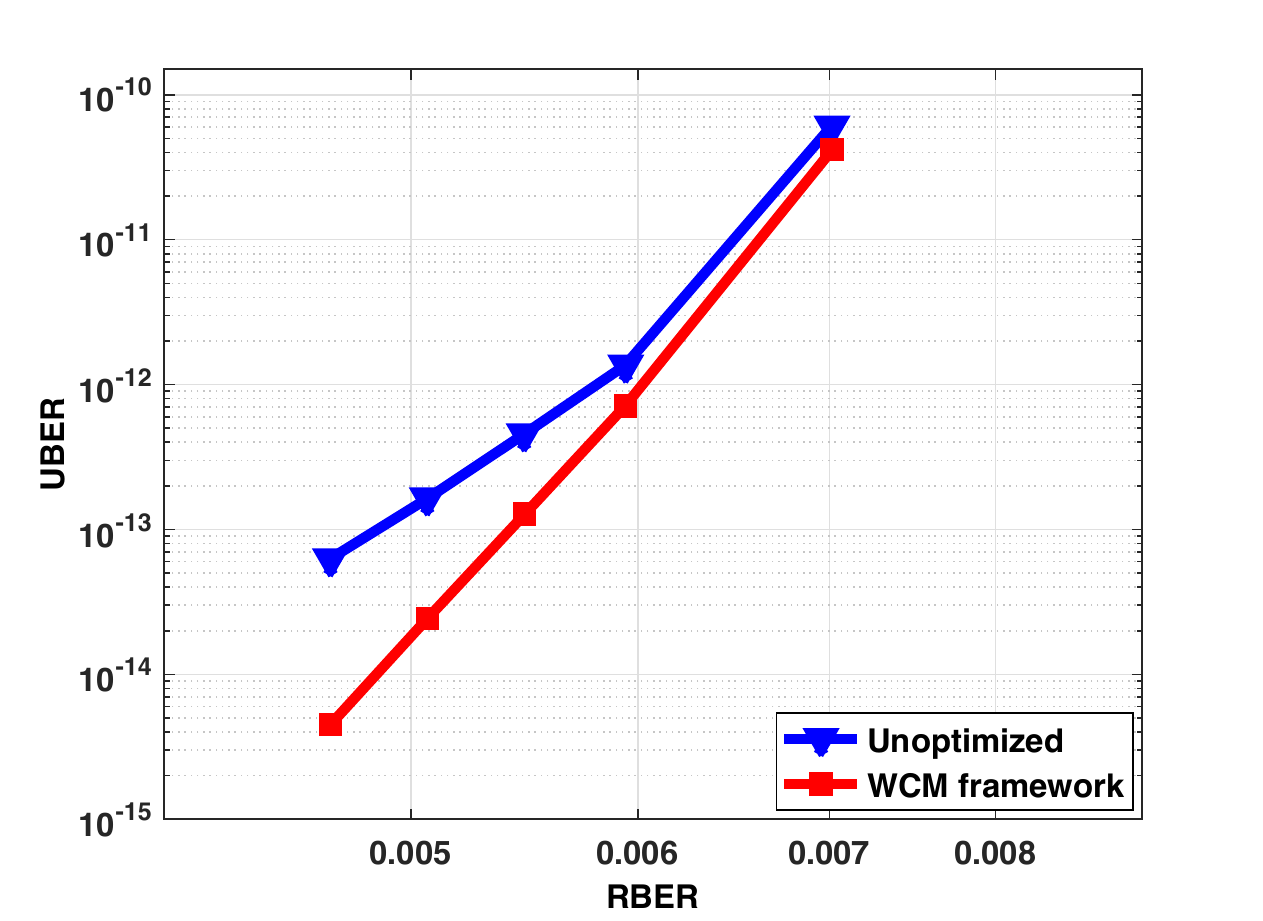}
\vspace{-1.5em}
\caption{Simulation results over the NLM channel with $3$ reads for Code~1 (unoptimized) and Code~2 (WCM framework). The two codes have $\gamma = 5$.}
\label{Figure_12}
\end{figure}

In this subsection, Code~1 is an NB-PB-LDPC code defined over GF($4$), with block length $= 6724$ bits, rate $\approx 0.88$, and $\gamma = 5$. Code~2 is the result of optimizing Code~1 for the asymmetric NLM channel by attempting to remove the GASTs in Table~\ref{Table3} using the WCM framework.

Fig.~\ref{Figure_12} shows that more than $1$ order of magnitude performance gain is achieved via optimizing Code~1 to arrive at Code~2 using the WCM framework. The figure also shows that using the WCM framework, an UBER of approximately $4.53 \times 10^{-15}$ is achievable at RBER of approximately $4.69 \times 10^{-3}$ on the NLM Flash channel (an aggressively asymmetric channel) with only $3$ reads.

\begin{table}
\caption{Error profile of Codes~1 and 2 over the NLM channel with $3$ reads, RBER $\approx 4.69 \times 10^{-3}$, UBER (unoptimized) $\approx 6.31 \times 10^{-14}$, and UBER (WCM framework) $\approx 4.53 \times 10^{-15}$ (see Fig.~\ref{Figure_12}).}
\vspace{-0.5em}
\centering
\scalebox{1.00}
{
\begin{tabular}{|c|c|c|c|}
\hline

\multirow{2}{*}{Error type} & \multicolumn{2}{|c|}{Count} \\
\cline{2-3}
{} & Code~1 & \makecell{Code~2} \\
\hline
$(4, 8, 8, 6, 0)$ & $18$ & $0$ \\
\hline
$(6, 8, 8, 11, 0)$ & $9$ & $0$ \\
\hline
$(6, 10, 8, 11, 0)$ & $11$ & $0$ \\
\hline
$(7, 5, 5, 15, 0)$ & $4$ & $0$ \\
\hline
$(7, 9, 9, 13, 0)$ & $4$ & $0$ \\
\hline
$(7, 10, 10, 9, 2)$ & $7$ & $1$ \\
\hline
$(8, 6, 6, 17, 0)$ & $23$ & $0$ \\
\hline
$(8, 8, 6, 17, 0)$ & $15$ & $0$ \\
\hline
Other & $9$ & $7$ \\
\hline
\end{tabular}}
\label{Table3}
\end{table}

\begin{figure}
\vspace{-0.5em}
\center
\includegraphics[trim={0.08in 0.0in 0.2in 0.0in},clip,width=3.5in]{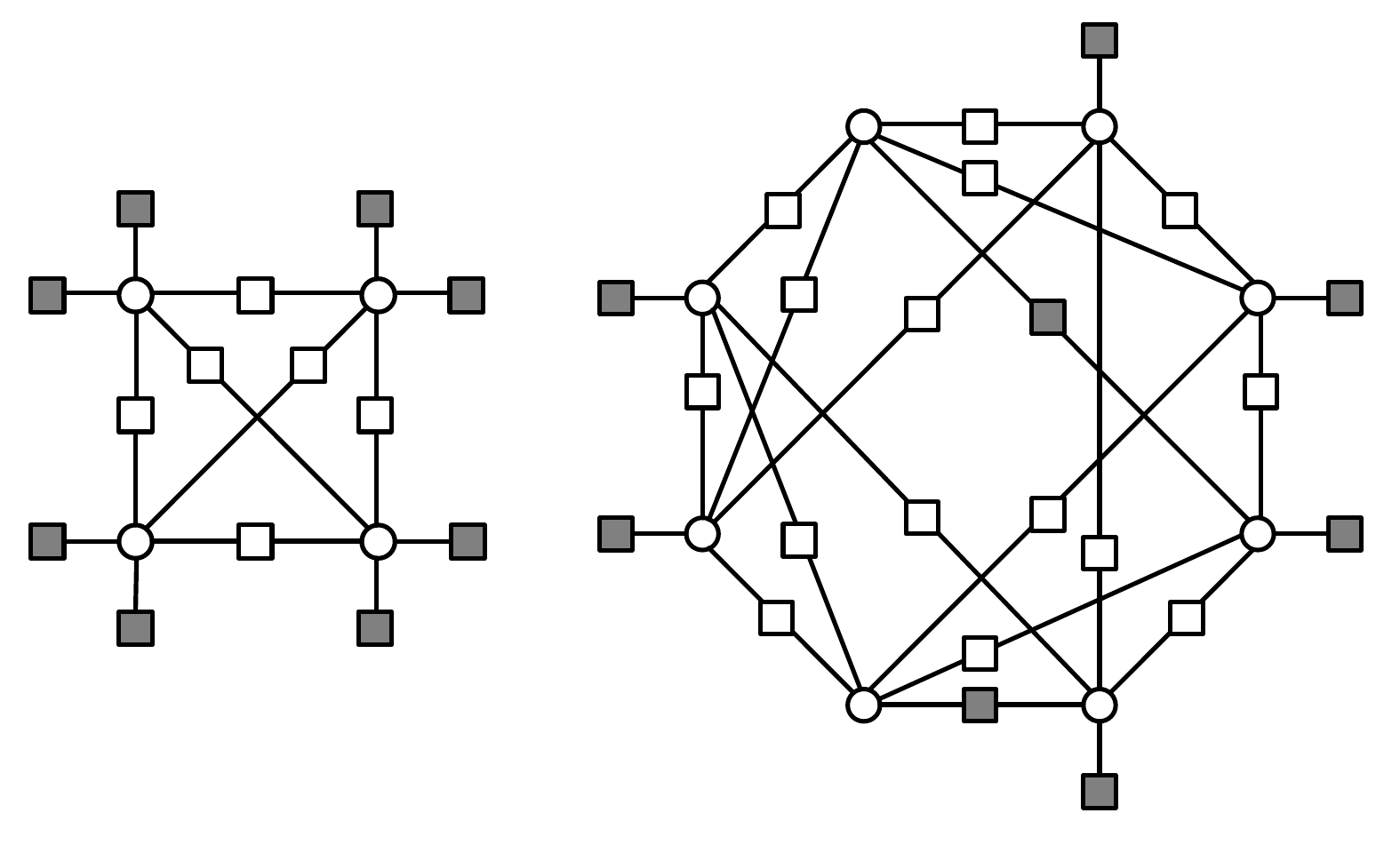}\vspace{-1.0em}
\text{\hspace{-1.5em} \footnotesize{(a) \hspace{15em} (b) \hspace{1em}}}
\caption{(a) A $(4, 8, 8, 6, 0)$ GAST for $\gamma=5$. (b) An $(8, 8, 6, 17, 0)$ GAST for $\gamma=5$. Appropriate non-binary edge weights are assumed.}
\label{Figure_13}
\vspace{-0.5em}
\end{figure}

Table~\ref{Table3} shows the error profiles of Codes~1 and 2 over the NLM channel with $3$ reads. The table reveals that $33\%$ of the errors in the error profile of Code~1 are non-elementary GASTs. The table also demonstrates the effectiveness of the WCM framework in removing the detrimental objects. Two of the GASTs that strongly contribute to the error profile of Code~1 are $(4, 8, 8, 6, 0)$ and $(8, 8, 6, 17, 0)$ GASTs, which are shown in Fig.~\ref{Figure_13}. The key difference between GASTs in codes with $\gamma=5$ (or $6$) and GASTs in codes with $\gamma \in \{3, 4\}$ is that for the former GASTs, $g=2$, while for the latter GASTs, $g=1$. In other words, a VN in an object in a code with $\gamma=5$ (or $6$) can be connected to a maximum of $2$ unsatisfied CNs while the object is classified as a GAST (see also Fig.~\ref{Figure_13} and Example~\ref{ex_2}); for $\gamma \in \{3, 4\}$, this maximum is $1$.

%%%%%%%%%%%%%%%%%%%%%%%%%%%%%%
\subsection{Achieving More Gain by Removing Oscillating Sets}\label{subsec_osc}

In this subsection, we demonstrate the additional gains that can be achieved for NB-LDPC codes with even column weights (particularly $\gamma = 4$) over practical asymmetric channels by removing OSTs as described in Section~\ref{sec_os}.

\begin{figure}
\vspace{-0.5em}
\center
\includegraphics[trim={0.0in 0.0in 0.0in 0.2in},clip,width=3.5in]{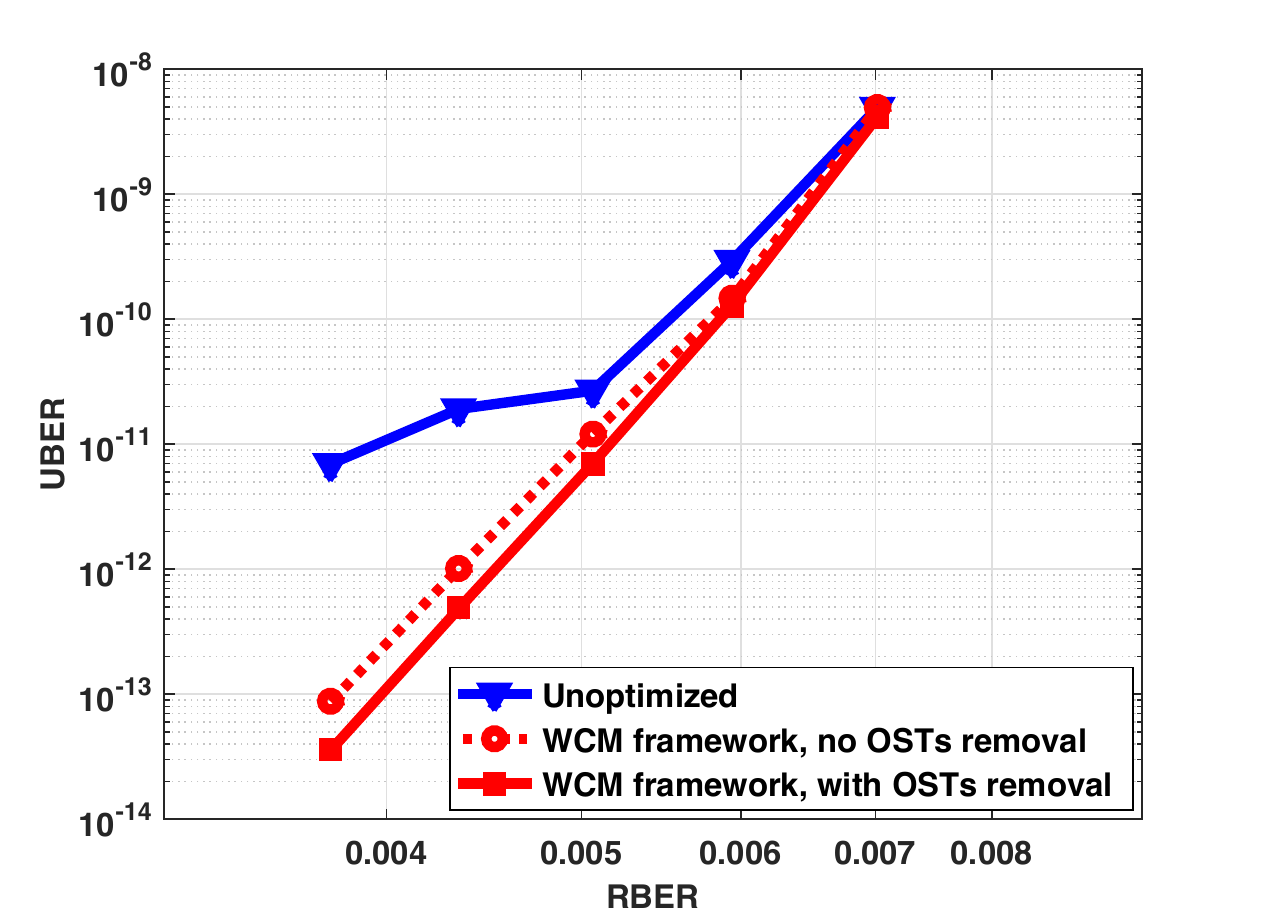}
\vspace{-1.5em}
\caption{Simulation results over the NLM channel with $3$ reads for Code~3 (unoptimized), Code~4 (WCM framework, no OSTs removal), and Code~5 (WCM framework, with OSTs removal). The three codes have $\gamma = 4$.}
\label{Figure_14}
\vspace{-0.2em}
\end{figure}

First, we present results for the NLM channel described in the previous subsection (still with $3$ reads). Code~3 is an NB-PB-LDPC code defined over GF($4$), with block length $= 8480$ bits, rate $\approx 0.90$, and $\gamma = 4$. Code~4 is the result of optimizing Code~3 by attempting to remove the dominant GASTs $(4, 4, 4, 6, 0)$, $(6, 4, 4, 10, 0)$, $(6, 5, 5, 8, 1)$, and $(8, 4, 2, 15, 0)$ using the WCM framework (see \cite{ahh_jsac}). Code~5 is the result of optimizing Code~4 for the asymmetric NLM channel by attempting to remove the OSTs in Table~\ref{Table4} using the WCM framework. The performance curves of Code~3 (unoptimized) and Code~4 (WCM framework, no OSTs removal) in Fig.~\ref{Figure_14} were introduced in \cite{ahh_jsac}.

\begin{figure}
\vspace{-0.5em}
\center
\includegraphics[trim={0.0in 0.0in 0.0in 0.2in},clip,width=3.5in]{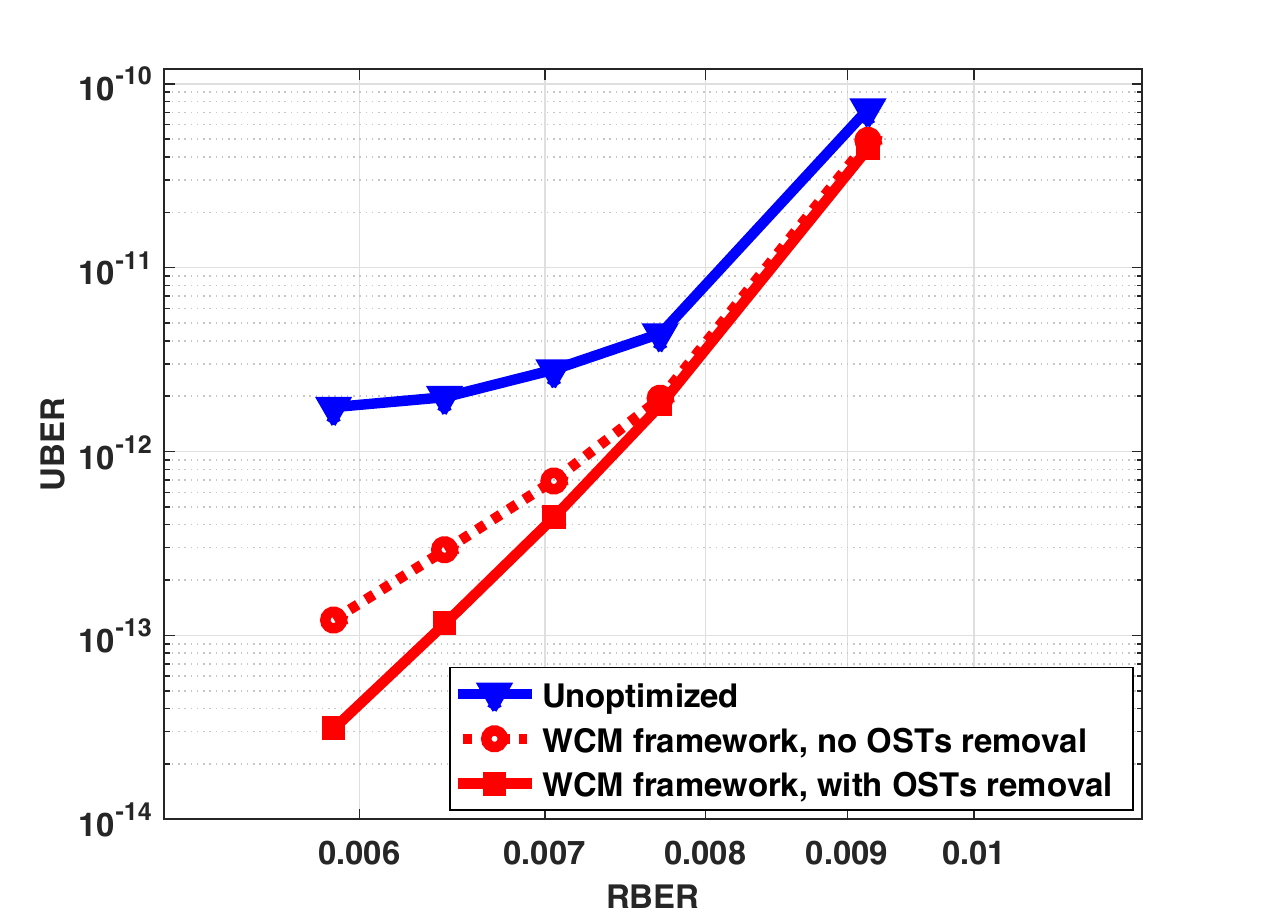}
\vspace{-1.5em}
\caption{Simulation results over the CHMM channel with $3$ reads for Code~6 (unoptimized), Code~7 (WCM framework, no OSTs removal), and Code~8 (WCM framework, with OSTs removal). The three codes have $\gamma=4$.}
\label{Figure_15}
\vspace{-0.2em}
\end{figure}

It is demonstrated by Fig.~\ref{Figure_14} that removing the dominant OSTs to generate Code~5 results in nearly $0.5$ of an order of magnitude gain in performance over Code~4 (for which only the dominant GASTs are removed) even though Code~4 is highly optimized (it outperforms Code~3 by about $2$ orders of magnitude). Thus, applying Algorithm~1 to remove OSTs after removing GASTs raises the gain to almost $2.5$ orders of magnitude for Code~5 compared with the unoptimized code (Code~3) over the NLM channel. Table~\ref{Table4} shows the significant reduction in the number of OSTs in the error profile of Code~5 compared with Code~3.

Second, we present results for another asymmetric Flash channel: the Cai-Haratsch-Mutlu-Mai (CHMM) Flash channel \cite{cai_fl}. The authors developed a model in \cite{cai_fl} for the threshold voltage distribution that is suitable for $20$nm and $24$nm MLC Flash memories. The four levels are modeled as different Gaussian distributions that are shifted and broadened with the increase in P/E cycles, resulting in limited asymmetry relative to the NLM channel. We implemented the CHMM channel based on the data and the model provided in \cite{cai_fl}. In this subsection, we use $3$ reads, and the sector size is $512$ bytes.

\begin{table}
\caption{OSTs error profile of Codes~3 and 5 over the NLM channel with $3$ reads, RBER $\approx 3.75 \times 10^{-3}$, UBER (unoptimized) $\approx 6.98 \times 10^{-12}$, and UBER (WCM framework, with OSTs removal) $\approx 3.58 \times 10^{-14}$ (see Fig.~\ref{Figure_14}).
}
\vspace{-0.5em}
\centering
\scalebox{1.00}
{
\begin{tabular}{|c|c|c|c|}
\hline

\multirow{2}{*}{Error type} & \multicolumn{2}{|c|}{Count} \\
\cline{2-3}
{} & Code~3 & \makecell{Code~5} \\
\hline
$(5, 5, 5, 6, 1)$ & $22$ & $0$ \\
\hline
$(6, 5, 4, 10, 0)$ & $29$ & $0$ \\
\hline
$(8, 4, 2, 12, 2)$ & $24$ & $0$ \\
\hline
$(8, 5, 3, 13, 1)$ & $25$ & $0$ \\
\hline
\end{tabular}}
\label{Table4}
\end{table}

\begin{table}
\caption{OSTs error profile of Codes~6 and 8 over the CHMM channel with $3$ reads, RBER $\approx 5.87 \times 10^{-3}$, UBER (unoptimized) $\approx 1.74 \times 10^{-12}$, and UBER (WCM framework, with OSTs removal) $\approx 3.11 \times 10^{-14}$ (see Fig.~\ref{Figure_15}).
}
\vspace{-0.5em}
\centering
\scalebox{1.00}
{
\begin{tabular}{|c|c|c|c|}
\hline

\multirow{2}{*}{Error type} & \multicolumn{2}{|c|}{Count} \\
\cline{2-3}
{} & Code~6 & \makecell{Code~8} \\
\hline
$(6, 5, 2, 11, 0)$ & $29$ & $0$ \\
\hline
$(6, 6, 2, 11, 0)$ & $11$ & $0$ \\
\hline
$(7, 5, 3, 11, 1)$ & $34$ & $0$ \\
\hline
$(8, 4, 3, 13, 1)$ & $15$ & $0$ \\
\hline
$(9, 4, 2, 14, 2)$ & $11$ & $1$ \\
\hline
\end{tabular}}
\label{Table5}
\end{table}

Here, Code~6 is an NB-PB-LDPC code defined over GF($4$), with block length $= 1840$ bits, rate $\approx 0.80$, and $\gamma = 4$. Code~7 is the result of optimizing Code~6 by attempting to remove the dominant GASTs $(4, 4, 4, 6, 0)$, $(6, 4, 2, 11, 0)$, $(6, 4, 4, 10, 0)$, $(7, 4, 3, 11, 1)$, $(8, 5, 5, 12, 1)$, and $(9, 5, 5, 14, 1)$ using the WCM framework (see also \cite{ahh_jsac}). Code~8 is the result of optimizing Code~7 for the asymmetric CHMM channel by attempting to remove the OSTs in Table~\ref{Table5} using the WCM framework. The performance curves of Code~6 (unoptimized) and Code~7 (WCM framework, no OSTs removal) in Fig.~\ref{Figure_15} were introduced in \cite{ahh_jsac}.

Fig.~\ref{Figure_15} reveals that removing the dominant OSTs to design Code~8 results in more than $0.5$ of an order of magnitude performance gain over Code~7 (for which only the dominant GASTs are removed). Consequently, applying Algorithm~1 to remove OSTs (after removing GASTs) raises the performance gain to more than $1.5$ orders of magnitude for Code~8 compared with the unoptimized code (Code~6) over the CHMM channel. Table~\ref{Table5} clarifies the significant reduction in the number of OSTs in the error profile of Code~8 compared with Code~6.

%%%%%%%%%%%%%%%%%%%%%%%%%%%%%%
\vspace{-0.22em}
\subsection{Effect of Soft Information in Flash Channels}\label{subsec_soft}

In this subsection, we show the performance of NB-LDPC codes optimized by the WCM framework over practical Flash channels with additional soft information. The NLM and CHMM Flash channels used in this subsection are as described in the previous two subsections, except that we now consider $6$ voltage reads instead of $3$. The additional reads increase the amount of soft information provided to the decoder from the Flash channel.

In the simulations of this subsection, Code~9 is an NB-PB-LDPC code defined over GF($4$), with block length $= 3996$ bits, rate $\approx 0.89$, and $\gamma = 3$. Code~10 is the result of optimizing Code~9 for the asymmetric NLM channel (with $6$ reads this time) by attempting to remove the dominant GASTs $(4, 2, 2, 5, 0)$, $(4, 3, 2, 5, 0)$, $(5, 2, 2, 5, 1)$, $(6, 0, 0, 9, 0)$, $(6, 1, 0, 9, 0)$, $(6, 1, 1, 7, 1)$, $(6, 2, 2, 5, 2)$, and $(6, 2, 2, 8, 0)$ using the WCM framework.

Furthermore, Code~11 is another NB-PB-LDPC code defined over GF($4$), with block length $= 3280$ bits, rate $\approx 0.80$, and $\gamma = 4$. Code~12 is the result of optimizing Code~11 for the asymmetric NLM channel (with $6$ reads) by attempting to remove the dominant GASTs $(4, 4, 4, 6, 0)$, $(6, 2, 2, 11, 0)$, $(8, 4, 3, 13, 1)$, and $(8, 5, 2, 15, 0)$ in addition to the dominant OSTs $(6, 5, 4, 10, 0)$, $(7, 6, 4, 12, 0)$, $(8, 4, 2, 12, 2)$, and $(9, 4, 2, 14, 2)$ using the WCM framework. We also reuse Code~6 in this subsection (its parameters are stated in the previous subsection). Code~13 is the result of optimizing Code~6 for the asymmetric CHMM channel (with $6$ reads) by attempting to remove the dominant GASTs $(4, 4, 4, 6, 0)$, $(6, 4, 4, 11, 0)$, and $(7, 4, 3, 11, 1)$ in addition to the dominant OSTs $(6, 5, 2, 11, 0)$, $(7, 5, 3, 11, 1)$, $(7, 5, 4, 9, 2)$, $(7, 6, 6, 8, 2)$, $(8, 6, 2, 15, 0)$, and $(10, 7, 5, 11, 4)$ using the WCM framework.

\begin{figure}
\vspace{-0.5em}
\center
\includegraphics[trim={0.0in 0.0in 0.0in 0.2in},clip,width=3.5in]{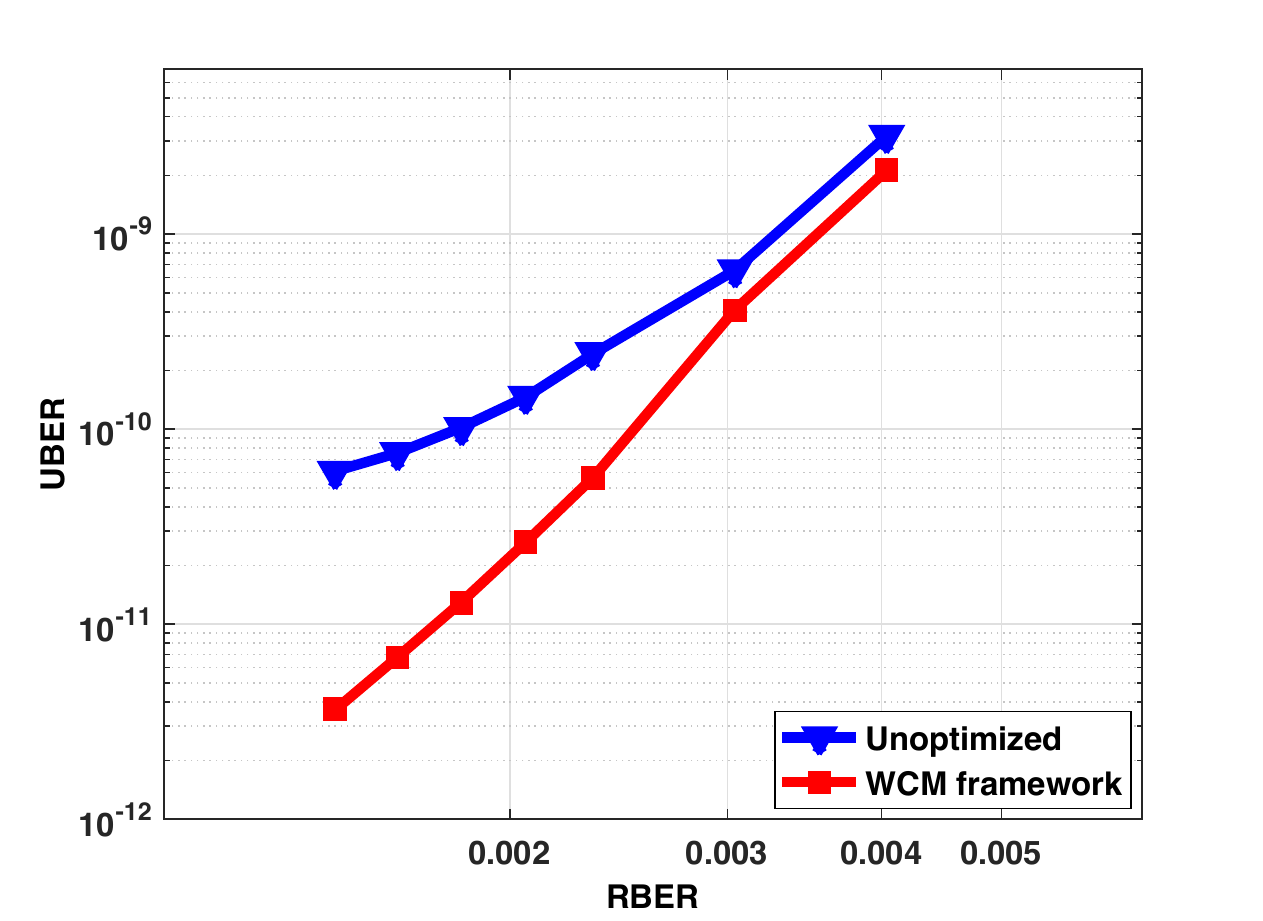}
\vspace{-1.5em}
\caption{Simulation results over the NLM channel with $6$ reads for Code~9 (unoptimized) and Code~10 (WCM framework). The two codes have $\gamma = 3$.}
\label{Figure_16}
\vspace{-0.3em}
\end{figure}

\begin{figure}
\vspace{-0.3em}
\center
\includegraphics[trim={0.0in 0.0in 0.0in 0.2in},clip,width=3.5in]{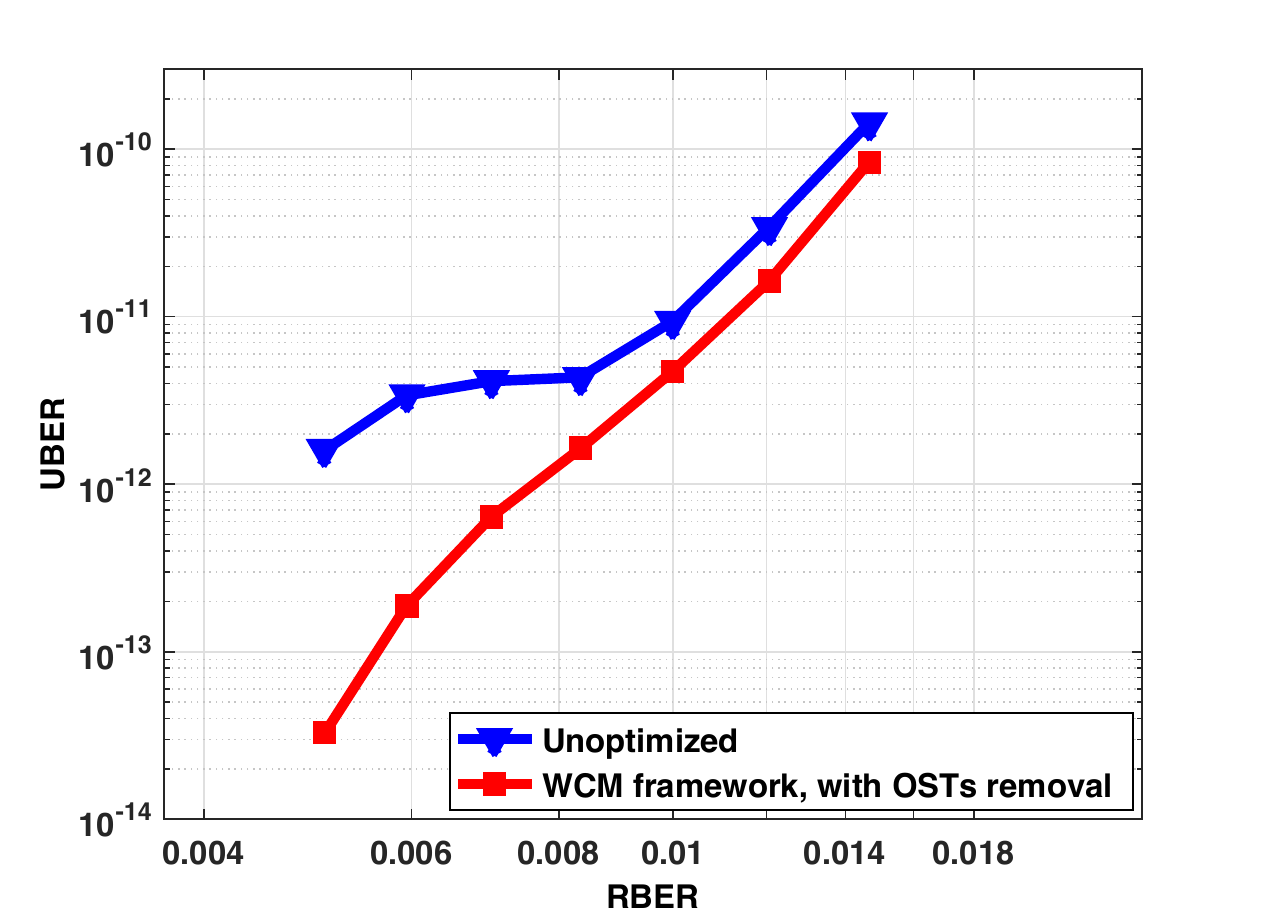}
\vspace{-1.5em}
\caption{Simulation results over the NLM channel with $6$ reads for Code~11 (unoptimized) and Code~12 (WCM framework, with OSTs removal). The two codes have $\gamma = 4$.}
\label{Figure_17}
\vspace{-0.1em}
\end{figure}

\begin{figure}
\vspace{-0.3em}
\center
\includegraphics[trim={0.0in 0.0in 0.0in 0.2in},clip,width=3.5in]{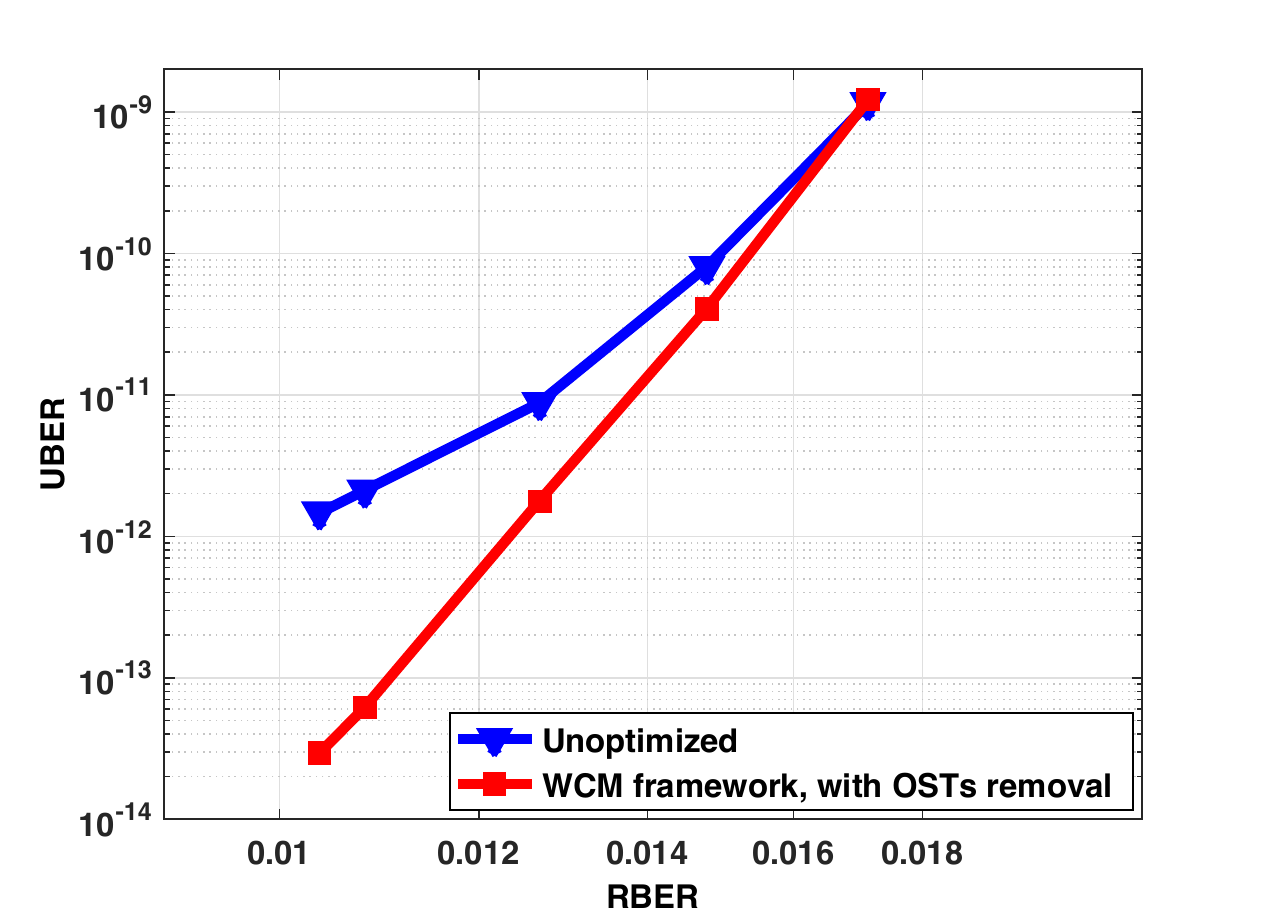}
\vspace{-1.5em}
\caption{Simulation results over the CHMM channel with $6$ reads for Code~6 (unoptimized) and Code~13 (WCM framework, with OSTs removal). The two codes have $\gamma=4$.}
\label{Figure_18}
\vspace{-0.3em}
\end{figure}

According to our simulations, the most dominant GASTs in the error floor of the unoptimized codes (Codes~9, 11, and 6) are hardly affected by the additional soft information (compare the dominant GASTs listed above for Codes~9, 11, and 6 with the dominant GASTs in \cite[Table~I]{ahh_jsac}, \cite[Table~II]{ahh_jsac}, and \cite[Table~IV]{ahh_jsac}, respectively). Moreover, Figures~\ref{Figure_16}, \ref{Figure_17}, and \ref{Figure_18} show that the performance gains achieved by applying the WCM framework over practical Flash channels with $6$ reads are in the same range as the gains achieved over the same channels with $3$ reads. In particular, more than $1$ order of magnitude gain is achieved in Fig.~\ref{Figure_16}, and more than $1.5$ orders of magnitude ($> 0.5$ of an order of magnitude is due to OSTs removal) gain is achieved in both Fig.~\ref{Figure_17} and \ref{Figure_18}. Furthermore, similar to the case of $3$ reads demonstrated in \cite{ahh_jsac}, the more asymmetric the Flash channel is, the higher the percentage of relevant non-elementary GASTs ($b > d_1$ or/and $d_3 > 0$) that appear in the error profile of the NB-LDPC code.

The major difference between the results over practical Flash channels with $3$ and $6$ reads is the gain achieved in RBER.  Consider the $\gamma=4$ codes simulated over the CHMM channel, and assume that the target UBER is $10^{-13}$. In Fig.~\ref{Figure_15}, Code~8 achieves the target UBER at RBER $\approx 6.5 \times 10^{-3}$. On the contrary, Code~13 achieves the target UBER at RBER $\approx 1.1 \times 10^{-2}$, as revealed by Fig.~\ref{Figure_18}. Thus, using $6$ reads achieves in this case about $70\%$ RBER gain compared with using only $3$ reads. This RBER gain is directly translated into P/E cycles gain, which means an extension in the lifetime of the Flash device. Similar gains are also observed for codes with different column weights over both the NLM and CHMM channels.

%%%%%%%%%%%%%%%%%%%%%%%%%%%%%%
\subsection{Optimizing Spatially-Coupled Codes}\label{subsec_sc}

In this subsection, we extend the scope of the WCM framework to irregular codes with fixed column weights (fixed VN degrees). In particular, we use the WCM framework to optimize non-binary spatially-coupled (NB-SC) codes with $\gamma \in \{3, 4\}$ for PR and AWGN channels, showing more than $1$ order of magnitude performance gain.

SC codes are a class of graph-based (LDPC) codes that have capacity-approaching asymptotic performance, and very good finite-length performance. Literature works studying the asymptotic performance of SC codes include \cite{kud_sc, lent_asy, naj_asy} for the binary case, and \cite{andr_asy, amna_asy} for the non-binary case. Recent results on finite-length constructions of SC codes include \cite{pus_sc, olm_sc, iye_sc, mitch_fl, yix_fl, scb_ext1, scb_ext2, scb_ext3, homa_boo} for the binary case, and \cite{irina_sc, homa_sc, ahh_nboo, ahh_nboo2} for the non-binary case. Most of these finite-length constructions are based on protographs. SC codes are constructed by partitioning an underlying block LDPC code, and then rewiring the partitioned components together multiple times \cite{scb_ext1, homa_boo, homa_sc, ahh_nboo}. We demonstrate the effectiveness of the WCM framework by optimizing the edge weights of NB-SC codes designed using two different finite-length construction techniques (both are also based on protographs). First, we show results for NB-SC codes partitioned using single cutting vectors (CVs) \cite{scb_ext1, homa_sc}, and the underlying block LDPC codes used are array-based LDPC (AB-LDPC) codes \cite{lara_as}. More details about the CV technique can be found in \cite{homa_sc}. Second, we show results for better NB-SC codes, designed using the optimal overlap, circulant power optimizer (OO-CPO) technique \cite{homa_boo, ahh_nboo, ahh_nboo2}. The partitioning here is derived through solving an optimization problem aiming at minimizing the total number of detrimental objects in the protograph of the SC code. Next, circulant powers of the underlying block code are optimized to further reduce the number of detrimental objects in the final unlabeled Tanner graph of the SC code. More details about the OO-CPO technique for AWGN and Flash channels can be found in \cite{homa_boo} and \cite{ahh_nboo}. In this subsection, we focus on the case of partitioning the underlying block code into only two component matrices (memory $=1$), and all the SC codes do not have cycles of length $4$ in their unlabeled graphs. Furthermore,

\begin{itemize}
\item The CV and the OO-CPO techniques mentioned above are chosen to design the underlying topologies (the binary images) of our NB-SC codes. SC codes designed using the OO-CPO technique have superior performance over AWGN \cite{homa_boo}, Flash \cite{ahh_nboo}, and PR channels \cite{ahh_nboo2}.

\item We use coupling lengths (see \cite{scb_ext1} and \cite{homa_sc}) of average values (namely, $5$ and $8$) in our SC codes. This is because for a fixed block length, the circulant size of an SC code is inversely proportional to the coupling length. Using a very small circulant size typically exacerbates the error floor problem.
\end{itemize}

The WCM framework requires the initial unoptimized code to have a fixed column weight (fixed VN degree) but not necessarily a fixed row weight (fixed CN degree). NB-SC codes that are based on the underlying structured and regular block codes incorporate irregularities in their CN degrees (different row weights), while having fixed VN degrees \cite{homa_boo, homa_sc}, making them suitable for optimization using the WCM framework for various applications.

We use the PR channel described in \cite{ahh_bas}. This PR channel incorporates inter-symbol interference (intrinsic memory), jitter, and electronic noise. The normalized channel density \cite{shafa, gl_tolga, tom_v} is $1.4$, and the PR equalization target is $[8 \text{ } 14 \text{ } 2]$. The receiver consists of filtering units followed by a Bahl Cocke Jelinek Raviv (BCJR) detector \cite{bcjr}, which is based on pattern-dependent noise prediction (PDNP) \cite{pdnp}, and an FFT-QSPA LDPC decoder \cite{dec_fft}. The number of global (detector-decoder) iterations is $10$, and the number of local (decoder only) iterations is $20$. Unless a codeword is reached, the decoder performs its prescribed number of local iterations for each global iteration. More details can be found in \cite{ahh_bas}.

Code~14 is an NB-SC code designed using the CV technique, and defined over GF($4$), with block length $= 8464$ bits, rate $\approx 0.85$, and $\gamma = 3$. The underlying block code is a non-binary AB-LDPC code defined over GF($4$), with circulant size $=23$ and $\gamma = 3$. The coupling length $L=8$ \cite{homa_sc}, and the underlying block code is partitioned using the optimal CV $[5 \text{ } 11 \text{ } 18]$ (see also \cite{homa_sc} for more details about determining the optimal CV). Code~15 is the result of optimizing Code~14 for the PR channel by attempting to remove the dominant BASTs $(6, 0, 0, 9, 0)$, $(6, 1, 0, 9, 0)$, $(6, 2, 0, 9, 0)$, $(8, 0, 0, 10, 1)$, and $(8, 0, 0, 12, 0)$ using the WCM framework.

\begin{figure}
\vspace{-0.5em}
\center
\includegraphics[trim={0.0in 0.0in 0.0in 0.2in},clip,width=3.5in]{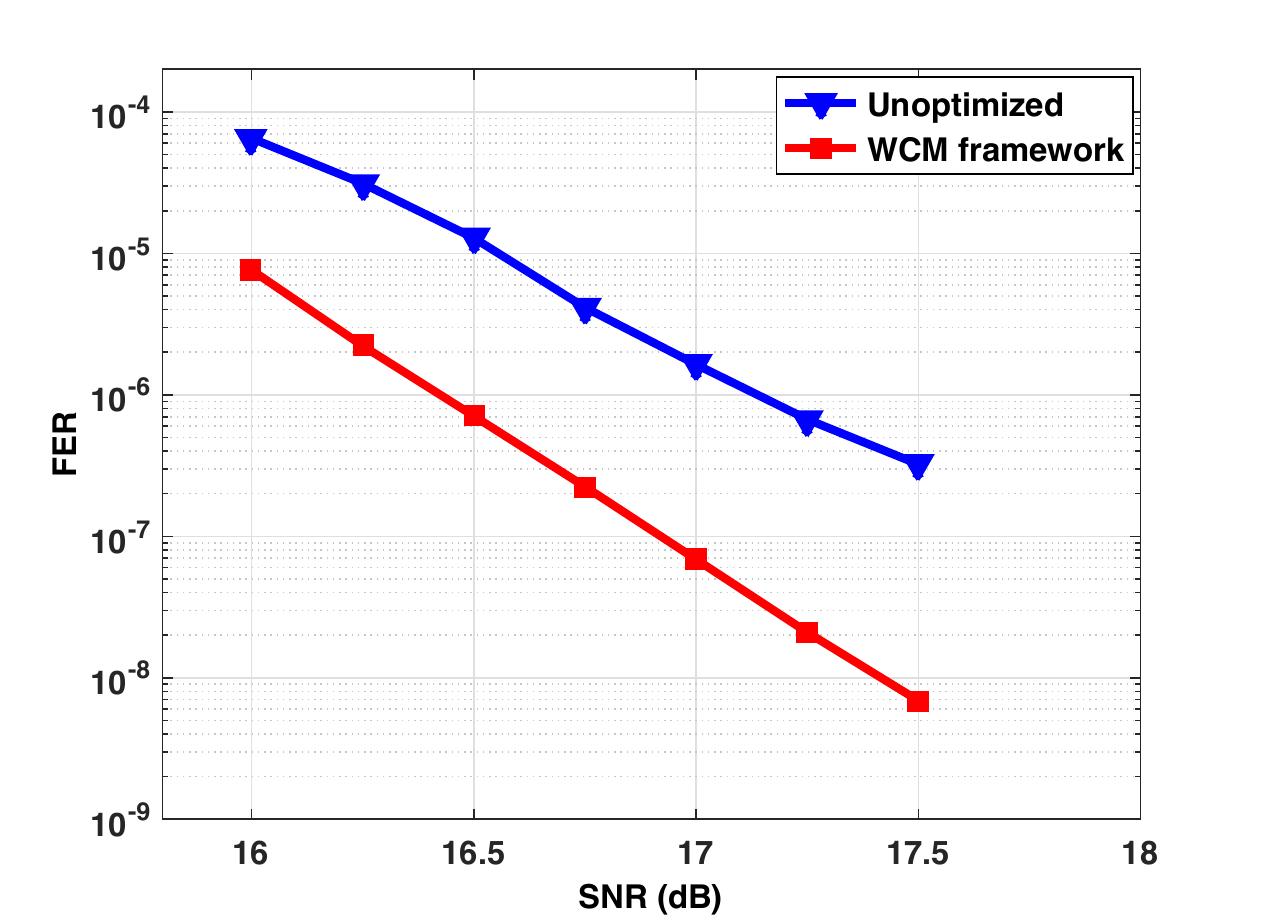}
\vspace{-1.5em}
\caption{Simulation results over the PR channel for SC~Code~14 (unoptimized) and SC~Code~15 (WCM framework). The two codes have $\gamma=3$.}
\label{Figure_19}
\vspace{-0.2em}
\end{figure}

Fig.~\ref{Figure_19} shows that the SC code optimized using the WCM framework (Code~15) outperforms the unoptimized SC code (Code~14) by more than $1.5$ orders of magnitude over the PR channel. Note that this significant performance gain is achieved despite the unlabeled Tanner graphs of Codes~14 and 15 both being designed using the optimal CV. In the caption of Fig.~\ref{Figure_19} we precede the names of Codes~14 and 15 with ``SC'' for clarity.

\begin{figure*}
\center
\includegraphics[trim={0.1in 0.8in 0.1in 0.0in},clip,width=6.0in]{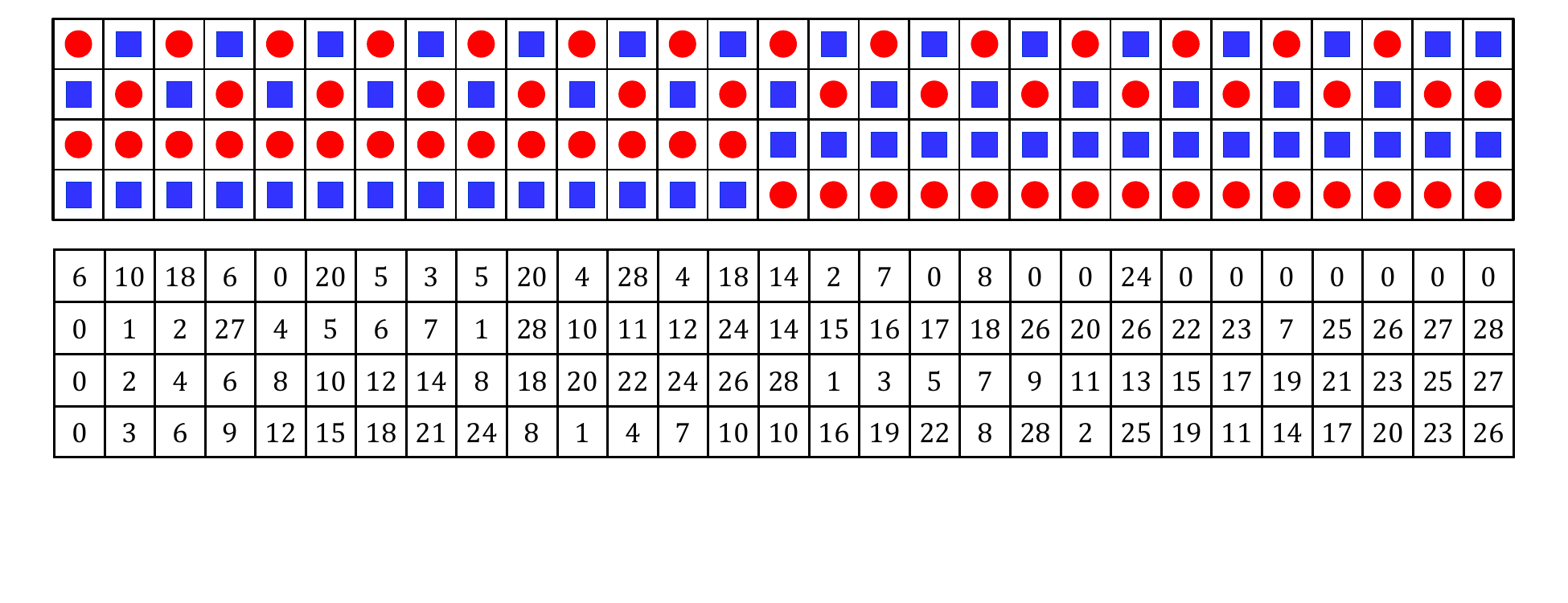}
\vspace{-0.7em}
\caption{Upper panel: the OO partitioning of the underlying block code of Code~17 (and Code~19). Entries with circles (resp., squares) are assigned to the first (resp., second) component matrix. Lower panel: the circulant power arrangement for the circulants in the underlying block code of Code~17 (and Code~19) after applying the CPO.}
\label{Figure_20}
\vspace{-0.5em}
\end{figure*}

In the AWGN simulations, Code~16 (resp., Code~17) is an NB-SC code designed using the CV (resp., OO-CPO) technique, and defined over GF($8$), with block length $= 12615$ bits, rate $\approx 0.83$, and $\gamma = 4$. The underlying block code is defined over GF($8$), with circulant size $=29$, $\gamma = 4$, and row weight $=29$. The coupling length $L=5$ \cite{homa_sc, ahh_nboo}. The underlying block code of Code~16 is partitioned using the CV $[5 \text{ } 11 \text{ } 18 \text{ } 24]$, and it is a non-binary AB-LDPC code. The underlying block code of Code~17 is partitioned according to Fig.~\ref{Figure_20}, upper panel, and its circulant power arrangement is given in Fig.~\ref{Figure_20}, lower panel (see also \cite{homa_boo} and \cite{ahh_nboo}). Code~18 (resp., Code~19) is the result of optimizing Code~16 (resp., Code~17) for the AWGN channel by attempting to remove the dominant EASs $(4, 4, 4, 6, 0)$ (only from Code~17), $(6, 4, 4, 10, 0)$, $(6, 6, 6, 9, 0)$, and $(8, 2, 2, 15, 0)$ using the WCM framework.

\begin{figure}
\vspace{-0.5em}
\center
\includegraphics[trim={0.1in 0.0in 0.0in 0.2in},clip,width=3.6in]{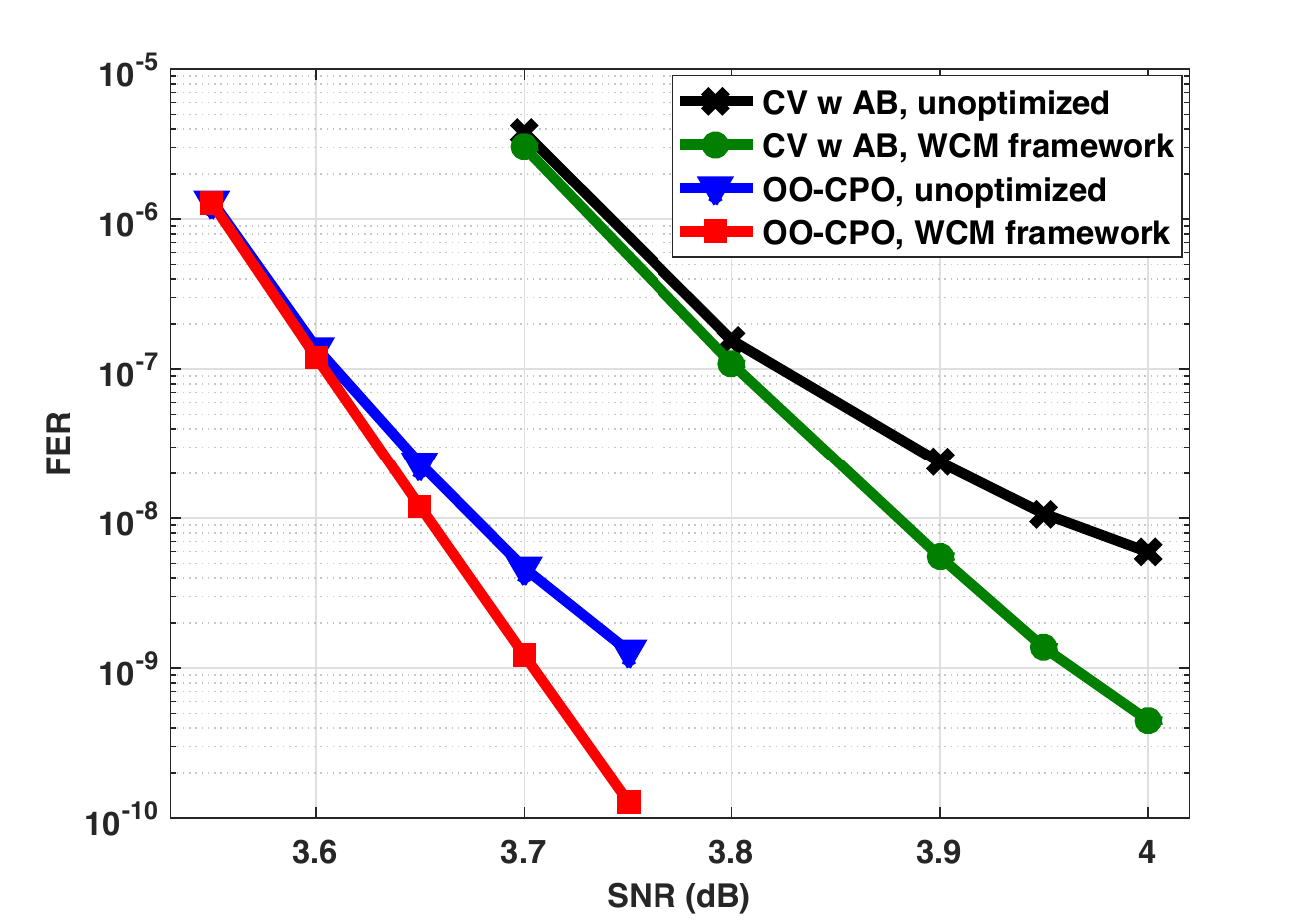}
\vspace{-1.5em}
\caption{Simulation results over the AWGN channel for SC~Codes~16 (CV, unoptimized), 17 (OO-CPO, unoptimized), 18 (CV, WCM framework), and 19 (OO-CPO, WCM framework). The four codes have $\gamma=4$.}
\label{Figure_21}
\vspace{-0.2em}
\end{figure}

Fig.~\ref{Figure_21} shows that the SC codes optimized using the WCM framework outperform the unoptimized SC codes by more than $1$ order of magnitude over the AWGN channel. Again, note that this significant performance gain is achieved despite the unlabeled Tanner graphs of Codes~16 and 18 (resp., Codes~17 and 19) both being designed using the same technique. An important observation is that despite the very good performance of Code~17 (the unoptimized code is designed using the OO-CPO technique), optimizing Code~17 using the WCM framework to reach Code~19 still achieves over $1$ order of magnitude performance gain. In the caption of Fig.~\ref{Figure_21} we precede the names of Codes~16, 17, 18, and 19 with ``SC'' for clarity.

%%%%%%%%%%%%%%%%%%%%%%%%%%%%%%
\section{Conclusion}\label{sec_conc}

In this paper, we have provided a theoretical analysis of a general combinatorial framework for optimizing non-binary graph-based codes. In particular, we proved the optimality of the WCM framework, and we demonstrated its efficiency by comparing the number of matrices it operates on with that number in a suboptimal idea. We have also detailed the theory behind the removal of a GAST; we discussed the dimension of the null space of a WCM and the minimum number of edge weight changes needed to remove a GAST. Furthermore, we proposed new combinatorial objects, OSTs, and showed how to extend the WCM framework to remove them and achieve additional performance gains for NB-LDPC codes with even column weights. On the applications side, the WCM framework was applied to different codes over a variety of channels with different characteristics, where performance gains of at least $1$ order, and up to nearly $2.5$ orders, of magnitude were achieved. A notable extension of the WCM framework was to use it in optimizing spatially-coupled codes for multiple channels. We believe that this framework will serve as an effective code optimization tool for emerging multi-dimensional storage devices, e.g., 3-D Flash and two-dimensional magnetic recording (TDMR) devices.

%%%%%%%%%%%%%%%%%%%%%%%%%%%%%%
\begin{appendices}
\section{Finding the WCMs of a Given GAST}\label{sec_appa}

The steps of \cite[Algorithm~1]{ahh_jsac} are:
\begin{enumerate}
\item \textbf{Input:} Tanner graph $G_s$ of the GAST $s$, with edge weights over GF($q$), from which the matrix $\bold{A}$ is formed.
\item Set the maximum number of nested \textbf{for} loops, loop\_max.
\item Mark all the CNs $\in (\mathcal{T} \cup \mathcal{H})$ as satisfied. \textit{(CNs $\in \mathcal{O}$ are always unsatisfied.)}
\item Check if $\exists$ in $G_s$ at least one degree-$2$ CN connecting two VNs, each is connected to $> \left \lceil \frac{\gamma+1}{2} \right \rceil$ CNs that are marked as satisfied.
\item \textbf{if} $\nexists$ any of them \textbf{then}
\item \hspace{2ex} $\exists$ only one $(\ell-d_1) \times a$ WCM. Extract it by removing all the rows corresponding to degree-$1$ CNs from the matrix $\bold{A}$.
\item \hspace{2ex} Go to 26.
\item \textbf{else}
\item \hspace{2ex} Count such CNs (that satisfy the condition in 4), save the number in $u^0$, and save their indices (the indices of their rows in $\bold{A}$) in $\bold{y}^0=[y^0(1) \text{ } y^0(2) \text{ } \dots \text{ } y^0(u^0)]^\textup{T}$.
\item \textbf{end if}
\item Compute $b_{\textup{ut}}$ from (\ref{eq_but}). If $b_{\textup{ut}}=1$, go to 25.
\item \textbf{for} $i_1 \in \{1, 2, \dots, u^0\}$ \textbf{do} \textit{ (Level $1$)}
\item \hspace{2ex} Remove the marking performed in levels $\geq 1$, and mark the selected CN $c_{y^0(i_1)}$ as unsatisfied.
\item \hspace{2ex} {Redo the counting in 9, but save in $u^1_{i_1}$ ($< u^0$) and $\bold{y}^1_{i_1}$ (instead of $u^0$ and $\bold{y}^0$, resp.).}
\item \hspace{2ex} If $b_{\textup{ut}}=2$ $\parallel$ $u^1_{i_1}=0$, go to 12.
\item \hspace{2ex} \textbf{for} $i_2 \in \{1, 2, \dots, u^1_{i_1}\}$ \textbf{do} \textit{ (Level $2$)}
\item \hspace{4ex} Remove the marking performed in levels $\geq 2$, and mark the selected CN $c_{y^1_{i_1}(i_2)}$ as unsatisfied.
\item \hspace{4ex} {Redo the counting in 9, but save in $u^2_{i_1,i_2}$ ($<u^1_{i_1}$) and $\bold{y}^2_{i_1,i_2}$.}
\item \hspace{4ex} If $b_{\textup{ut}}=3$ $\parallel$ $u^2_{i_1,i_2}=0$, go to 16.
\item \hspace{4ex} $\dots$
\item \hspace{4ex} {The lines from 16 to 19 are repeated (loop\_max$-2$) times, with  the nested (loop\_max$-2$) \textbf{for} loops executed over the running indices $i_3, i_4, \dots, i_{\text{loop\_max}}$.}
\item \hspace{4ex} $\dots$
\item \hspace{2ex} \textbf{end for}
\item \textbf{end for}
\item Obtain the WCMs via the indices in the $\bold{y}$ arrays. In particular, by removing permutations of the rows corresponding to  $c_{y^0(i_1)}, c_{y^1_{i_1}(i_2)}, \dots, c_{y^{b_{\textup{ut}}-1}_{i_1,i_2, \dots, i_{b_{\textup{ut}}-1}}(i_{b_{\textup{ut}}})}$, and the degree-$1$ CNs from $\bold{A}$, all the WCMs are reached.
\item Eliminate all the repeated WCMs to reach the final set of WCMs, $\mathcal{W}$, where $t=\vert{\mathcal{W}}\vert$.
\item \textbf{Output:} The set $\mathcal{W}$ of all WCMs of the GAST.
\end{enumerate}

%%%%%%%%%%%%%%%%%%%%%%%%%%%%%%
\section{Optimizing NB-LDPC Codes by Reducing the Number of GASTs}\label{sec_appb}

The steps of \cite[Algorithm~2]{ahh_jsac} are:
\begin{enumerate}
\item \textbf{Input:} Tanner graph $G_{\textup{C}}$ of the NB-LDPC code with edge weights over GF($q$).
\item Using initial simulations and combinatorial techniques (e.g., \cite{bani_cycle}), determine $\mathcal{G}$, the set of GASTs to be removed.
\item Let $\mathcal{X}$ be the set of GASTs in $\mathcal{G}$ that cannot be removed, and initialize it with $\varnothing$.
\item Let $\mathcal{P}$ be the set of GASTs in $\mathcal{G}$ that have been processed, and initialize it with $\varnothing$.
\item Sort the GASTs in $\mathcal{G}$ according to their sizes (parameter $a$) from the smallest to the largest.
\item Start from the smallest GAST (smallest index).
\item {\textbf{for} every GAST $s \in \mathcal{G} \setminus \mathcal{P}$ \textbf{do}}
\item \hspace{2ex} If the unlabeled configuration of $s$ does not satisfy the {unlabeled GAST} conditions in Definitions~\ref{def_ugas} and \ref{def_gast}, skip $s$ and go to 7.
\item \hspace{2ex} Determine the minimum number of edge weight changes needed to remove the GAST $s$, $E_{\textup{GAST},\textup{min}}$, by using \cite[Lemma~2]{ahh_bas} (see also Lemma~\ref{lem_emin_gas} in this paper).
\item \hspace{2ex} Extract the subgraph $G_s$ of the GAST $s$, from $G_{\textup{C}}$.
\item \hspace{2ex} Use \cite[Algorithm~1]{ahh_jsac} to determine the set $\mathcal{W}$ of all WCMs of $s$ ($\vert{\mathcal{W}}\vert=t$).
\item \hspace{2ex} \textbf{for} $h \in \{1, 2, \dots, t\}$ \textbf{do}
\item \hspace{4ex} Find the null space $\mathcal{N}(\bold{W}^{\textup{cm}}_h)$ of the $h$th WCM.
\item \hspace{4ex} \textbf{if} (\ref{eq_rem_cond}) is satisfied (i.e., the WCM already has broken weight conditions) \textbf{then}
\item \hspace{6ex} Go to 12.
\item \hspace{4ex} \textbf{else}
\item \hspace{6ex} Keep track of the changes already performed in $G_s$. \textit{(The total number of changes to remove the GAST should be as close as possible to $E_{\textup{GAST},\textup{min}}$.)}
\item \hspace{6ex} Determine the smallest set of edge weight changes in $G_s$ needed to achieve (\ref{eq_rem_cond}) for the $h$th WCM, without violating (\ref{eq_rem_cond}) for WCMs prior to the $h$th.
\item \hspace{6ex} If this set of edge weight changes does not undo~the removal of any GAST $\in \mathcal{P} \setminus \mathcal{X}$, perform these changes in $G_s$ and go to 12.
\item \hspace{6ex} \textbf{if} $\nexists$ more edge weights to execute 18 and 19 \textbf{then}
\item \hspace{8ex} Add GAST $s$ to the set $\mathcal{X}$ and  go to 27.
\item \hspace{6ex} \textbf{else} Go to 18 to determine a new set of changes.
\item \hspace{6ex} \textbf{end if}.
\item \hspace{4ex} \textbf{end if}
\item \hspace{2ex} \textbf{end for}
\item \hspace{2ex} Update $G_{\textup{C}}$ by the changes performed in $G_s$.
\item \hspace{2ex} Add GAST $s$ to the set $\mathcal{P}$.
\item \hspace{2ex} If $\mathcal{P} \neq \mathcal{G}$, go to 7 to pick the next smallest GAST.
\item \textbf{end for}
\item If $\mathcal{X}=\varnothing$, then all the GASTs in $\mathcal{G}$ have been removed. Otherwise, only the remaining GASTs in $\mathcal{X}$ cannot be removed.
\item \textbf{Output:} Updated Tanner graph $G_{\textup{C}}$ of the optimized NB-LDPC code with edge weights over GF($q$).
\end{enumerate}

%%%%%%%%%%%%%%%%%%%%%%%%%%%%%%
\section{Null Spaces of WCMs of GASTs with $b=d_1$}\label{sec_appc}

In this appendix, we investigate the null spaces, along with their dimensions, of WCMs that belong to GASTs with $b=d_1$.

\begin{remark}\label{rem_6}
There are few configurations that can be categorized as $(a, b_{\textup{g}}, d_1, d_2, d_3)$ GASTs, with $b_{\textup{g}} \in \{b_i, b_{ii}, \dots\}$ ($b_{\textup{g}}$ is not unique). In other words, it is possible to have a configuration which is an $(a, b_i, d_1, d_2, d_3)$ GAST for some set of VN vectors, and it is an $(a, b_{ii}, d_1, d_2, d_3)$ GAST for another set of VN vectors, where $b_i \neq b_{ii}$. For example, the configuration in Fig.~\ref{Figure_9}(a), with $w_{1,1}=w_{6,1}=1$, is a $(6, 0, 0, 9, 0)$ GAST for the vector $[\alpha \text{ } 1 \text{ } 1 \text{ } \alpha \text{ } 1 \text{ } 1]^\textup{T}$ (along with others), while the same configuration is a $(6, 3, 0, 9, 0)$ GAST for the vector $[\alpha^2 \text{ } 1 \text{ } 1 \text{ } \alpha \text{ } \alpha \text{ } \alpha]^\textup{T}$ (along with others). In cases like these, we identify the configuration with its \textbf{smallest} $b_{\textup{g}}$ in the set $\{b_i, b_{ii}, \dots\}$. Thus, we identify the configuration in Fig.~\ref{Figure_9}(a) as a $(6, 0, 0, 9, 0)$ GAST as mentioned in Example~\ref{ex_9}. Note that this situation is not problematic for our GAST removal process, as our goal is to convert the GAST into another object $\notin \mathcal{Z}$, where $\mathcal{Z}$ is the set of all $(a, b', d_1, d_2, d_3)$, $d_1 \leq b' \leq b_{\textup{max}}$.
\end{remark}

\begin{corollary}\label{cor_b_d1}
An $(a, d_1, d_1, d_2, d_3)$ GAST, which is a GAST with $b=d_1$, has unbroken weight conditions for all its WCMs.
\end{corollary}

\begin{IEEEproof}
From \cite[Lemma~1]{ahh_jsac} (see Section~\ref{sec_sum}), a GAST that has $b=d_1$ must have the particular $\bold{W}^{\textup{z}}$ matrix of size $(\ell-d_1) \times a$ (extracted by removing the rows of all degree-$1$ CNs from $\bold{A}$) with unbroken weight conditions, i.e., $\exists \text{ } \bold{v}=[v_1 \text{ } v_2 \text{ } \dots \text{  } v_a]^\textup{T} \in \mathcal{N}(\bold{W}^{z})$ s.t. $v_f \neq 0$, $\forall f \in \{1, 2, \dots, a\}$. Since by definition of WCMs, each $\bold{W}^{\textup{cm}}_h$, $\forall h \in \{1, 2, \dots, t\}$, is a submatrix of this particular $\bold{W}^{\textup{z}}$, it follows that $\mathcal{N}(\bold{W}^{z}) \subseteq \mathcal{N}(\bold{W}^{\textup{cm}}_h)$, $\forall h$. In other words, $\exists \text{ } \bold{v}=[v_1 \text{ } v_2 \text{ } \dots \text{  } v_a]^\textup{T} \in \mathcal{N}(\bold{W}^{\textup{cm}}_h)$, $\forall h$, s.t. $v_f \neq 0$, $\forall f \in \{1, 2, \dots, a\}$.
\end{IEEEproof}

Corollary~\ref{cor_b_d1} highlights that each WCM of an $(a, d_1, d_1, d_2, d_3)$ GAST has $\textup{dim}\left (\mathcal{N}(\bold{W}^{\textup{cm}}_h) \right ) > 0$, $\forall h$, which is a consequence of all of them having unbroken weight conditions. The following example further discusses the null spaces of WCMs belonging to GASTs with $b=d_1$.

\begin{example}\label{ex_11}
We once more return to the $(6, 0, 0, 9, 0)$ GAST in Fig.~\ref{Figure_9}(a), with $w_{1,1}=w_{6,1}=1$. The configuration is a $(6, 0, 0, 9, 0)$ GAST because the vector $\bold{v} = [\alpha \text{ } 1 \text{ } 1 \text{ } \alpha \text{ } 1 \text{ } 1]^\textup{T}$, for example, is in the null space of the $9 \times 6$ matrix $\bold{W}^{\textup{z}}=\bold{A}$ (note that there are no degree-$1$ CNs in this configuration). The null spaces of the $10$ WCMs, extracted according to Example~\ref{ex_9}, of that GAST are detailed below:
\begin{align}\label{eq_ex11_1}
\mathcal{N}(\bold{W}^{\textup{cm}}_1) &= \mathcal{N}(\bold{W}^{\textup{cm}}_3) = \mathcal{N}(\bold{W}^{\textup{cm}}_4) = \mathcal{N}(\bold{W}^{\textup{cm}}_5) \nonumber \\ &= \mathcal{N}(\bold{W}^{\textup{cm}}_6) = \mathcal{N}(\bold{W}^{\textup{cm}}_7) = \mathcal{N}(\bold{W}^{\textup{cm}}_8) = \mathcal{N}(\bold{W}^{\textup{cm}}_9) \nonumber \\ &= \mathcal{N}(\bold{W}^{\textup{cm}}_{10}) = \textup{span}\{[\alpha \text{ } 1 \text{ } 1 \text{ } \alpha \text{ } 1 \text{ } 1]^\textup{T}\} \text{ and} \nonumber \\
\mathcal{N}(\bold{W}^{\textup{cm}}_2)&=\textup{span}\{[\alpha \text{ } 0 \text{ } 0 \text{ } 0 \text{ } 1 \text{ } 1]^\textup{T}, [0 \text{ } 1 \text{ } 1 \text{ } \alpha \text{ } 0 \text{ } 0]^\textup{T}\}.
\end{align}
We now turn our attention to the $(6, 2, 2, 5, 2)$ GAST in Fig.~\ref{Figure_10}(a), with $w_{1,2}=\alpha^2$. The configuration is a $(6, 2, 2, 5, 2)$ GAST because the vector $\bold{v} = [\alpha^2 \text{ } 1 \text{ } 1 \text{ } 1 \text{ } \alpha \text{ } \alpha]^\textup{T}$, for example, is in the null space of the $7 \times 6$ matrix $\bold{W}^{\textup{z}}$, extracted by removing the rows of the $2$ degree-$1$ CNs (that are $c_8$ and $c_9$) from $\bold{A}$. The null spaces of the $2$ WCMs, extracted according to Example~\ref{ex_10}, of that GAST are:
\begin{align}\label{eq_ex11_2}
\mathcal{N}(\bold{W}^{\textup{cm}}_1)&= \textup{span}\{[\alpha^2 \text{ } 1 \text{ } 1 \text{ } 1 \text{ } \alpha \text{ } \alpha]^\textup{T}\} \text{ and} \nonumber \\
\mathcal{N}(\bold{W}^{\textup{cm}}_2)&=\textup{span}\{[0 \text{ } 1 \text{ } 1 \text{ } \alpha^2 \text{ } 1 \text{ } 0]^\textup{T}, [1 \text{ } 1 \text{ } 1 \text{ } 0 \text{ } 0 \text{ } \alpha^2]^\textup{T}\}.
\end{align}
It is clear that, all the WCMs for both GASTs have unbroken weight conditions, which is expected according to Corollary~\ref{cor_b_d1} (both GASTs have $b=d_1$). 
\end{example}

Note that in Example~\ref{ex_11}, all the WCMs have $p_h = \textup{dim}\left (\mathcal{N}(\bold{W}^{\textup{cm}}_h) \right ) = \delta_h$, except for one WCM; $\bold{W}^{\textup{cm}}_2$ of the $(6, 2, 2, 5, 2)$ GAST, which is a short WCM, has $\textup{dim}\left (\mathcal{N}(\bold{W}^{\textup{cm}}_2) \right ) = 2 > \delta_2 = 1$. As mentioned before, it is typically the case that $p_h=\textup{dim}\left (\mathcal{N}(\bold{W}^{\textup{cm}}_h) \right ) = \delta_h$.

\end{appendices}

%%%%%%%%%%%%%%%%%%%%%%%%%%%%%%
\section*{Acknowledgement}

The authors thank the associate editor, Prof. Michael Lentmaier, for the handling of the paper and for the constructive feedback that has improved the paper.

The research was supported in part by UCLA dissertation year fellowship, in part by a grant from ASTC-IDEMA, in part by NSF Grant CCF-CAREER 1150212, and in part by NSF Grant CCF-BSF 1718369.

%%%%%%%%%%%%%%%%%%%%%%%%%%%%%%


\begin{thebibliography}{13}

\bibitem{gal_th}
R. G. Gallager, \textit{Low-Density Parity-Check Codes.} Cambridge, MA: MIT Press, 1963.

\bibitem{rich_urb1}
T. J. Richardson and R. L. Urbanke, ``The capacity of low-density parity-check codes under message-passing decoding,'' \emph{IEEE Trans. Inf. Theory}, vol. 47, no. 2, pp. 599--618, Feb. 2001.

\bibitem{rich_urb2}
T. J. Richardson, M. A. Shokrollahi, and R. L. Urbanke, ``Design of capacity-approaching irregular low-density parity-check codes,'' \emph{IEEE Trans. Inf. Theory}, vol. 47, no. 2, pp. 619--637, Feb. 2001.

\bibitem{ahh_smm}
A. H. Hareedy and M. M. Khairy, ``Selective max-min algorithm for low-density parity-check decoding,''  \emph{IET Commun.}, vol. 7, no. 1, pp. 65--70, Jan. 2013.

\bibitem{ahh_jsac}
A. Hareedy, C. Lanka, and L. Dolecek, ``A general non-binary LDPC code optimization framework suitable for dense Flash memory and magnetic storage,''  \emph{IEEE J. Sel. Areas Commun.}, vol. 34, no. 9, pp. 2402--2415, Sep. 2016.

\bibitem{aslm_fl}
C. A. Aslam, Y. L. Guang, and K. Cai, ``Read and write voltage signal optimization for multi-level-cell (MLC) NAND Flash memory,'' \emph{IEEE Trans. Commun.}, vol. 64, no. 4, pp. 1613--1623, Apr. 2016.

\bibitem{maeda_fl}
Y. Maeda and H. Kaneko, ``Error control coding for multilevel cell Flash memories using nonbinary low-density parity-check codes,'' in \emph{Proc. 24th IEEE Int. Symp. Defect and Fault Tolerance in VLSI Systems (DFT)}, Chicago, IL, USA, Oct. 2009, pp. 367--375.

\bibitem{jia_sym}
J. Wang, K. Vakilinia, T.-Y. Chen, T. Courtade, G. Dong, T. Zhang, H. Shankar, and R. Wesel, ``Enhanced precision through multiple reads for LDPC decoding in flash memories,''  \emph{IEEE J. Sel. Areas Commun.}, vol. 32, no. 5, pp. 880--891, May 2014.

\bibitem{kin_sym}
K. Ho, C. Chen, and H. Chang, ``A 520k (18900, 17010) array dispersion LDPC decoder architectures for NAND Flash memory,'' \emph{IEEE Trans. VLSI Systems}, vol. 24, no. 4, pp. 1293--1304, Apr. 2016.

\bibitem{ahh_bas}
A. Hareedy, B. Amiri, R. Galbraith, and L. Dolecek, ``Non-binary LDPC codes for magnetic recording channels: error floor analysis and optimized code design,'' \emph{IEEE Trans. Commun.}, vol. 64, no. 8, pp. 3194--3207, Aug. 2016.

\bibitem{shafa}
S. Srinivasa, Y. Chen, and S. Dahandeh, ``A communication-theoretic framework for 2-DMR channel modeling: performance evaluation of coding and signal processing methods,'' \emph{IEEE Trans. Magn.}, vol. 50, no. 3, pp. 6--12, Mar. 2014.

\bibitem{yuval1}
R. Cohen and Y. Cassuto, ``Iterative decoding of LDPC codes over the $q$-ary partial erasure channel,'' \emph{IEEE Trans. Inf. Theory}, vol. 62, no. 5, pp. 2658--2672, May 2016.

\bibitem{yuval2}
Y. Cassuto and A. Shokrollahi, ``LDPC codes for 2D arrays,'' \emph{IEEE Trans. Inf. Theory}, vol. 60, no. 6, pp. 3279--3291, Jun. 2014.

\bibitem{lara_floor}
L. Dolecek, P. Lee, Z. Zhang, V. Anantharam, B. Nikolic, and M. Wainwright, ``Predicting error floors of structured LDPC codes: deterministic bounds and estimates,'' \emph{IEEE J. Sel. Areas Commun.}, vol. 27, no. 6, pp. 908--917, Aug. 2009.

\bibitem{bani_est}
H. Xiao, A. Banihashemi, and M. Karimi, ``Error rate estimation of low-density parity-check codes decoded by quantized soft-decision iterative algorithms,'' \emph{IEEE Trans. Commun.}, vol. 61, no. 2, pp. 474--484, Feb. 2013.

\bibitem{b_ryan}
Y. Han and W. Ryan, ``Low-floor detection/decoding of LDPC-coded partial response channels,'' \emph{IEEE J. Sel. Areas Commun.}, vol. 28, no. 2, pp. 252--260, Feb. 2010.

\bibitem{hu_sim}
X. Hu, Z. Li, B. V. K. V. Kumar, and R. Barndt, ``Error floor estimation of long LDPC codes on magnetic recording channels,'' \emph{IEEE Trans. Magn.}, vol. 46, no. 6, pp. 1836--1839, Jun. 2010.

\bibitem{tom_ferr}
A. Tomasoni, S. Bellini, and M. Ferrari, ``Thresholds of absorbing sets in low-density parity-check codes,'' \emph{IEEE Trans. Commun.}, vol. 65, no. 8, pp. 3238--3249, Aug. 2017.

\bibitem{lara_as}
L. Dolecek, Z. Zhang, V. Anantharam, M. Wainwright, and B. Nikolic, ``Analysis of absorbing sets and fully absorbing sets of array-based LDPC codes,'' \emph{IEEE Trans. Inf. Theory}, vol. 56, no. 1, pp. 181--201, Jan. 2010.

\bibitem{behzad_elem}
B. Amiri, J. Kliewer, and L. Dolecek, ``Analysis and enumeration of absorbing sets for non-binary graph-based codes,'' \emph{IEEE Trans. Commun.}, vol. 62, no. 2, pp. 398--409, Feb. 2014.

\bibitem{decl_nb}
C. Poulliat, M. Fossorier, and D. Declercq, ``Design of regular $(2,d_c)$-LDPC codes over GF($q$) using their binary images,'' \emph{IEEE Trans. Commun.}, vol. 56, no. 10, pp. 1626--1635, Oct. 2008.

\bibitem{olgica1}
O. Milenkovic, E. Soljanin, and P. Whiting, ``Asymptotic spectra of trapping sets in regular and irregular LDPC code ensembles,'' \emph{IEEE Trans. Inf. Theory}, vol. 53, no. 1, pp. 39--55, Jan. 2007.

\bibitem{vas_trap1}
M. Ivkovic, S. K. Chilappagari, and B. Vasic, ``Eliminating trapping sets in low-density parity-check codes by using Tanner graph covers,'' \emph{IEEE Trans. Inf. Theory}, vol. 54, no. 8, pp. 3763--3768, Aug. 2008.

\bibitem{jia_cyc}
J. Wang, L. Dolecek, and R. Wesel, ``The cycle consistency matrix approach to absorbing sets in separable circulant-based LDPC codes,'' \emph{IEEE Trans. Inf. Theory}, vol. 59, no. 4, pp. 2293--2314, Apr. 2013.

\bibitem{olgica2}
A. McGregor and O. Milenkovic, ``On the Hardness of Approximating Stopping and Trapping Sets,'' \emph{IEEE Trans. Inf. Theory}, vol. 56, no. 4, pp. 1640--1650, Apr. 2010.

\bibitem{siegel_flr}
B. K. Butler and P. H. Siegel, ``Error floor approximation for LDPC codes in the AWGN channel,'' \emph{IEEE Trans. Inf. Theory}, vol. 60, no. 12, pp. 7416--7441, Dec. 2014.

\bibitem{bani_trap}
M. Karimi and A. Banihashemi, ``On characterization of elementary trapping sets of variable-regular LDPC codes,'' \emph{IEEE Trans. Inf. Theory}, vol. 60, no. 9, pp. 5188--5203, Sep. 2014.

\bibitem{vas_trap2}
D. V. Nguyen, S. K. Chilappagari, M. W. Marcellin, and B. Vasic, ``On the construction of structured LDPC codes free of small trapping sets,'' \emph{IEEE Trans. Inf. Theory}, vol. 58, no. 4, pp. 2280--2302, Apr. 2012.

\bibitem{shu_flr1}
Q. Diao, Y. Y. Tai, S. Lin, and K. Abdel-Ghaffar, ``LDPC codes on partial geometries: construction, trapping set structure, and puncturing,'' \emph{IEEE Trans. Inf. Theory}, vol. 59, no. 12, pp. 7898--7914, Dec. 2013.

\bibitem{shu_flr2}
Q. Huang, Q. Diao, S. Lin, and K. Abdel-Ghaffar, ``Cyclic and quasi-cyclic LDPC codes on constrained parity-check matrices and their trapping sets,'' \emph{IEEE Trans. Inf. Theory}, vol. 58, no. 5, pp. 2648--2671, May 2012.

\bibitem{ahh_glc}
A. Hareedy, B. Amiri, R. Galbraith, S. Zhao, and L. Dolecek, ``Non-binary LDPC code optimization for partial-response channels,'' in \emph{Proc. IEEE Global Commun. Conf. (GLOBECOM)}, San Diego, CA, USA, Dec. 2015, pp. 1--6.

\bibitem{vas_prc}
B. Vasic and E. Kurtas, \textit{Coding and Signal Processing for Magnetic Recording Systems.} CRC Press, 2005.

\bibitem{cola_pr}
G. Colavolpe and G. Germi, ``On the application of factor graphs and the sum-product algorithm to ISI channels,'' \emph{IEEE Trans. Commun.}, vol. 53, no. 5, pp. 818--825, May 2005.

\bibitem{mit_nl}
T. Parnell, N. Papandreou, T. Mittelholzer, and H. Pozidis, ``Modelling of the threshold voltage distributions of sub-20nm NAND flash memory,'' in \emph{Proc. IEEE Global Commun. Conf. (GLOBECOM)}, Austin, TX, USA, Dec. 2014, pp. 2351--2356.

\bibitem{cai_fl}
Y. Cai, E. Haratsch, O. Mutlu, and K. Mai, ``Threshold voltage distribution in MLC NAND flash memory: Characterization, analysis, and modeling,'' in \emph{Proc. Design, Autom., Test Eur. Conf. Exhibition (DATE)}, Grenoble, France, Mar. 2013, pp. 1285--1290.

\bibitem{kud_sc}
S. Kudekar, T. J. Richardson, and R. L. Urbanke, ``Spatially coupled ensembles universally achieve capacity under belief propagation,'' \emph{IEEE Trans. Inf. Theory}, vol. 59, no. 12, pp. 7761--7813, Dec. 2013.

\bibitem{lent_asy}
M. Lentmaier, A. Sridharan, D. J. Costello, and K. S. Zigangirov, ``Iterative decoding threshold analysis for LDPC convolutional codes,'' \emph{IEEE Trans. Inf. Theory}, vol. 56, no. 10, pp. 5274--5289, Oct. 2010.

\bibitem{andr_asy}
I. Andriyanova and A. Graell i Amat, ``Threshold saturation for nonbinary SC-LDPC codes on the binary erasure channel,'' \emph{IEEE Trans. Inf. Theory}, vol. 62, no. 5, pp. 2622--2638, May 2016.

\bibitem{pus_sc}
A. E. Pusane, R. Smarandache, P. O. Vontobel, and D. J. Costello, ``Deriving good LDPC convolutional codes from LDPC block codes,'' \emph{IEEE Trans. Inf. Theory}, vol. 57, no. 2, pp. 835--857, Feb. 2011.

\bibitem{olm_sc}
P. M. Olmos and R. L. Urbanke, ``A scaling law to predict the finite-length performance of spatially-coupled LDPC codes,'' \emph{IEEE Trans. Inf. Theory}, vol. 61, no. 6, pp. 3164--3184, Jun. 2015.

\bibitem{iye_sc}
A. R. Iyengar, M. Papaleo, P. H. Siegel, J. K. Wolf, A. Vanelli-Coralli, and G. E. Corazza, ``Windowed decoding of protograph-based LDPC convolutional codes over erasure channels,'' \emph{IEEE Trans. Inf. Theory}, vol. 58, no. 4, pp. 2303--2320, Apr. 2012.

\bibitem{bani_cycle}
M. Karimi and A. H. Banihashemi, ``Efficient algorithm for finding dominant trapping sets of LDPC codes,'' \emph{IEEE Trans. Inf. Theory}, vol. 58, no. 11, pp. 6942--6958, Nov. 2012.

\bibitem{dec_fft}
D. Declercq and M. Fossorier,  ``Decoding algorithms for nonbinary LDPC codes over GF($q$),'' \emph{IEEE Trans. Commun.}, vol. 55, no. 4, pp. 633--643, Apr. 2007.

\bibitem{lara_prot}
L. Dolecek, D. Divsalar, Y. Sun, and B. Amiri, ``Non-binary protograph-based LDPC codes: enumerators, analysis, and designs,'' \emph{IEEE Trans. Inf. Theory}, vol. 60, no. 7, pp. 3913--3941, Jul. 2014.

\bibitem{baz_qc}
A. Bazarsky, N. Presman, and S. Litsyn, ``Design of non-binary quasi-cyclic LDPC codes by ACE optimization,'' in \emph{Proc. IEEE Inf. Theory Workshop (ITW)}, Sevilla, Spain, Sep. 2013, pp. 1--5.

\bibitem{dec_yuta}
Y. Toriyama and D. Markovic,  ``A 2.267-Gb/s, 93.7-pJ/bit non-binary LDPC decoder with logarithmic quantization and dual-decoding algorithm scheme for storage applications,'' \emph{IEEE J. Solid-State Circuits}, vol. 53, no. 8, pp.  2378--2388, Aug. 2018.

\bibitem{cai_defn}
Y. Cai, G. Yalcin, O. Mutlu, E. Haratsch, A. Cristal, O. Unsal, and K. Mai, ``Flash correct-and-refresh: Retention-aware error management for increased Flash memory lifetime,'' in \emph{Proc. IEEE 30th IEEE Int. Conf. Comput. Des. (ICCD)}, Montreal, Quebec, Canada, Oct. 2012, pp. 94--101.

\bibitem{naj_asy}
N. Ul Hassan, M. Lentmaier, I. Andriyanova, and G. P. Fettweis, ``Improving code diversity on block-fading channels by spatial coupling,'' in \emph{Proc. IEEE Int. Symp. Inf. Theory (ISIT)}, Honolulu, HI, USA, Jun. 2014, pp. 2311--2315.

\bibitem{amna_asy}
A. Piemontese, A. Graell i Amat, and G. Colavolpe, ``Nonbinary spatially-coupled LDPC codes on the binary erasure channel,'' in \emph{Proc. IEEE Int. Conf. Commun. (ICC)}, Budapest, Hungary, Jun. 2013, pp. 3270--3274.

\bibitem{mitch_fl}
D. G. M. Mitchell, M. Lentmaier, and D. J. Costello, ``Spatially coupled LDPC codes constructed from protographs,'' \emph{IEEE Trans. Inf. Theory}, vol. 61, no. 9, pp. 4866--4889, Sep. 2015.

\bibitem{yix_fl}
Y. Xie, L. Yang, P. Kang, and J. Yuan, ``Euclidean geometry-based spatially coupled LDPC codes for storage,'' \emph{IEEE J. Sel. Areas Commun.}, vol. 34, no. 9, pp. 2498--2509, Sep. 2016.

\bibitem{scb_ext1}
D. G. M. Mitchell, L. Dolecek, and D. J. Costello, Jr., ``Absorbing set characterization of array-based spatially coupled LDPC codes,'' in \emph{Proc. IEEE Int. Symp. Inf. Theory (ISIT)}, Honolulu, HI, USA, Jun. 2014, pp. 886--890.

\bibitem{scb_ext2}
D. G. M. Mitchell and E. Rosnes, ``Edge spreading design of high rate array-based SC-LDPC codes,'' in \emph{Proc. IEEE Int. Symp. Inf. Theory (ISIT)}, Aachen, Germany, Jun. 2017, pp. 2940--2944.

\bibitem{scb_ext3}
A. Beemer, S. Habib, C. A. Kelley, and J. Kliewer, ``A generalized algebraic approach to optimizing SC-LDPC codes,'' in \emph{Proc. 55th Annual Allerton Conf. Commun., Control, and Computing}, Monticello, IL, USA, Oct. 2017, pp. 672--679.

\bibitem{homa_boo}
H. Esfahanizadeh, A. Hareedy, and L. Dolecek, ``Finite-length construction of high performance spatially-coupled codes via optimized partitioning and lifting,'' accepted at \emph{IEEE Trans. Commun.}, doi: 10.1109/TCOMM.2018.2867493, Aug. 2018.

\bibitem{irina_sc}
I. E. Bocharova, B. D. Kudryashov, and R. Johannesson, ``Searching for binary and nonbinary block and convolutional LDPC codes,'' \emph{IEEE Trans. Inf. Theory}, vol. 62, no. 1, pp. 163--183, Jan. 2016.

\bibitem{homa_sc}
H. Esfahanizadeh, A. Hareedy, and L. Dolecek, ``Spatially-coupled codes optimized for magnetic recording applications,'' \emph{IEEE Trans. Magn.}, vol. 53, no. 2, pp. 1--11, Feb. 2016.

\bibitem{ahh_nboo}
A. Hareedy, H. Esfahanizadeh, and L. Dolecek, ``High performance non-binary spatially-coupled codes for flash memories,'' in \emph{Proc. IEEE Inf. Theory Workshop (ITW)}, Kaohsiung, Taiwan, Nov. 2017, pp. 229--233.

\bibitem{ahh_nboo2}
A. Hareedy, H. Esfahanizadeh, A. Tan, and L. Dolecek, ``Spatially-coupled code design for partial-response channels: optimal object-minimization approach,'' accepted at \emph{IEEE Global Commun. Conf. (GLOBECOM)}, Abu Dhabi, UAE, Dec. 2018. [Online]. Available: http://arxiv.org/abs/1804.05504

\bibitem{gl_tolga}
T. Duman and E. Kurtas, ``Comprehensive performance investigation of turbo codes over high density magnetic recording channels,'' in \emph{Proc. IEEE Global Telecommun. Conf. (GLOBECOM)}, Rio de Janeiro, Brazil, Dec. 1999, pp. 744--748.

\bibitem{tom_v}
T. Souvignier, M. \"{O}berg, P. Siegel, R. Swanson, and J. Wolf, ``Turbo decoding for partial response channels,'' \emph{IEEE Trans. Commun.}, vol. 48, no. 8, pp. 1297--1308, Aug. 2000.

\bibitem{bcjr}
L. Bahl, J. Cocke, F. Jelinek, and J. Raviv, ``Optimal decoding of linear codes for minimizing symbol error rate,''  \emph{IEEE Trans. Inf. Theory}, vol. 20, pp. 284--287, Mar. 1974.

\bibitem{pdnp}
J. Moon and J. Park, ``Pattern-dependent noise prediction in signal dependent noise,''  \emph{IEEE J. Sel. Areas Commun.}, vol. 19, no. 4, pp. 730--743 , Apr. 2001.\\

\end{thebibliography}
\end{document}